
\magnification=\magstep1
\baselineskip=15pt
\overfullrule=0pt

\input epsf

\def\C{{\bf C}}

\def\G{{\cal G}}
\def\H{{\cal H}}
\def\K{{\cal K}}
\def\M{{\cal M}}
\def\N{{\cal N}}

\def\R{{\cal R}}
\def\half{{1 \over 2}}
\def\13{{\scriptstyle{1\over 3}}}
\def\23{{\scriptstyle{2\over 3}}}

\def\Im{{\rm Im}}
\def\tr{{\rm tr}}
\def\cn{{\rm cn}}
\def\sn{{\rm sn}}
\def\dn{{\rm dn}}
\def\tp{{\bf T} _ {\bf p}}
\def\uq{{\bf U} _ {\bf q}}
\def\f27{{{\bf 26}$\oplus${\bf 1}}}
\def\a{\alpha}

\def\notin{\in \!\!\!\!\! / ~}
\def\next{\hfill\break}

\rightline{UCLA/98/TEP/9}
\rightline{Columbia/Math/98}
\rightline{NSF-ITP-98-060}

\bigskip

\centerline{{\bf CALOGERO-MOSER LAX PAIRS WITH SPECTRAL PARAMETER}}

\medskip

\centerline{{\bf FOR GENERAL LIE ALGEBRAS}
\footnote*{Research supported in part by
the National Science Foundation under grants PHY-95-31023, PHY-94-07194 
and DMS-95-05399.}}

\bigskip
\bigskip
\bigskip

\centerline{{\bf Eric D'Hoker}${}^1$ 
            {\bf and D.H. Phong} ${}^2$}
\bigskip
\centerline{${}^1$ Department of Physics}
\centerline{University of California, Los Angeles, CA 90024, USA}
\centerline{Institute for Theoretical Physics}
\centerline{University of California, Santa Barbara, CA 93106, USA}
\bigskip
\centerline{${}^2$ Department of Mathematics}
\centerline{Columbia University, New York, NY 10027, USA}

\bigskip
\bigskip
\bigskip

\centerline{\bf ABSTRACT}

\bigskip

We construct a Lax pair with spectral parameter for the elliptic
Calogero-Moser Hamiltonian systems associated with each of the 
finite dimensional Lie algebras, of the classical and of the exceptional 
type. When the spectral parameter equals one of the three half periods of the
elliptic curve, our result for the classical Lie algebras reduces to one of 
the Lax pairs without spectral parameter that were known previously. These
Calogero-Moser systems are invariant under the Weyl group of the
associated untwisted affine Lie algebra. For non-simply laced Lie algebras, 
we introduce new integrable systems, naturally associated with twisted affine 
Lie algebras, and construct their Lax operators with spectral parameter
(except in the case of $G_2$).
 
\bigskip

\vfill\break

\centerline{\bf I. INTRODUCTION}

\bigskip

The recent exact solutions of four-dimensional supersymmetric gauge theories 
[1] (for a review, see for example [2]) have revealed a deep correspondence 
between these theories and integrable models [3-8].
While the existence of such a correspondence can now be established
on very general grounds [4] (see also [9]), the fundamental nature of the 
correspondence itself
has remained largely elusive. Precise matches between specific
gauge theories and integrable models have been identified in
special cases, on an {\it ad hoc} basis, and it is clearly
desirable to have a more systematic classification. Since four-dimensional
${\cal N}=2$ supersymmetric gauge theories are labelled by their gauge Lie 
algebra $\G$ and the representation $R$ of their matter hypermultiplets,
it is natural to look for integrable models associated with Lie 
algebras to obtain such a classification.

\medskip

Integrable models associated to the root system of a Lie algebra
$\G$ include the Toda and affine Toda systems [10]
$$
\eqalign{
\ddot X&=-\half\sum_{\a\in\R^*({\G})}M_{|\a|}^2e^{-\a\cdot X}\alpha,\cr
\ddot X&=-\half\sum_{\a\in\R^*(\G^{(1)})}M_{|\a|}^2e^{-\a\cdot X}\alpha,\cr}
\eqno(1.1)
$$
the elliptic, trigonometric and rational Calogero-Moser
systems [11,12]
$$
\eqalign{
\ddot x&=\half\sum_{\a\in\R({\G})}m_{|\a|}^2\wp'(\a\cdot 
x)\alpha,\cr
\ddot x&=-\sum_{\a\in\R({\G})}m_{|\a|}^2
{{\rm cosh}\over{\rm sinh}^3}(\a\cdot x)\alpha,\cr
\ddot x&=-\sum_{\a\in\R({\G})}m_{|\a|}^2(\a\cdot x)^{-3}\alpha,\cr}
\eqno(1.2)
$$ 
and the Hitchin systems [13]. 
Here $\R({\G})$ and $\R^*({\G})$ denote respectively the set of roots and the 
set of simple roots of $\G$, and $\G^{(1)}$ is the untwisted affine Lie 
algebra associated with $\G$.
For general reviews, see [14]. Very early on, the spectral curves for the pure 
$SU(2)$ Yang-Mills theory and for
the affine $SU(2)$ Toda system were recognized as identical [3]. Since then, 
many more correspondences have been established. Of particular interest are 
the one between pure Yang-Mills theories with gauge algebra $\G$ and 
affine Toda systems for $(G^{(1)})^{\vee}$ [5], 
and the one between the $SU(N)$ Yang-Mills theory 
with matter in the adjoint representation and the $SU(N)$ Hitchin system [4],
(or equivalently, the elliptic $SU(N)$ Calogero-Moser system [8]).
The advantage of an explicit correspondence with integrable models is much in 
evidence throughout [8], where an exact renormalization group type relation 
is derived for the effective prepotential function of the gauge theory in 
terms of the Calogero-Moser Hamiltonian.

\medskip

The purpose of the present paper and the two companion papers [15,16] is to 
identify/construct the integrable models corresponding to ${\cal N}=2$ 
supersymmetric Yang-Mills 
with gauge algebra $\G$ and matter in the adjoint representation of $\G$, 
where $\G$ is an arbitrary simple Lie algebra. In analogy with the $SU(N)$ 
case, a natural candidate for the integrable model corresponding to the gauge 
theory with gauge algebra $\G$ is the elliptic Calogero-Moser system defined 
by the root 
system of $\G$, with the hypermultiplet mass $m$ and the coupling constants 
$g,\theta$ of the gauge theory corresponding respectively to the coupling 
parameter $m_{|\a|}$ and the modulus parameter 
$$
\tau={4\pi i\over g^2}+i{\theta\over 2\pi}
$$ 
of the Calogero-Moser system. 
Our main results are

\medskip

$\bullet$ This is indeed the case when $\G$ is simply laced, i.e., all roots
of $\G$ have equal length, but not otherwise;

\medskip

$\bullet$ When $\G$ is non-simply laced, i.e. has both long and short roots, 
denoted respectively by $\R_l(\G)$ and $\R_s(\G)$, the correct model
is given rather by the following {\it twisted} version of the elliptic
Calogero-Moser system
$$
\ddot x=\half\bigg(\sum_{\a\in\R_s(\G)}m_{|\alpha|}^2\wp_{\nu}'(\a\cdot x)
\alpha+\sum_{\a\in\R_l(\G)}
m_{|\alpha|}^2\wp'(\a\cdot x)\alpha\bigg),
\eqno(1.3)
$$
where $\nu$ is the ratio of the length squared of the long to the short roots 
of 
$\G$, and $\wp_{\nu}(z)$ is the twisted version of the Weierstrass 
$\wp(z)$-function defined by (2.4) below. When $\G$ is one of the classical 
algebras $B_n$ or $C_n$, the system (1.3) can be  re-expressed as one of the
systems introduced by Inozemtsev [17-18]. For the exceptional Lie algebras, 
the 
twisted Calogero-Moser systems appear not to have been considered before.

\medskip 

$\bullet$ Our considerations are based on
two crucial consistency checks, corresponding to the limits $m\rightarrow0$ 
and 
$m\rightarrow\infty$.
When $m$ tends to $0$, the ${\cal N}=2$ gauge theory with matter in the 
adjoint
representation acquires an ${\cal N}=4$ supersymmetry. Since this theory 
receives no quantum corrections, its prepotential is the classical 
prepotential
and we can verify directly that it agrees with the
prepotential predicted by the Calogero-Moser systems at zero coupling.
The consistency check when $m\rightarrow\infty$ is more subtle. From the 
viewpoint of the four-dimensional gauge theory, this limit corresponds to the 
decoupling of the hypermultiplet as it becomes infinitely massive. On the 
basis 
of instanton considerations, the correct scaling law as $m\rightarrow\infty$ 
is 
given by
$$
m=Mq^{-\half\delta^{\vee}},\ q=e^{2\pi i\tau},
\eqno(1.4)
$$
where $\delta^{\vee}$ is the {\it dual} Coxeter number of $\G$. The
limiting theory is pure Yang-Mills, and the corresponding
integrable model is the affine Toda system for $(\G^{(1)})^{\vee}$.
Now the $SU(N)$ elliptic Calogero-Moser system is known to scale under (1.4) 
to 
the affine  $SU(N)$ Toda system [18]. More generally, we find that
the elliptic Calogero-Moser systems always tend to a finite limit under the 
rule
$$
m=Mq^{-\half\delta}\eqno(1.5)
$$
where $\delta$ is the Coxeter number, and that the limit is the affine Toda
system defined by $\G^{(1)}$. When $\G$ is simply-laced, the Coxeter number 
and
the dual Coxeter numbers are the same, $\G^{(1)}=(\G^{(1)})^{\vee}$, and 
the elliptic Calogero-Moser system satisfies the $m\rightarrow\infty$ 
consistency check.
However, when $\G$ is not simply laced (i.e. when $\G=B_n$, $C_n$, $F_4$, or 
$G_2$), the limit of the $\G$ elliptic Calogero-Moser system under (1.4) is 
infinite. Thus new Calogero-Moser systems are required. The twisted systems 
defined by (1.3) are systems which admit finite limits under (1.4). These 
limits 
also turn out to be precisely the desired $(\G^{(1)})^{\vee}$ affine Toda 
systems, and thus the twisted systems qualify as the models solving the $\G$ 
gauge theory with matter in the adjoint representation.
   
\medskip

$\bullet$ Although we have repeatedly referred to both the twisted and the 
untwisted $\G$ Calogero-Moser systems as {\it integrable} models, the 
integrability of these models for general Lie algebras is a complex issue far 
less understood (see e.g. [31] for a recent discussion) than their Toda and 
affine Toda counterparts. (We discuss this 
in greater detail below). Nevertheless, we have succeeded in producing a Lax 
pair of operators $L(z)$, $M(z)$, {\it with spectral parameter} $z$, 
satisfying 
the Lax equation $\dot L(z)=[L(z),M(z)]$ for each of these models. 
(For the case of $E_8$, we have to make an extra assumption on the existence 
of 
a certain $\pm 1$-valued cocyle. For the case of twisted $G_2$, we have been 
able to complete the proof only partially.) It is quite important for
our considerations that the Lax pair be allowed to depend on a free
external parameter. In particular, this is required for the construction
of the corresponding spectral curves $\Gamma$ and
differentials
$d\lambda$ 
$$
\Gamma=\{(k,z);\ \det(kI-L(z))=0\},\qquad
 d\lambda=kdz,
 \eqno(1.6)
$$
Taking the Lax pair for the untwisted Calogero-Moser when $\G=A_n$, $D_n$, 
$E_6$, $E_7$, and the Lax pair for the twisted Calogero-Moser when $\G=B_n$, 
$C_n$, $F_4$, we obtain in this way the candidate Seiberg-Witten spectral 
curves 
and differentials for the corresponding ${\cal N}=2$ $\G$ gauge theory
with matter in the adjoint representation. 
In principle, the prepotential ${\cal F}(a,\tau)$ can then be evaluated
explicitly, as was done in [8] for the $SU(N)$ theory, and in [19][20] for 
theories with classical gauge groups and matter in the fundamental 
representation.  (A sample calculation will be given in [16] for $\G =D_n$.)

\medskip

In this first paper of the series, we shall concentrate on the construction 
of 
the Lax pairs with spectral parameter. A second paper [15] will be devoted to a 
detailed study of scaling limits for twisted and untwisted Calogero-Moser 
systems. We have already described above the outcome for the limits of the 
Hamiltonians of these  systems.
It turns out that the Lax pairs we construct in this paper
also scale to appropriate finite limits, and produce a Lax pair for the
affine Toda systems. We discuss the spectral curves
themselves and the resulting physics of ${\cal N}=2$ supersymmetric
gauge theories in a third paper [16]. Finally, more severe divergences for 
Calogero-Moser systems than (1.4) and (1.5) will be discussed in a fourth 
paper, together with certain decoupling limits in the spirit of the results of 
[8].
 
\medskip

We return to a detailed description of the main topic of the
present paper, namely the construction of Lax pairs with spectral
parameters for Calogero-Moser systems. For affine Toda systems, a very 
general 
prescription is
available, with $L(Z)$ and $M(Z)$ given by [10][5]
$$
\eqalign{
L & = \sum_{i=1}^nP_ih_i + \sum _{\alpha \in \R _* (\G)} 
M_{|\alpha|}
e^{-\half \alpha \cdot X} \bigl (E_\alpha - E_{-\alpha} \bigr ) 
+ M_{|\alpha _0|} 
e^{+\half \alpha_0 \cdot X} \bigl (- Z^{-1} E_{\alpha _0} 
       + Z   E_{-\alpha_0} \bigr )
\cr
M & = -\half  \sum _{\alpha \in \R _* (\G)} 
 M_{|\alpha|}
e^{-\half \alpha \cdot X} \bigl (E_\alpha + E_{-\alpha} \bigr ) 
 +
\half M_{|\alpha _0|} 
e^{+\half \alpha_0 \cdot X} \bigl ( Z^{-1} E_{\alpha _0} + Z E_{-\alpha_0}
\bigr ).
\cr}
\eqno(1.7)
$$
Here $\a_0$ is the highest root of $\G$ (so that $-\a_0$ is the additional
simple root for the affine algebra $\G^{(1)}$ (Kac [21], Goddard-Olive [22]),
$Z$ is the spectral parameter, identifiable in this case with the
loop variable of the loop group, and $E_{\a}$ are generators for $\G$ in a
Cartan-Weyl basis. The operators $L(Z)$, $M(Z)$ become $N\times N$ matrices 
$\rho(L(Z))$, $\rho(M(Z))$ upon choosing a $N$-dimensional representation 
$\rho$ 
of $\G$, and a ($\rho$-dependent) spectral curve
$\Gamma$ can then be defined by
$$
\Gamma=\{(k,Z);\ \det\bigg(kI-\rho(L(Z))\bigg)=0\}.
\eqno(1.8)
$$
Different curves corresponding to different choices of representations $\rho$
are related by syzygies, and it has been advocated in [5] that the
corresponding prepotentials, and hence physics, should be the same.  
At the present time, there is no general formula of type (1.7) for 
Calogero-Moser systems. In fact, only in the $SU(N)$ case was a Lax pair 
$L(z)$, 
$M(z)$ with spectral
parameter known (this is a classical result going back to Krichever [23]).
For the classical finite dimensional (simple) Lie algebras $\G$ other than
$A_{N-1}$, only a Lax pair {\it without spectral parameter} appears to be 
known. Its construction was due to Olshanetsky and Perelomov [14],
and it was formulated in terms of the geometry and group theory of
symmetric spaces. For the exceptional Lie algebras 
$E_8, \ E_7, \ E_6, \ F_4$ and $G_2$, no Lax pair appears to be known at all.
These statements refer to the untwisted Calogero-Moser systems. The twisted 
Calogero-Moser systems had of course not even been considered.

\medskip

Our approach is based on a general Ansatz for Lax
operators $L(z)$ and $M(z)$ with spectral parameter, which is applicable to 
any
finite-dimensional Lie algebra in an arbitrary representation. This Ansatz
is a natural extension of the original Ansatze of Calogero-Moser, Krichever,
and Olshanetsky-Perelomov. 
The conditions
that the proposed Lax operators close onto the Calogero-Moser system for $\G$
are reduced to purely algebraic equations. Nevertheless these algebraic
equations can be quite difficult to solve,
and they depend in an intricate manner on the choice of representation.  

\medskip

We show that these equations can be solved in the following cases. 
For the untwisted
Calogero-Moser systems,

\medskip

\item{$\bullet$} in the case of $BC_n$, $B_n$, $C_n$, and $D_n$,
by imbedding $B_n$ into $GL(2n+1,{\bf C})$; 

\item{$\bullet$} in the case of $A_n$,
by taking the fundamental and the anti-symmetric rank $p$, $1\leq p\leq n-1$,
tensor representations; 

\item{$\bullet$} in the case
of $B_n$ and $D_n$, by taking the spinor representation;

\item{$\bullet$} in the case of $G_2$, by taking the ${\bf 7}$ of $G_2$;
 
\item{$\bullet$} in the case of $F_4$, by taking the ${\bf 26+1}$ of $F_4$;

\item{$\bullet$} in the case
of $E_6$, by taking the ${\bf 27}$ of $E_6$; 

\item{$\bullet$} in the case of $E_7$,
by taking the ${\bf 56}$ of $E_7$.

\item{$\bullet$} in the case of $E_8$, by taking the ${\bf 248}$ 
representation, we obtain a solution for the Lax pair upon making an extra 
assumption on the existence of a $\pm 1$-values cocycle. Thus we expect our 
Ansatz will also produce a Lax pair, but we have no full proof at this time.

\medskip

\noindent
The dimension of the Lax pair is then the dimension of the imbedding. For the 
twisted Calogero-Moser systems, the equations can be solved to produce

\medskip

\item{$\bullet$} in the case of $B_n$, a $2n$-dimensional Lax pair;

\item{$\bullet$} in the case of $C_n$, a $2n+2$-dimensional Lax pair;

\item{$\bullet$} in the case of $F_4$, a $24$-dimensional Lax pair.

\item{$\bullet$} in the case of $G_2$, the Ansatz for a $6$-dimensional
Lax pair seems consistent, but we have not yet found any full proof of it.

\medskip

In the case of the classical Lie algebras $BC_n$, $D_n$, when the
spectral parameter $z$ equals one of the three half periods of the torus
$\Sigma={\bf C}/(2\omega_1{\bf Z}+2\omega_2{\bf Z})$, our 
Lax pairs coincide with the Lax pairs {\it without spectral 
parameter} which were formulated previously for these groups
by Olshanetsky and Perelomov [14]. Thus these previously known solutions
have been in effect imbedded in a whole family of solutions. As mentioned
previously, the Olshanetsly-Perelomov admit a symmetric space interpretation
which depends heavily on the fact that the crucial function $\Phi(x,z)$
introduced by Krichever [23] becomes odd in $x$. It may be worth noting
that a similar geometric interpretation of our Lax pair for general
values of the spectral parameter $z$ is still possible, upon replacing
the naive reflection symmetries $x\leftrightarrow -x$, $z\leftrightarrow-z$,
by a twisted version (c.f. (4.17) below).

\medskip

We also observe that the dimensions of the Lax pairs in the twisted
$\G$ Calogero-Moser case may not even be dimensions of representations
of $\G$. This is one of the difficulties in finding a systematic
representation theoretic construction of the particular Lax pairs
we found, say a construction analogous to (1.7) for affine Toda systems.
It is also remarkable that the Lax pair in the spinor of $SO(2n+1)$ has two 
free 
couplings, while the Lax pair in the fundamental of this group only had one. 
In practice, to find the Lax pairs, we are forced at this moment to proceed 
case by case, and it would certainly be very valuable to have a more general
or more conceptual approach.  

\bigskip

The remainder of the paper is organized as follows.
The basic material is introduced in \S II. This includes
the description of both the twisted and untwisted elliptic Calogero-Moser
Hamiltonians, their symmetries, and the explicit form
of the Lax pair in the case of $A_n$.

In \S III, we
propose a general construction for the Lax operators $L(z)$ and 
$M(z)$, for any Lie algebra in an arbitrary representation. The conditions 
under
which the Lax equation is the Hamilton-Jacobi equation of a Calogero-Moser
system are reduced to a set of purely algebraic equations, the solutions
of which are the central issue of the paper.
 
In \S IV, we present the solutions, and thus the
Lax pairs with spectral parameter for the classical Lie algebra series $B_n$,
$C_n$ and $D_n$, corresponding to the complexified Lie algebras of 
$SO(2n+1)$,
$Sp(2n)$, and $SO(2n)$. In some cases, solutions in several representations
are possible, as listed above. By introducing twisted actions
and twisted Lie algebras,
we show how our formulas can be viewed as natural extensions
of the Olshanetsky-Perelomov formulas when $z$ is at a half period.
 
In \S V, we apply our construction to the case of the exceptional Lie
algebras. For the cases of $E_6$ and $E_7$ a Lax pair with
spectral parameter is constructed in the representations of dimension {\bf 
27} 
and
{\bf 56} respectively, each with a single free coupling.  For the case of 
$G_2$, a Lax pair is constructed in the representations of dimensions {\bf 7} 
(the
fundamental) and {\bf 8} (the fundamental plus a singlet), each with a
spectral parameter, and with one and two free couplings respectively. 
For the case of $F_4$, no Lax pair appears to exist in the representation of
dimension 26 (the fundamental), but we find a Lax pair with spectral 
parameter
and two couplings in the representation of dimension {\bf 27} 
(the fundamental plus 
a singlet). For $E_8$, we summarize the key features of the ${\bf 248}$
Lax pair we expect, leaving the detailed verification of some 
consistency checks to the appendices.

Finally, the twisted Calogero-Moser systems are treated
in \S VI. An important new feature of the Lax pairs is the emergence
of new twisted versions of the function $\Phi(x,z)$, notably
the functions $\Lambda(x,z)$, $\Phi_2(x,z)$, and $\Phi_2(x\pm\omega_2,z)$.
These functions satisfy striking functional equations generalizing Landen's
doubling identities [24], without which the algebraic equations 
characterizing 
the 
Lax pair
would be intractable. It is likely that similar functions
should exist which satisfy the {\it tripling} identities needed to solve the
$G_2$ case, but we have not succeeded in constructing them, and this
case remains open.
  
\medskip

In Appendix \S A, we have collected for the convenience
of the reader all the necessary group theoretical information, including 
weight systems and the Weyl orbit decompositions of various representations 
which we need for the analysis of exceptional Lie algebras.
In Appendix \S B, we derive all the identities for elliptic functions
we need, including the doubling identities we mentioned above.
In Appendix \S C, we present the details of the consistency checks for $E_8$.

\medskip

We conclude the introduction by pointing out that the Seiberg-Witten curves and 
differential may be constructed by string theory techniques. One method is by 
exploiting the appearance of enhanced gauge symmetries (of the A-D-E type) at 
certain singular compactifications. (See for example [26].) A second method is 
by obtaining supersymmetric Yang-Mills theory as an effective theory on a 
configuration of branes in string theory or in M-theory. This approach was 
pioneered in [27] (see also [28]) for SU(N) gauge group (and products thereof), 
and was extended to other classical groups in [29]. The relation between the 
string theory and M-theory approaches and integrable systems were proposed in 
[27] and [30]. Their interplay should produce 
substantial advances in both fields.

\bigskip
\bigskip

\centerline{\bf II. THE ELLIPTIC CALOGERO-MOSER SYSTEMS}

\bigskip

The elliptic Calogero-Moser systems are integrable Hamiltonian models with 
$n$
dynamical degrees of freedom $x_i$, $i=1, \cdots , n$ and associated 
canonical
momenta $p_i$, $i = 1, \cdots , n$, which are complex valued, and denoted 
simply by vectors $x$ and $p$ respectively. They are 
parametrized by the periods $2\omega _1$ and $2 \omega _2$ of an elliptic 
curve
(or torus) $\Sigma$,  whose modulus is $\tau = \omega _2 /\omega  _1$.
Each system is naturally associated with a finite dimensional simple Lie 
algebra $\G$ of rank $n$, whose set of roots is denoted by $\R (\G) $. 
We are led to distinguish between two types of elliptic Calogero-Moser 
systems. 
The {\it ordinary} or {\it untwisted elliptic Calogero-Moser system} was 
introduced long ago in [14] for all simple Lie algebras $\G$. The {\it 
twisted 
Calogero-Moser system} will be introduced in (b) below for all simple Lie 
algebras $\G$. The twisted system coincides with the untwisted system when 
$\G$ 
is simply laced (i.e. all roots of $\G$ have the same length), but differs 
from 
it when $\G$ is non-simply laced.

\medskip

\noindent
{\bf (a) The Untwisted Elliptic Calogero-Moser Systems}

\medskip

The untwisted elliptic Calogero-Moser system associated with a simple Lie 
algebra $\G$ and with periods $2\omega _1$, $2 \omega _2$ is defined by the 
Hamiltonian
$$
H = \half p\cdot p - \half \sum _{ \alpha \in \R (\G) } 
m_{|\alpha|} ^2 \ \wp ( \alpha \cdot  x; 2 \omega _1 , 2 \omega _2 ).
\eqno (2.1)
$$
Here, $\wp (z; 2\omega _1, 2 \omega _2)$ is the Weiestrass elliptic function
of $\Sigma$, whose definition and properties may be found in Appendix \S B.
Henceforth we shall suppress the $2\omega _1$ and $2 \omega _2$ dependence
of $\wp$.
The inner product of vectors in ${\bf R}^n$ or ${\bf C}^n$,
such as $\alpha$ and $x$, is denoted by $\alpha \cdot x$. The Hamiltonian is
required  to be invariant under the Weyl group of the Lie algebra $\G$, so 
that the  constants $m_{|\alpha|}$ only depend upon the Weyl orbit, denoted 
by  
$|\alpha |$, of the root $\alpha$. When the orbits are uniquely labeled by 
the 
length of the roots, we set simply $|\alpha |= \alpha ^2$.

\medskip

The Hamiltonian of (2.1) is invariant under a large discrete symmetry group, 
generated by the  following transformations on the periods $2 \omega _1$, $2 
\omega _2$ and on $x$ and $p$. \next 
(1) The Weyl group $W_\G$ of the finite dimensional Lie algebra $\G$,
leaving $\omega _{1,2}$ unchanged, and acting on $x$ and $p$ by Weyl 
reflections $W_\alpha$
$$
x \ \to \ W_\alpha (x) = x - 2  \alpha {x\cdot \alpha \over \alpha
^2},
\qquad 
p \ \to \ W_\alpha (p), 
\qquad \alpha \in \R (\G).
\eqno (2.2a)
$$ 
(2) The modular group $SL(2, {\bf Z})$, leaving $x$ and $p$ unchanged 
and acting on $\omega _{1,2}$ by
$$
\eqalign{
\omega _1 \ \to & \ a \omega _1 + b \omega _2, \cr
\omega _2 \ \to & \ c \omega _1 + d \omega _2,
\qquad \qquad a,b,c,d \in {\bf Z}; 
\quad ad-bc=1. \cr}
\eqno (2.2b)
$$ 
(3) Affine transformations ${\bf Z}^{2n}$ leaving $\omega _1$, $\omega _2$ 
and 
$p$ unchanged while shifting $x$ by
$$ 
x \ \to \ x + \sum _{i=1} ^n (2\omega _1 n_1 ^i + 2 \omega _2 
n_2 ^i) \lambda ^i,
\qquad \qquad
n_{1,2} ^i \in {\bf Z},
\eqno (2.2c)
$$
where $\lambda ^i$ are a set 
of generators of the dual lattice to $\R (\G)$. (If $\alpha ^i$ is a set of 
simple roots of $\R (\G)$, then $\lambda ^j$ may be defined by $\alpha ^i 
\cdot 
\lambda ^j = \delta ^{ij}$, $i,j=1,\cdots ,n$.)

\medskip

The combined set of transformations (2.2a,b,c) contains the action of
the Weyl  group $W_{\hat \G}$ of the affine extension $\hat \G$ of
$\G$. This affine extension is untwisted, and will be denoted by $\G ^{(1)}$, 
following standard notation. 
Thus, it is natural to associate the Calogero-Moser system not just with a 
finite dimensional Lie algebra $\G$, but instead with its full affine 
extension 
$\G ^{(1)}$.
The automatic appearence of the affine Weyl group in Calogero-Moser systems 
seems to have gone unnoticed so far.

\bigskip

\noindent
{\bf (b) The Twisted Elliptic Calogero-Moser Systems}

\medskip

The twisted elliptic Calogero-Moser system associated with a 
finite-dimensional 
simple Lie algebra $\G$ and with periods $2 \omega _1$, $2 \omega _2$ is 
defined by the Hamiltonian
$$
H = \half \sum _{i=1} ^n  p _i ^2 
- \half \sum _{ \alpha \in \R (\G) } m_{|\alpha |} ^2 \ \wp _{\nu (\alpha)} ( 
\alpha \cdot  x ).
\eqno (2.3)
$$
The function $\nu (\alpha)$ depends upon the length of the root $\alpha$ 
only. 
For any simply laced $\G$, we set $\nu(\alpha )=1$ on all roots. For 
non-simply 
laced $\G$, roots of only two different lengths appear : long roots and short 
roots. We set $\nu (\alpha)=1$ for all long roots, $\nu (\alpha) =2$ for the
short roots of $B_n$, $C_n$ and $F_4$, and $\nu (\alpha )=3$ for the short 
roots of $G_2$. $\wp_\nu (z)$ is a {\it twisted} Weierstrass function 
$$
\wp_\nu (z) 
= \sum _{\sigma =0} ^{\nu -1} 
\wp (z +  2 \omega _a {\sigma \over \nu} ) .
\eqno (2.4)
$$
Here, $\omega _a$ is any one of the half periods $\omega _1$, $\omega _2$ or
$\omega _3 = \omega _1 + \omega _2$. Since we have $\wp _1(z) = \wp (z)$, the 
twisted system coincides with the untwisted one for any simply laced $\G$.  
It will turn out that the twisted elliptic Calogero-Moser systems for the Lie 
algebra $\G$ is naturally associated with the affine Lie algebra $(\G 
^{(1)})^\vee$. When $\G$ is simply laced, we have $(\G ^{(1)})^\vee = \G 
^{(1)}$, and we recover the untwisted elliptic Calogero-Moser system. When 
$\G$ 
is non-simply laced, $(\G ^{(1)})^\vee$ equals a twisted affine Lie algebra :
$ (B_n  ^{(1)})^\vee = A_{2n-1}^{(2)}$, $(C_n ^{(1)})^\vee = D_{n+1} ^{(2)} 
$,
$(F_4 ^{(1)} )^\vee = E_6 ^{(2)} $  and $(G_2 ^{(1)}) ^\vee = D_4 ^{(3)} $.
\footnote{$\dagger$}{Dynkin diagrams of affine Lie algebras, as well as other 
group theoretic information is collected in Appendix \S A; for general sources, 
see [21,22,25].}

\bigskip

\noindent
{\bf (c) Limits of the Elliptic Calogero-Moser Systems}

\medskip

Various limits of the elliptic Calogero-Moser system yield related integrable
systems. By taking one of the periods $2\omega _1$ or $2 \omega _2$ to 
infinity, we obtain the trigonometric Calogero-Moser system, with $\wp (x) $ 
replaced by $1/\sin ^2 x $. 
By taking both of the periods to infinity, we obtain the rational 
Calogero-Moser system, with $\wp (x)$ replaced by $1/x ^2$. 
Finally, a simultaneous limit of the periods and of $x$ yields the Toda 
integrable system. Specifically, the scaling of the untwisted and twisted 
elliptic Calogero-Moser systems yields the non-periodic (or ordinary) Toda 
system associated with the Lie algebras $\G$ and $\G^\vee$ respectively. A 
certain critical scaling of the untwisted and twisted elliptic Calogero-Moser 
systems yields the periodic Toda system associated with the affine Lie 
algebras 
$\G ^{(1)}$ and $(\G ^{(1)})^\vee$ respectively. These results are derived in 
a companion  paper [15].

\bigskip

\noindent
{\bf (d) Existence of a Lax Pair with Spectral Parameter}

\medskip

The Hamilton-Jacobi equations are $\dot x  =  p$ and 
$$
\dot p =  \half \sum _{\alpha \in \R (\G) } m_{|\alpha |} ^2 \ \alpha  \ 
\wp _{\nu (\alpha)} ' (\alpha \cdot x ).
\eqno (2.5)
$$
Here, a dot denotes time derivation, and $\wp _\nu '(z)$ denotes the 
derivative with respect to $z$ of $\wp _\nu (z)$. 

\medskip

Complete integrability of the (twisted or untwisted) elliptic Calogero-Moser 
system is guaranteed by the existence of a {\it Lax pair} of $N \times N$ 
matrix-valued functions of $x$ and $p$ (and also of $\omega _1$ and $\omega 
_2$), denoted by $L,\ M$,  such 
that the {\it Lax equation} 
$$
\dot L = [L,M]
\eqno (2.6)
$$
is equivalent to the Hamilton-Jacobi equations of (2.5). One manifestation of
integrability is the existence of a maximal number of conserved integrals of
motion. Assuming that we have a Lax pair, it is immediately clear that the
quantities $\tr L^r$, for $r=0,1,2,\cdots , \infty$ are conserved integrals 
of
motion, since they are time independent by (2.6). On general grounds, at most 
$N$ of these quantities are functionally independent of one another. In 
practice, we always have $N\geq n$, so that there are enough integrals of 
motion to completely separate the dynamics of the $n$ degrees of freedom in 
(2.5). 

\medskip

Many integrable systems admit a generalized Lax pair in which $L$ and
$M$ are allowed to depend upon an arbitrary complex valued {\it spectral
parameter} $z$. The spectral parameter $z$ does not enter the Hamilton-Jacobi 
equations of the system, and modifies the Hamiltonian by at most an added 
term 
that depends on $z$ but is independent of $x$ and $p$. When this is
the case, the Lax equation (2.6) for $L(z)$ and $M(z)$ yields (2.5) for every
value of $z$. The existence of a Lax pair with spectral parameter allows one 
to associate a time independent {\it spectral curve} with the Calogero-Moser 
system (twisted or untwisted) for each finite dimensional Lie algebra $\G$, 
given by the following equation
$$
\det (k I- L(z)) =0.
\eqno (2.7)
$$
The Weyl group of $\G$ leaves the spectral curve invariant and acts by
permutation on the various leafs of the Riemann surface defined by the
various roots for $k$ of (2.7). 

\medskip

For any given Hamiltonian, the Lax pair is not unique. The Lax equation (2.6) 
is covariant under conjugation of the operator $L$ by an arbitrary $N\times 
N$ 
matrix valued function $S$ of $x$, $p$, $z$ and $\omega _{1,2}$ and an 
accompagnying gauge transformation on $M$. We have $\dot L^S = [L^S, M ^S]$ 
with
$$
\eqalign{
L ^S = & S L S^{-1} \cr
M ^S = & S M S ^{-1} - \dot S S^{-1}. \cr
}
\eqno (2.8)
$$
The conserved integrals of motion $\tr L^m$ and the spectral curve of (2.7)
are invariant under these gauge transformations. 

\bigskip

\noindent
{\bf (e) The Lax Pair for the A$_{{\bf n}}$ System}

\medskip
    
A generalized Lax pair was found for the elliptic Calogero-Moser system
associated with the Lie algebras $A_{N-1}$, in
terms of a spectral parameter $z$ that lives on the elliptic curve $\Sigma$.
We parametrize the roots of $A_{N-1}$ by an orthonormal basis in
$\C^N$ of vectors $e_i$, $i=1,\cdots , N$. The set of all roots is then $\R
(A_{N-1})=\{ e_i - e_j;\ i\not=j\}$. We set $m_{|\alpha |} = m_2$, 
since all roots have the same length squared, 2. 
In this basis, the Lax pair is in terms of
$N\times N$  matrices (with $i,j= 1, \cdots , N$)
$$
\eqalign{
L_{ij} (z) = \ 
& p_i \delta _{ij} - m_2 (1 - \delta _{ij}) \Phi (x_i - x_j , z)
\cr
M_{ij} (z) = \
& m_2 \delta _{ij} \ \sum _{k\not=i} \wp (x_i - x_k) + m_2 (1 - \delta _{ij} 
)
\Phi ' (x _i - x_j, z).
\cr}
\eqno (2.9)
$$
Here, $\Phi '(x,z)$ stands for the $x$-derivative of $\Phi (x,z)$. 
The function $\Phi(x,z)$ obeys
$$
\Phi (x,z) \Phi ' (y,z) - \Phi (y,z) \Phi '(x,z) 
= \left \{ \matrix{
(\wp (x) - \wp (y) ) \Phi (x+y,z), 
 &  x+y \not=0; \cr
\wp ' (x),  & x+y=0. \cr} \right .
\eqno (2.10)
$$
$\Phi (x,z)$ is doubly periodic in $z$ with periods $2 \omega _1$ and $2 
\omega
_2$, has monodromy in $x$, and has an essential singularity at $z=0$. 
Properties of elliptic functions are given in Appendix \S B.

\bigskip
\bigskip

\centerline{{\bf  III. THE GENERAL CONSTRUCTION OF LAX PAIRS}}

\bigskip

In this section, we present a general formalism for the construction of Lax
pairs for the untwisted and twisted elliptic Calogero-Moser systems 
associated 
with an arbitrary simple Lie algebra $\G$. We begin with a discussion of the 
Lie algebra theory needed to formulate the Ansatz for the Lax pairs in 
Theorems 
1 and 2.

\bigskip

\noindent
{\bf (a) Decomposing Roots of GL(N) under the Action of a Subalgebra $\G$}

\medskip

Let $\G$ be a finite dimensional, complex simple Lie algebra of rank $n$,
and dimension $d$.
Let $\Lambda$ be a representation of $\G$ with dimension $N < \infty$, and 
generators $\Lambda _a$, $a=1, \cdots, d$. 
The representation $\Lambda$ embeds $\G$ into the fundamental 
($N$-dimensional) 
representation of $GL(N,\C)$ as a subalgebra; 
the generators $\Lambda _a$ are then linear combinations of the
$N^2$ generators of $GL(N,\C)$. 

\medskip

Since $\G$ is a subalgebra of $GL(N,\C)$, we choose a Cartan subalgebra 
$\H$ for $GL(N,\C)$ that contains the Cartan subalgebra $\H _\G$ chosen for 
$\G$, so that $\H_\G = \H \cap \G$. 
This choice allows us to split the set of Cartan generators of $\H$ into a 
set 
of Cartan generators
$h_i, \ i=1,\cdots n$ of $\H _\G$ and a complementary set of generators
$\tilde h_j, \ j=n+1, \cdots ,N$ in $\H$ that commute with all $h_i$
and mutually commute : $[h_i, \tilde h_j] = 0$ and $[\tilde h_j, \tilde 
h_{j'}]=0$.
Without loss of generality, we shall choose the $h$ and $\tilde h$ generators 
to be mutually orthogonal under the Cartan-Killing inner product, $\tr
(Ad_{h_i}  Ad _{\tilde h_j})=0$.

\medskip

The centralizer in $GL(N,{\bf C})$ of the Cartan subalgebra $\H _{\G}$ of 
$\G$  
may be larger than $\H$. We denote it by $\H \oplus GL_0$. The subspace 
$GL_0$ 
consists of the roots of $GL(N, {\bf C})$ which project to zero 
under the orthogonal projection of the weights of $GL(N, {\bf C})$ to $\G$.
When the representation $\Lambda$ has at most one zero
weight under $\G$, we have $GL_0=0$. This is the case for all Lie algebras in
their lowest dimensional faithful representation, except for $F_4$ and  
$E_8$,
where $\dim GL_0 =2$ and 56 respectively. 

\medskip

We now develop a general description for the embedding of the
representation $\Lambda$  of $\G$ into the fundamental representation of
$GL(N,\C)$. The weights of the fundamental representation of $GL(N,\C)$ may
be chosen to be $N$ orthonormal vectors in $\C ^N$, which we denote by $u_I$,
$I=1,\cdots , N$. As usual, the root vectors of $GL(N,\C)$ are then given by 
the differences $u_I - u_J$. The fundamental representation of $GL(N,\C)$ 
restricts to the representation $\Lambda$, as $GL(N,\C) $ is restricted to 
the
subalgebra $\G$. The weight vectors of the fundamental representation of
$GL(N,\C)$ admit a decomposition into weight vectors of $\G$, (corresponding 
to 
the eigenvalues of the Cartan generators $h_i$, $i=1,\cdots , n$ of $\H _\G$) 
and 
into vectors in the complement of $\G$ in $GL(N,\C)$, (corresponding to the
eigenvalues of the Cartan generators $\tilde h_j$, $j=n+1, \cdots , N$).
This decomposition is orthogonal in view of the orthogonality of $h$ and 
$\tilde h$.

\medskip

The weight vectors of the representation $\Lambda $ of $\G$ are $N$ vectors 
in
$\C^n$, which we shall view as vectors in $\C ^N \supseteq \C^n$, by 
assigning 
coordinates 0 to the extra generators $\tilde h$. According to
the preceding discussion, we have the orthogonal decomposition
$$
s u_I = \lambda _I + v_I 
\qquad \qquad
I=1,\cdots , N
\eqno (3.1)
$$
where any $v_I$, $I=1,\cdots , N$ is orthogonal to any $\lambda _J$,
$J=1,\cdots ,N$, and $s$ is a normalization factor. The vector
space generated by all $\lambda _I$ is of dimension $n$, while that generated
by all $v_I$ is of dimension $N-n$. The normalization
factor $s$ is related to the second Dynkin index of the
representation $\Lambda $ :
$$
s^2 = I_2 (\Lambda ) = { 1 \over n} \sum _{I=1}^N  \lambda _I \cdot \lambda 
_I.
\eqno (3.2)
$$ 
Once $s$ has been determined, we parametrize the weight vectors $\lambda _I$ 
as well as the vectors $v_I$ in terms of the $N$-dimensional basis vectors
$u_I$ in an explicit and unique way by
$$
\eqalign{
\lambda _I = & \ {1 \over s} \sum _{J=1}^N (\lambda _I \cdot \lambda _J) u_J
\cr
v_I = & \ s ~u_I - {1 \over s} \sum _{J=1}^N (\lambda _I \cdot \lambda _J) 
u_J.
\cr}
\eqno (3.3)
$$

\medskip

The decomposition of the weight vectors of the fundamental representation
of the algebra $GL(N,\C)$ under the action of $\G$ allows us to obtain the
corresponding decomposition for any representation of $GL(N,\C)$. In 
particular,
the generators of $GL(N,\C)$ corresponding to the roots may be decomposed in 
this
way. We shall label the generators associated with the roots of $GL(N,\C)$ by
$E_{IJ}$, $I\not=J$. The root decomposition is then obtained by evaluating 
the
commutators of $h$ and $\tilde h$ with $E_{IJ}$. We find
$$
\eqalign{
[h         , E_{IJ}] = & \ (\lambda _I  - \lambda _J ) E_{IJ} \cr
[\tilde h  , E_{IJ}] = & \ (v_I  - v_J ) E_{IJ}. \cr}
\eqno (3.4)
$$
Under the action of $\G$, the adjoint representation of $GL(N,\C)$ decomposes
into the adjoint representation of $\G$, plus other representations, 
according
to the tensor product $\Lambda \otimes \Lambda ^* $. 

\vfill\eject

\noindent
{\bf (b) Construction of Lax pairs for arbitrary Lie Algebra $\G$}

\medskip

We make use of the results of (a) to provide a Lax pair construction for the 
elliptic Calogero-Moser systems of (2.1) and (2.3) associated with any Lie 
algebra $\G$ in an
arbitrary representation $\Lambda$ of dimension $N$, with weight system $\{
\lambda _I \} _{I=1,\cdots ,N}$. The Lax operators $L$ and $M$ are $N\times 
N$
dimensional matrices, given by the following Ansatz,
$$
\eqalign{
L=  P+X, 
\qquad & \qquad
P =  p \cdot h,  \cr
M= D +Y,
\qquad & \qquad
D = d \cdot (h \oplus \tilde h)  + \Delta .
\cr}
\eqno (3.5)
$$
Here, $P \in \H _{\G}$, $\Delta \in GL_0$ and $ D \in \H \oplus GL_0$, so 
that 
$[P,D]=0$, while $X$ and $Y$ are given by
$$
\eqalign{
X= & \sum _{I,J=1;I\not= J} ^N C_{I,J}  
                             \Phi _{IJ}(\alpha _{IJ} \cdot x, z) E_{IJ} \cr
Y= & \sum _{I,J=1;I\not= J} ^N C_{I,J}  
                             \Phi _{IJ} ' (\alpha _{IJ} \cdot x, z) E_{IJ}, 
\cr}
\eqno (3.6)
$$
The combination $\alpha _{IJ} \equiv \lambda _I - \lambda _J$ is the weight
under
$\G$ associated with the root $u_I - u_J$ of $GL(N, \C)$, $C_{I,J}$ are
constants, $\Phi _{IJ} '(x,z)$ are the $x$-derivatives of $\Phi _{IJ}(x,z)$ 
and 
the 
functions $\Phi _{IJ}$ are certain elliptic functions of the type of $\Phi$,
which  remain to be determined. 

\medskip

\noindent
{\bf Theorem 1 : The General Case}

The Lax  equation $\dot L= [L,M]$ implies the elliptic Calogero-Moser system 
(2.1) or (2.3) if and only if the following conditions hold

\item{(1)} 
$$
s^2 \sum _{\alpha \in \R (\G)} ~m_{|\alpha |} ^2 \wp _{\nu (\alpha)} ' 
(\alpha
\cdot x) ~\alpha  =\sum _{I,J=1; ~I\not=J } ^N C_{I,J} C_{J,I} 
\wp _{IJ} ' (\alpha _{IJ} \cdot x) ~ \alpha _{IJ} .
\eqno (3.7)
$$
\item{(2)} 
$$
0=\sum _{I,J=1; ~I\not=J} ^N C_{I,J} C_{J,I} 
\wp _{IJ} ' (\alpha _{IJ} \cdot x) (v_I  - v_J ) . 
\eqno (3.8)
$$
\item{(3)} There exists a vector $d \in \C ^N$ and a matrix $\Delta$ with $D= 
d 
\cdot (h \oplus \tilde h) + \Delta ~  \in \H \oplus GL_0$, such that for all 
$I\not=J$, we have
$$
\eqalign{
 s C_{I,J} \Phi _{IJ} (\alpha _{IJ} \cdot x) d \cdot (u_I - u_J)
+ & \sum _{K\not= I,J} \Delta _{IK} C_{K,J} \Phi _{KJ} (\alpha _{KJ} \cdot x) 
\cr
- & \sum _{K\not= I,J} C_{I,K} \Phi _{IK} (\alpha _{IK} \cdot x) \Delta _{KJ}
 \cr
=\sum _{K \not= I,J} C_{I,K} C_{K,J} 
\{  \Phi _{IK}  (\alpha _{IK} & \cdot x) \Phi ' _{KJ} (\alpha _{KJ} \cdot x) 
-  \Phi ' _{IK} (\alpha _{IK} \cdot x) \Phi  _{KJ} (\alpha _{KJ} \cdot x)
\}
\cr }
\eqno (3.9)
$$
Here, the Weierstrass functions are defined by
$$
\Phi _{IJ} (x,z) \Phi _{JI} ' (-x,z) -  \Phi _{IJ} ' (x,z)  \Phi _{JI} (-x,z) 
= 
\wp _{IJ} ' (x).
\eqno (3.10)
$$

\medskip

To prove this Theorem, we use the fact that $[P,D]=0$,
and that the Lax equation (2.6) decomposes into three parts upon separating 
out
the $\dot x$ and $p$  dependence of various terms.
$$
\eqalign{
\dot X = & \ [P,Y] \cr
\dot P = & [X,Y] _{\H} \cr 
[D,X]  = & [X,Y] _{\M}. \cr }
\eqno (3.11)
$$
Here, ${\cal M}$ is the complement to $\H$ in $GL(N, \C) = \H  \oplus {\cal 
M}$, and
the symbols $[,]_\H$ and $[,]_\M$ denote the  projections of the commutator
$[,]$ onto $\H$ and $\M$ respectively.

\medskip

The first equation in (3.11) is guaranteed by the form of the Ansatz for 
$X$ and $Y$, and by the fact that $\alpha _{IJ} = \lambda _I - \lambda _J$.    
The second equation in (3.11) may be reduced to conditions (1) and
(2), by using the algebra of $GL(N, \C)$ generators
$$
[E_{IJ}, E_{KL} ] = \delta _{JK} E_{IL} - \delta _{IL} E_{KJ},
\eqno (3.12)
$$
as well as the identities between the $\Phi _{IJ}$ and $\wp _{IJ}$ functions 
of 
(3.10). Indeed, the second equation in (3.11) is equivalent to
$$
\dot p \cdot h  = \sum _{I,J=1;I\not= J} ^N 
  C_{I,J}  C_{J,I} \wp _{IJ} '( \alpha _{IJ} \cdot x) E_{II},
\eqno (3.13)
$$
which decomposes into two parts. Its part along 
$\H _{\G}$ yields the Hamilton-Jacobi equation for the Calogero-Moser system,
while the remaining part is a further constraint on the couplings
$C_{I,J}$. To disentangle the two, it suffices  to obtain the decomposition 
of
the generators $E_{II}$ in terms  of $h$ and $\tilde h$. By comparing (3.4) 
and 
(3.12) for $I=J$, one finds
$$
E_{II} = { 1 \over s^2} \bigl (
     \lambda _I \cdot h + v_I \cdot \tilde h  \bigr ),
\eqno (3.14)
$$
where $s$ and $v_I$ were introduced in (3.1) and $s$ was evaluated in (3.2). 
Thus, (3.13) splits into
$$
\eqalign{
\dot p  = & \sum _{I,J=1;I\not= J} ^N {1 \over 2s^2} C_{I,J}  C_{J,I}
\wp _{IJ} ' (\alpha _{IJ} \cdot x) \alpha _{IJ}  \cr
0         = & \sum _{I,J=1;I\not= J} ^N C_{I,J}  C_{J,I}
\wp _{IJ} ' (\alpha _{IJ} \cdot x) (v _I  - v_J ). \cr}
\eqno (3.15)
$$
The first equation in (3.15) manifestly reproduces the Calogero-Moser
Hamilton-Jacobi equations (2.5), provided (1) in (3.7) holds. The second 
equation
is purely algebraic and coincides with (2) in (3.8).
Finally, the third equation in (3.11) is easily reduced to condition (3),
using (3.12).
                            
\medskip

\noindent
{\bf Theorem 2 : Untwisted Elliptic Calogero-Moser Systems}

\medskip

For each of the untwisted Calogero-Moser systems with the Hamiltonians of
(2.1), a Lax pair may obtained in which the functions $\Phi _{IJ}$ are all 
identical
$$
\Phi _{IJ} (x,z) = \Phi (x,z)
\qquad \qquad I,J = 1, \cdots ,N
\eqno (3.16)
$$
and the matrix of constants $C_{I,J}$ is symmetric : $C_{I,J} = C_{J,I}$.
Except for the Lie algebra $E_8$, we may also set $\Delta =0$. 
Under these conditions, very considerable simplifications take place, and the 
statements of Theorem 1 may be simplified, as follows.

\medskip

The Lax  equation $\dot L= [L,M]$ implies the untwisted elliptic 
Calogero-Moser 
system (2.1) if and only if conditions (1), (2) and (3) below are
satisfied.

\item{(1)} The constants $C_{IJ}$ are related to the coupling constants
$m_{|\alpha|}$ of the Calogero-Moser system by
$$
s^2 ~m_{|\alpha |} ^2 =  \sum _{{I,J=1 ; \atop \alpha _{IJ} = \alpha }} ^N
C_{I,J} ^2.
\eqno (3.17)
$$
\item{(2)} The constants $C_{I,J}$ for
each  weight $\alpha$ of $\Lambda \otimes \Lambda ^*$ satisfy :
$$
0=\sum _{{I,J=1; \atop \alpha _{IJ} = \alpha}} ^N C_{I,J} ^2 
(v_I  - v_J ). 
\eqno (3.18)
$$
\item{(3)} For all $\G$, except $E_8$, there exists a vector $d$ with $D= d 
\cdot (h \oplus \tilde h) ~ 
\in \H$, such that for all $I\not=J$, we have
$$
 s C_{I,J}  d \cdot (u_I - u_J)
=\sum _{K \not= I,J} C_{I,K} C_{K,J} 
\bigl \{ \wp (\alpha _{IK} \cdot x  ) - \wp (\alpha _{KJ} \cdot x) 
\bigr \}.
\eqno (3.19)
$$
For the case of $E_8$, $\Delta \not=0$, and we should retain here the full 
statement of Theorem 1, (3), but with $\Phi _{IJ}$ given by (3.16).
\item{(4)} Conditions (1) and (2) imply that whenever  $\alpha _{IJ} \notin 
\R 
(\G)$, we have
$$
 C_{IJ} =0.
\eqno (3.20)
$$

\medskip

Conditions (1), (2) and (3) of Theorem 2 are readily derived from Theorem 1 
by
using the fact that all $\Phi _{IJ}$ are equal as given in (3.16), that
$C_{I,J}$ is symmetric, that $\Delta =0$ (for $\G \not=E_8$) and that the 
functions $\wp (\alpha \cdot x)$ and $\wp (\beta \cdot x)$ are linearly 
independent when $\beta \not= \pm \alpha$.   

\medskip

To show (4), we use the fact that when $\alpha \notin \R(\G)$, we have
$m_{|\alpha|}=0$. Projecting condition (2) onto a vector $u_L$, and using 
(3.3),
conditions (1) and (2) then reduce to
$$
\eqalign{
0= & \sum _{I,J;~\alpha _{IJ} = \alpha} C_{I,J}^2 \cr
0= & \sum _{I,J;~\alpha _{IJ} = \alpha} C_{I,J}^2 (s^2 (\delta _{I,L} - 
\delta
_{J,L} ) - \alpha \cdot \lambda _L ) \cr}
\eqno (3.21)
$$
The term proportional to $\alpha \cdot \lambda _L$ in the second equation
vanishes in view of the first one. In the remaining equation, let $\alpha
\notin \R (\G)$ be such that it can be written as $\alpha = \lambda _I -
\lambda _J$ for some weights $\lambda _I$ and $\lambda _J$. Now choose $L=I$ 
: 
it follows that $C_{I,J}=0$. Clearly this result holds for any $I,J$
such that $\alpha = \lambda _I - \lambda _J$, so that (4) immediately 
follows.

\bigskip

For the twisted elliptic Calogero-Moser systems associated with non-simply 
laced $\G$, the functions $\Phi _{IJ}$ in (3.6) cannot be all equal. 
One is left to using the general Theorem 1, although we shall find that a 
solution may be obtained with $C_{I,J} = C_{J,I}$, which we shall henceforth 
assume. The precise expressions will depend upon each case and will be 
obtained 
in \S VI, for $\G =B_n, C_n, F_4$. For $\G = G_2$, we have been able to solve 
the conditions of Theorem 1 only partially.

\bigskip
\bigskip

\centerline{{\bf IV. THE CLASSICAL LIE ALGEBRAS : UNTWISTED CASES}}

\bigskip

We make use of Theorem 2, presented in \S III, to construct Lax pairs with 
spectral parameter for the untwisted elliptic Calogero-Moser systems 
associated 
with the classical Lie algebras $B_n$, $C_n$ and $D_n$. (The
case of the algebra $A_n$ was already discussed in \S II.) In
\S (a), we find the Lax pairs for the $BC_n$ root system (to be defined 
below), 
and deduce the Lax pairs for $B_n$, $C_n$ and $D_n$ in their fundamental 
representations. The results are summarized in Theorem 3. In \S (b), we 
indicate how the symmetric space construction of the Lax pairs, given in 
Perelomov [14], is recovered as a special case of our results. The general 
formalism developed in \S III is flexible enough to describe Lax pairs in
representations other than the fundamental. To illustrate this result, we 
present three examples, in
\S (c) : $A_n$ in a rank $p$ totally anti-symmetric tensor
representations; in \S (d) : $B_n$ and $D_n$ in a spinor
representation.  

\bigskip

\noindent
{\bf (a) Untwisted Elliptic Calogero-Moser Systems for Classical Lie 
algebras}

\medskip

To obtain the Lax pairs for the classical Lie algebras $B_n$, $C_n$ and 
$D_n$, 
it is convenient to derive the Lax pair for the untwisted elliptic 
Calogero-Moser system associated with the root system $\R (BC_n) = \R(B_n) 
\cup 
\R(C_n)$. While this root system is not properly associated with a 
finite-dimensional simple Lie algebra, Theorem 2 nonetheless still applies.
We then deduce the Lax pair for each of the classical Lie algebras by setting 
one of the independent couplings in the $BC_n$ system to
zero. 

\medskip

As a starting point, we take $\G =B_n$ viewed as a subgroup of $GL(N,\C)$,
with $N=2n+1$, by embedding the fundamental representation of $B_n$ into the
fundamental representation of $GL(N, \C)$. The weights of $\G$ obtained by 
the
decomposition of the adjoint representation of $GL(N, \C)$ under $\G$ then
automatically contains all the root vectors of the $BC_n$  system.
We denote the weights of the fundamental representation of $GL(N, \C)$ by
$u_I$, $I=1,\cdots, N$, and the weigths of the fundamental repesentation
of $B_n$ by $\lambda _I$, just as in (3.1). Since the rank of the weight 
system $\lambda$ is $n$, we may express all weights in an orthonormal basis 
of  
vectors $e_i$, $i=1,\cdots ,n$, so that
$$
\eqalign{
\lambda _i = & ~+e_i, \qquad         i=1,\cdots ,n \cr 
\lambda _{n +i} =&~ - e_i, \qquad  i=1,\cdots ,n \cr
\lambda _N = & ~0,  \quad \ \qquad N = 2n+1 \cr
}
\eqno (4.1)
$$
 It is straightforward to work out the decomposition (3.1) : we find $s^2=2$, 
$v_{n+i}=v_i$, for all $i=1,\cdots ,n$ and
$$
\eqalign{
\sqrt {2} u_i     = & + \ e_i   + v_i
\cr
\sqrt {2} u_{n+i} = & \ - e_i + v_i 
\cr
\sqrt 2 u_N     = &  ~ v_N
\cr
}
\eqno (4.2)
$$
The decomposition of the roots of $GL(N, \C)$ into weights of $B_n$ 
immediately
follows from (4.2) and yields three orbits. We have weights of $B_n$ of 
length$^2= 2$ (which may be viewed as roots of $D_n$) 
$$
\eqalign{
\sqrt 2 (u_i - u_j)            = &  +e_i - e_j + v_i - v_j\qquad i\not=j 
\cr
\sqrt 2 (u_{n +j} - u_{n +i} ) = &  +e_i - e_j - v_i + v_j\qquad i\not=j 
\cr
\sqrt 2 (u_i - u_{n +j})       = &  +e_i + e_j + v_i - v_j\qquad i\not=j 
\cr 
\sqrt 2 (u_{n +i} - u_j)       = &  -e_i - e_j + v_i - v_j\qquad i\not=j,
\cr}
\eqno (4.3a)
$$
weights of length$^2= 4$ (additional roots for $C_n$) 
$$
\eqalign{ 
\sqrt 2 (u_i - u_{n +i})     = & +2e_i \cr
\sqrt 2 (u_{n +i} - u_i)     = & -2e_i, \cr}
\eqno (4.3b)
$$
and weights of length$^2 = 1$ (additonal roots for $B_n$)
$$
\eqalign{
\sqrt 2 (u_i - u_N)         = &  +e_i + v_i -  v_N \cr
\sqrt 2 (u_N - u_{n +i})    = &  +e_i - v_i +  v_N \cr
\sqrt 2 (u_{n +i} - u_N)    = &  -e_i + v_i -  v_N \cr
\sqrt 2 (u_N - u_i)         = &  -e_i - v_i +  v_N. \cr
}
\eqno (4.3c)
$$
Following the general construction of the Lax pair for this system in (3.5),
(3.6) and (3.16), we have three couplings $m_2$, $m_4$ and $m_1$, namely one 
for each of the above orbits under the Weyl group of $B_n$, and the Lax pair 
is 
given by (3.5), (3.6) and (3.16).
It remains to satisfy the three conditions of Theorem 2 in order to guarantee
closure of  the Lax equation onto the Calogero-Moser system.

\medskip

We begin by verifying conditions (1) and (2) of Theorem 2. There are two
distinct roots of $GL(N,\C)$  that project to $e_i-e_j$ for each $i\not=j$ :
$\sqrt 2 (u_i - u_j)$ and $\sqrt 2 (u_{\nu +j} - u_{\nu +i})$. Conditions 
(1) and (2) on these roots read respectively
$$
\eqalign{
2m_2 ^2 = & C_{i,j} ^2 + C_{n + j, n +i} ^2 \cr
0       = & C_{i,j} ^2 (v_i - v _j)  
          + C_{n + j, n +i} ^2 (-v_i + v_j). \cr}
\eqno (4.4)
$$
Proceeding analogously for the roots $e_i + e_j$, $i<j$, $2e_i$ and $e_i$,
and solving by using the linear independence of the vectors $v_i$, we find
$$
\eqalign{
m_2 ^2 = & ~C_{i,j} ^2 = C_{n + i , n +j} ^2
       = C_ {n +i , j} ^2  
\qquad \qquad i\not=j \cr
2m_4 ^2 = & ~C_{i, n +i} ^2 \cr
m_1 ^2 = & ~C_{i,N} ^2 = C_{n +i, N} ^2 \cr}
\eqno (4.5)
$$
We shall obtain a solution for the Lax pair in Theorem 2 by choosing all 
square
roots of the above relations with the same sign,
$$
\eqalign{
m_2          = & ~C_{i,j}= C_{n +i,n +j}=C_ {n +i,j}, \qquad i\not=j, 
\cr
\sqrt 2 m_4  = & ~C_{i, n +i},  \cr
m_1  = & ~C_{i,N}  = C_{n +i, N}. \cr}
\eqno (4.6)
$$

\medskip

Next, we verify condition (3) of Theorem 2, using the results of (4.6). 
Because 
of antisymmetry of (3.19) under the interchange of $I$ and $J$, there are 6 
cases to be analyzed (here, $i,j=1,\cdots ,n$) :
(1) $I=i,~J=j$; (2) $I=i, ~J=N$; (3) $I=i, ~J=n+j$ with $i\not=j$; (4) 
$I=i, ~J=n+i$; (5) $I=N, ~J=n+j$; and (6) $I=n +i,~J=n+j$. A
solution may be obtained in which $d \cdot e_i=0$. Then, case (4) is 
satisfied
automatically, cases (3) and (6) yield the same equations as case (1), and 
case
(5) yields the same equation as case (2). Thus, there remain just two cases :
(1) and (2), which yield the following equations
$$
\eqalign{
m_2 d \cdot (v_i - v_j)
= & \sum _{k\not= i,j} m_2 ^2 \big [ \wp (x_i - x_k) -
\wp (x_k - x_j)  + \wp (x_i + x_k) - \wp (x_k + x_j) \big ] \cr
& \qquad \qquad 
+ m_1 ^2 \big [\wp (x_i) - \wp (x_j)\big ] 
+ \sqrt 2  m_2 m_4 \big [\wp (2x_i) - \wp (2x_j) \big ] 
\cr
& \cr
m_1 d \cdot (v_i -  v_N)
= & \sum _{k\not= i} m_1 m_2  \big [ \wp (x_i - x_k) 
    + \wp (x_i + x_k) - 2\wp (x_k ) \big ] \cr
& \qquad \qquad
+\sqrt 2 m_1 m_4 \big [\wp (2x_i) - \wp (x_i)\big ] \cr
}
\eqno (4.7)
$$

\medskip

Without loss of generality, we assume that $m_2\not=0$, since for $m_2=0$, 
the
system would decompose into a set of non-interacting one dimensional systems.
Thus, the first equation in (4.7) is non-trivial, and its most general 
solution is given by
$$
d\cdot v_i =  d_0 + {m_1 ^2 \over m_2} \wp (x_i) + \sqrt 2 m_4 \wp (2x_i)
+  \sum _{k\not=i} m_2 \big [ \wp (x_i - x_k) + \wp (x_i + x_k) \big ],
\eqno (4.8)
$$
where $d_0$ is an arbitrary function of $x$ which is independent of $i$. 
Substituting this solution for $d\cdot v_i$ into the second equation in (4.7) 
yields
$$
m_1   d\cdot v_N 
= m_1 d _0 +  m_1(-2m_2 + \sqrt 2 m_4 + {m_1 ^2 \over m_2}) \wp (x_i) +  \sum 
_k
2 m_1 m_2 \wp (x_k) . 
\eqno (4.9)
$$
The left-hand side is independent of $i$, hence the right-hand side must
also be independent of $i$, which requires that $ m_1(m_1 ^2 - 2m_2^2 + \sqrt 
2
m_2m_4) =0$. Once this condition is
satisfied, (4.9) yields the component of the vector $d$ along $v_N$, and
integrability of the associated Calogero-Moser system is guaranteed. 
Henceforth, we choose $d_0$ so that $d\cdot v_N=0$.
The result may be summarized in the following Theorem 3, in which we make the
matrix form of the Lax operators completely explicit.

\medskip

\noindent
{\bf Theorem 3 : Lax pair for the $BC_n$, $B_n$, $C_n$ and $D_n$ Systems}

\medskip

The untwisted elliptic Calogero-Moser Hamiltonian of (2.1) with root system 
$\R (BC_n)$ is integrable if  
$$
m_1(m_1 ^2 - 2 m_2 ^2 + \sqrt 2 m_2 m_4) =0.
\eqno (4.10)
$$
It reduces to the untwisted elliptic Calogero-Moser systems for the classical 
Lie algebras $B_n$,
$C_n$ and $D_n$ by the following choice of couplings
$$
\eqalign{
B_n & \qquad \qquad m_4=0, \ \ m_1^2 = 2m_2 ^2 \cr
C_n & \qquad \qquad m_1=0, \ \  \cr
D_n & \qquad \qquad m_1=0, \ \ m_4=0. \cr}
\eqno (4.11)
$$
In all of these cases, the Calogero-Moser system admits a Lax pair with
spectral parameter, given by the following $(2n+1) \times (2n+1)$ 
matrix-valued
functions  
$$
\eqalign{
L= & \ P + X 
\qquad \qquad 
P = {\rm diag} (p_1, \cdots , p_n;-p_1, \cdots , -p_n;0)
\cr
M= & \ D + Y
\qquad \qquad
D = {\rm diag} (d_1, \cdots , d_n; +d_1, \cdots , +d_n;0)
\cr}
\eqno (4.12a)
$$
The matrices $X$ and $Y$ are given by
$$
X = \left ( \matrix{  A     & B_1    & C_1 \cr
                     B_2    & A^T    & C_2 \cr
                     C_2 ^T & C_1 ^T & 0   \cr } \right )
\qquad \qquad
Y = \left ( \matrix{  A'       & B'_1      & C'_1 \cr
                     B'_2      & A'^T       & C'_2 \cr
                     C'_2{} ^T & C'_1{} ^T & 0    \cr } \right ),
\eqno (4.12b)
$$
where the superscript $T$ stands for transposition.
The entries of the matrix $X$ are defined by (with $i,j = 1 , \cdots , n$)
$$
\eqalign{
A_{ij}  = & \ m_2 (1-\delta _{ij}) \Phi (+x_i-x_j, z) \cr
B_{1ij} = & \ m_2 (1-\delta _{ij}) \Phi (+x_i+x_j,z )                        
+ \sqrt 2 m_4 \delta _{ij} \Phi (2x_i,z) \cr
B_{2ij} = & \ m_2 (1-\delta _{ij}) \Phi (-x_i-x_j,z ) 
+ \sqrt 2 m_4 \delta _{ij} \Phi
(-2x_i,z) \cr
C_{1i}  = & \ m_1 \Phi (+x_i,z) \cr
C_{2i}  = & \ m_1 \Phi (-x_i,z), 
\cr}
\eqno (4.12c)
$$
while the entries of the matrix $Y$  are as in (4.12c), but with $A,\ B, C$
replaced by $A',\ B' , \ C'$ and $\Phi$ replaced by $\Phi '$.  The
entries $d_i= d\cdot v_i$ of the matrix $D$ are as given in (4.8).

\bigskip

\noindent
{\bf Remarks}

\medskip

\noindent (1) 
The Lax operators for the Lie algebras $C_n$ and $D_n$ are
effectively of dimension $2n$ instead of $2n+1$, as is seen in (4.12) by
setting $m_1=0$, so that $C_1=C_2=C_1'=C_2'=0$.

\medskip

\noindent (2) 
The Lax pairs with spectral parameter for the Calogero-Moser system
associated with the root system $BC_n$ and hence with the Lie algebras $B_n$,
$C_n$ and $D_n$ coincide with the known Lax pairs without spectral parameter
[14] at the three half periods $\omega _1$, $\omega _2$ and $\omega _3$. 
This is seen by evaluating $\Phi$ at these values
$$
\eqalign{
\Phi (x, \omega _1) = & \ \rho {\cn (\rho x) \over \sn (\rho x)} \cr
\Phi (x, \omega _2) = & \ \rho {\dn (\rho x) \over \sn (\rho x)} \cr
\Phi (x, \omega _3) = & \ \rho {1 \over \sn (\rho x)}, \cr}
\eqno (4.13)
$$
where $\rho ^2 = \wp (\omega _1) - \wp (\omega _2)$, and substituting these
expressions into (4.12). 

\medskip

\noindent (3) 
The three half periods $\omega _a$, $a=1,2,3$ are the only values of the
spectral parameter for which the function $\Phi$ has an extra reflection
symmetry : $\Phi (-x,\omega _a) = - \Phi (x,\omega _a)$, and  $\Phi '
(-x,\omega _a) =  \Phi ' (x,\omega_a)$, so that
$$
\eqalign{
A^T = -A, \qquad B_2 =& \ - B_1  \qquad C_2 = - C_1 \cr
A'^T = +A', \qquad B'_2 =& + B'_1,   \qquad C'_2 = + C'_1. \cr}
\eqno (4.14)
$$
At these values, $X$ is anti-symmetric, and thus belongs to the Lie algebras
$B_n$, $C_n$ or $D_n$ respectively. 

\bigskip

\noindent
{\bf (b) Construction from Symmetric Spaces of Affine Lie Algebras}

\medskip

A natural setting for the construction of the Lax pairs {\it without spectral
parameter} was obtained in Perelomov [14] in terms of symmetric spaces of 
classical Lie algebras. That construction uses specific reflection symmetry
properties  of the Lax pair that do not continue to hold for arbitrary values 
of
the spectral parameter. However, a generalization of these reflection 
symmetry
properties, in which the sign of the spectral parameter $z$ is reversed,
does hold for all values of $z$. To establish this, we use the reflection
property of  $\Phi$ (see Appendix \S B) :
$$
\Phi (-x,z) = - \Phi (x,-z).
\eqno (4.15)
$$
Upon introducing the conjugation matrix $S$, with $S^2=I_{2n+1}$,
$$
S\equiv \left ( \matrix{
0   & I_n & 0 \cr
I_n & 0   & 0 \cr
0   & 0   & 1 \cr} \right ),
\eqno (4.16)
$$
we have the following reflection relations 
$$
\eqalign{
P(z) ^T = & \ +P(-z) = -S P(z) S \cr
D(z) ^T = & \ +D(-z) = +S D(z) S \cr
X(z) ^T = & \ -X(-z) = +S X(z) S \cr
Y(z) ^T = & \ +Y(-z) = +S Y(z) S. \cr}
\eqno (4.17)
$$
Here, we have included a fictitious $z$-dependence for $P$, for the sake of
respecting the pattern exhibited by $D, \ X$, and $Y$.

\medskip

We shall now formulate the Lax pair in more geometrically intrinsic terms.  
We 
begin by defining the ring $F[z]$ of elliptic functions that are holomorphic,
except for a singularity at $z=0$. This ring is generated by the functions 
$\Phi 
(x,z)$ and its $x$-derivatives. We define an affine Lie algebra $\G [z]$ by
$$
\G [z] \equiv \G \otimes F[z].
\eqno (4.18)
$$
$\G [z]$ may be viewed as the space of elliptic functions with values in the
Lie algebra $\G$, that are  holomorphic
except for a singularity at $z=0$ .

\medskip

The Lax operators $L(z)$ and $M(z)$ belong to $\G [z] = SL(2n+1)[z]$ for 
$B_n$,
and  $ \G [z] = SL(2n)[z]$ for $C_n$ and $D_n$, which may be decomposed into
$\G [z] = \K \oplus \N$, where 
$$
\eqalign{ 
M(z) \in \K & = \{ g(z) \in \G[z] : g(-z) = + Sg(z)S \} \cr
L(z) \in \N & = \{ g(z) \in \G[z] : g(-z) = - Sg(z)S \} .\cr} 
\eqno (4.19)
$$
The coset $\N$ is the tangent space to a symmetric space, since we have 
$$
[\K, \K ] \in \K;
\qquad
[\K, \N ] \in \N;
\qquad
[\N, \N ] \in \K.
\eqno (4.20)
$$
At one of the three half periods, this symmetric space may be identified
in terms of classical Lie algebras and is given by
$$
\N \sim {SL(2n+1) \over SL(n) \times SL(n+1)}.
\eqno (4.21)
$$
This form of the Lax pair is familiar from the symmetric space construction 
of
the $BC_n$ root system in Perelomov [14]. However, at generic values of the
spectral parameter, the space $\N$ is characterized by a reflection symmetry
relation (4.19) that reverses the sign of the spectral parameter.

\bigskip

\noindent
{\bf (c) Calogero-Moser for $A_n$ in Anti-symmetric Tensor 
Representations}

\medskip

The rank $p$ anti-symmetric tensor representation of the Lie algebra $A_n$
has dimension 
$$
N=\left ( \matrix{n +1 \cr p} \right ) \equiv {(n+1) !\over p! (n+1-p)!}
\eqno (4.22)
$$ 
and will be denoted here by $\tp$. Its complex conjugate is ${\bf
T}_{\bf n+1-p}$ and will also be denoted by $\tp^*$.  In the
standard basis for the weight space of $A_n$ of
$n+1$ orthonormal vectors $e_i$, $i=1, \cdots , n+1$, (with $e_0 = (e_1 + e_2 
+
\cdots + e_{n+1}) /(n+1)$), the weights of $\tp$ are 
$$
\eqalign{
\tp = \{ \lambda  = e_{i_1} + e_{i_2} + & \cdots + e_{i_p} - p e_0, 
\quad   
i_1 < i_2 < \cdots < i_p \}  \cr
\lambda  \cdot \lambda  = &  p(n+1-p) / (n+1). \cr }
\eqno (4.23)
$$
All weights of $\tp$ have the same length and lie in a single Weyl orbit. The
precise correspondence between the labels $I$ and
$(i_1 ~i_2 ~ \cdots ~ i_p)$ is immaterial,
\footnote{*}{For any representation $\Lambda$ in which each weight vector
$\lambda$ occurs with multiplicity precisely 1, one may parametrize the 
labels 
$I$ directly by the weights $\lambda$ themselves, since the correspondence is 
one to one.}
 since the weights are permuted into
one another under the action of the Weyl group of $A_n$. 

\medskip

\noindent
{\bf Theorem 4 : Lax pairs for A$_n$ in anti-symmetric tensor
representations}

\medskip

The Calogero-Moser system for the Lie algebra $A_n$ admits a Lax pair with
spectral parameter in the anti-symmetric tensor representation $\tp$ given by
(3.5), (3.6) and (3.16) with
$$
C_{\lambda , \mu } =  \left \{ \matrix{ m_2 & (\lambda - \mu )^2 =2 \cr
                               0 & {\rm otherwise,} \cr} \right . 
\qquad \qquad
s d \cdot u_\lambda =    \sum _{\lambda \cdot \delta =1} m_2 \wp (\delta
\cdot x),
\eqno (4.24)
$$
where $\lambda$ and $\mu$ run over the weights of $\tp$, as explained in the 
last footnote.

\medskip

To prove this Theorem, we show that the conditions of Theorem 2 are 
satisfied.
We begin by describing the necessary group theory.
The weights of $GL(N,{\bf C})$ are denoted by the $N$ orthonormal vectors
$u_\lambda$, where $\lambda$ runs over all the weights of $\tp$. The
decomposition (3.1) is given by 
$$
s u_\lambda = \lambda + v _\lambda
\qquad {\rm with} \qquad  s^2 = \left ( \matrix{ n-1 \cr p-1} \right )
\eqno (4.25)
$$
where the vectors $v_\lambda $ are orthogonal to the root space of $A_n$.
The roots of $GL(N, {\bf C})$ decompose under $A_n$ as follows
$$
s(u_\lambda - u _\mu) = \lambda - \mu + v_\lambda - v_\mu
\qquad \lambda \not= \mu.
\eqno (4.26)
$$
Under the Weyl group of $A_n$, the roots transform in different orbits 
$[\uq]$,
which are precisely the Weyl orbits occurring in the tensor product $\tp
\otimes \tp ^*$. The orbits $[\uq]$ are defined by
$$
[\uq] \equiv \{ 
\alpha = e_{i_1} + \cdots + e_{i_q}  - e_{j_1} - \cdots - e_{j_q},
\ {\rm all} ~ i_k , \ j_l ~ {\rm different}\}
\qquad
\alpha ^2 =  2q.
\eqno (4.27)
$$
The representation $({\bf U} _{\bf 1})$ is the adjoint representation of 
$A_n$.
The decomposition of the roots of $GL(N, \C)$ into orbits of the Weyl group 
of 
$A_n$ is given by
$$
\tp \otimes \tp ^* = \bigoplus _{q=0} ^p 
\left ( \matrix{ n+1-2q \cr p-q} \right )
[\uq].
\eqno (4.28)
$$
In order to reproduce the Calogero-Moser system for $A_n$, only the coupling 
of
the roots of $A_n$ can be non-zero. We denote this unique coupling by $m_2$.
Thus, by (4) of Theorem 2, we see right away that $C_{\lambda, \mu}=0$ unless
$\lambda - \mu \in \R(A_n)$.

\medskip

Conditions (1) and (2) for a root $\alpha \in \R (A_n)$, are given by
$$
\eqalign{
s^2 m_{2}^2 = & ~ \sum _{\alpha = \lambda - \mu} C_{\lambda , \mu} ^2 \cr
0= & ~ \sum _{\alpha = \lambda - \mu} C_{\lambda , \mu} ^2 (v_\lambda - v
_\mu) \cr
}
\eqno (4.29)
$$
If a given root $\alpha$ can be written as $\alpha = \lambda - \mu$, we take
the inner product of the second equation in (4.29) with $u_\lambda$. To
evaluate this, we use (3.3) and we find $C_{\lambda , \mu }^2 =  m_{2 } ^2 $.
Using the expression for $s^2$ of (4.25) and of the multiplicity of $[\uq]$
for $q=1$ of (4.28), we see that the first equation in (4.29) holds for all
roots. It remains to verify that the second condition in (4.29) holds  when
projected onto a vector $u_\sigma$ for an arbitrary weights $\sigma \in \tp$.
Using again (3.3), we find 
$$
\alpha \cdot \sigma ~ m_2 ^2  = \sum _{\alpha = \lambda - \mu}
C_{\lambda , \mu }^2 (\delta _{\lambda , \sigma} - \delta _{\mu , \sigma}).
\eqno (4.30)
$$
Since $\alpha $ is now a root of $A_n$, and in view of (4.23), $\alpha \cdot
\sigma$ can take only the values $\alpha \cdot
\sigma = -1, ~0,~+1$. When $\alpha \cdot \sigma =0$, $\sigma \pm \alpha$ is 
not
a root, and all terms in (4.30) vanish separately. When $\alpha \cdot \sigma 
=
+ 1$, $\sigma - \alpha $ is a weight $\beta$, while $\sigma + \alpha $ is not 
a
weight. As a result, $\alpha = \sigma - \beta$, and we recover $C_{\sigma,
\beta} ^2 = m_2 ^2$. Similar reasoning for $\alpha \cdot \sigma =-1$ yields
again the same result, and (4.30) is satisfied in all cases. It is possible 
to
find a Lax pair by choosing all the square roots to have the same sign, so 
that 
$C_{\lambda, \mu}= m_2$ when $\lambda - \mu \in \R
(A_n)$, whence the first equation of (4.24).

\medskip

It remains to satisfy condition (3) of Theorem 2, i.e. equation (3.19), which 
may be re-expressed as
$$
s C_{\lambda, \mu} d \cdot (u_\lambda - u _\mu)
= \sum _{\kappa \not= \lambda , \mu}
C_{\lambda , \kappa} C_{\kappa, \mu} \{ 
\wp ((\lambda - \kappa) \cdot x) - \wp ((\kappa - \mu)\cdot x) \}.
\eqno (4.31)
$$
Here, the weights $\lambda,~\kappa,~\mu$ all belong to $\tp$. Because
$C_{\lambda, \mu}$ is non-zero only when $\lambda - \mu$ is a root of $A_n$,
we may replace the summation over $\kappa $ in (4.31) by a summation over
roots : $\delta = \lambda - \kappa$ in the first sum, and $\delta = \kappa -
\mu$ in the second sum. 
It is convenient to express the conditions that $\lambda -\delta$ and
$\mu + \delta$ are weights while $\lambda - \mu - \delta$ is a root, in terms
of inner products of these weights :
$$
s C_{\lambda, \mu} d \cdot (u_\lambda - u _\mu)
= \sum _{{\delta ^2 =2; ~\lambda \cdot \delta = 1; \atop \mu \cdot \delta = 1 
-
\half (\lambda - \mu)^2}} m_2 ^2 \wp (\delta \cdot x) -
\sum _{{\delta ^2 =2; ~\mu \cdot \delta = -1; \atop \lambda \cdot \delta = -1 
+
\half (\lambda - \mu)^2}} m_2 ^2 \wp (\delta \cdot x). 
\eqno (4.32)
$$
We now analyze this equation for $\lambda - \mu$ belonging to each of
the possible Weyl orbits $\uq$ occurring in (4.28), i.e. for each of the
possible value of $(\lambda - \mu )^2 = 2q$, $q=1,\cdots , p$. 
For $q\geq 3$, the l.h.s. vanishes because $C_{\lambda, \mu}=0$ in this
case by the first equation in (4.24). The r.h.s. of (4.32) also vanishes :
no roots $\delta$ can satisfy the inner product conditions since 
$\mu \cdot \delta \leq -2$ in the first sum and $\lambda \cdot \delta \geq 2$ 
in
the second sum. Indeed, if $\delta$ is a root, then it is clear from (4.23) 
that
$\lambda \cdot \delta$ and $\mu \cdot \delta$ can only take on the values
$-1,~0,~+1$. For $q=2$, the left hand side vanishes, and so does the right
hand side since the first and second sum cancel one another. Finally,
for $q=1$, (4.32) is readily solved and we obtain the last equation in 
(4.24).

\bigskip

\noindent
{\bf (d) Calogero-Moser for $B_n$ and $D_n$ in Spinor 
Representations }

\medskip

The spinor representation of $B_n$ (denoted by {\bf S}, of dimension
$N=2^n$) has a single Weyl orbit of weights $\lambda$ 
$$
{\bf S} = \{ \lambda = \half \sum _{i=1} ^n \epsilon _i e _i ~\},
\qquad \qquad 
\lambda ^2 = { n \over 4}.
\eqno (4.33)
$$
Here and below, $\epsilon _i$ can take the values $\pm 1$.
The spinor representations of $D_n$ (denotes by {\bf S}$_\pm$, of dimension
$N=2^{n-1}$) each have a single Weyl orbit of weights of $D_n$, given by
$$
{\bf S}_\pm = \{ \lambda   = \half \sum _{i=1} ^n \epsilon _i e_i,
\quad  \prod _{i=1} ^n \epsilon _i =\pm 1~\},
\qquad\qquad
\lambda  ^2 = {n \over 4}.
\eqno (4.34)
$$
The Lax pairs in these spinor representations are described by the following 
result.

\medskip

\noindent
{\bf Theorem 5 : Lax pairs for $B_n$ and $D_n$ in spinor representations}

\item{(1)} The Calogero-Moser system for $B_n$ admits a Lax pair, with 
spectral
parameter and {\it two independent couplings $m_1$ and $m_2$}, in the spinor
representation {\bf S} given by (3.5), (3.6), (3.16) and
$$
C_{\lambda , \mu } =  \left \{ \matrix{ 
m_1/\sqrt 2 & (\lambda - \mu )^2 =1 \cr
m_2 & (\lambda - \mu )^2 =2 \cr
  0 & {\rm otherwise,} \cr} \right . 
\qquad \qquad
sd\cdot u_\lambda =  \sum _{\delta ^2=2;~ \lambda \cdot \delta =1} 
      m_2\wp (\delta \cdot x).
\eqno (4.35)
$$
 
\item{(2)} The Calogero-Moser system for $D_n$ admits a Lax pair with 
spectral
parameter in the spinor representations {\bf S}$_\pm$ given by (3.5), (3.6),
(3.16)  and
$$
C_{\lambda , \mu } =  \left \{ \matrix{ 
m_2 & (\lambda - \mu )^2 =2 \cr
  0 & {\rm otherwise,} \cr} \right . 
\qquad \qquad
sd\cdot u_\lambda =  \sum _{ \lambda \cdot \delta =1} 
      m_2\wp (\delta \cdot x).
\eqno (4.36)
$$

\medskip

We begin by proving (1) in detail.
 The weights of $GL(N,\C )$ are denoted
by $u _ \lambda$ and the decomposition (3.1) is given by
$$
su_\lambda = \lambda + v_\lambda 
\qquad {\rm with} \qquad
s^2 = 2^{n-2}.
\eqno (4.37)
$$
The tensor product ${\bf S} \otimes {\bf S}$ decomposes into the Weyl orbits 
of the anti-symmetric tensor representations of rank
$p$, which we denote by $[\tp]$, and which have weight system
$$
[\tp] = \{ \alpha = \pm e_{i_1} \pm e_{i_2} \pm \cdots \pm e_{i_p}, \
 i_1 < i_2 < \cdots < i_p \}; 
 \qquad \alpha ^2 = p.
\eqno (4.38)
$$
The decomposition of the roots of $GL(N, \C)$ into orbits of the Weyl group 
of
$A_n$ is given by the tensor productdecomposition of ${\bf S} \otimes {\bf 
S}$ 
$$
{\bf S} \otimes {\bf S} = \bigoplus  _{p=0} ^n 2^{n-p} [\tp].
\eqno (4.39)
$$

\medskip

In order to reproduce the Calogero-Moser system for $B_n$, only the couplings
of the roots of $B_n$ can be non-zero. We denote the couplings associated 
with
short and long roots by $m_1$ and $m_2$ respectively. Thus, by (4) of Theorem 
2,
we see right away that $C_{\lambda , \mu }=0$ unless $\lambda - \mu \in \R
(B_n)$.

\medskip

Conditions (1) and (2) for a root $\alpha \in [\tp]$, with $p=1,2$ are given 
by
$$
\eqalign{
s^2 m_p^2 = &  \sum _{\alpha = \lambda - \mu} C_{\lambda , \mu} ^2 \cr
0= &  \sum _{\alpha = \lambda - \mu} C_{\lambda , \mu} ^2 (v_\lambda - v
_\mu). \cr
}
\eqno (4.40)
$$
If a given root $\alpha \in [\tp]$ can be written as $\alpha = \lambda -
\mu$, then taking the inner product of the second equation in (4.40) with 
$u_\lambda$ and using (3.3) yields 
$$
C_{\lambda , \mu }^2 = \half p m_p ^2, 
\eqno (4.41)
$$
for any $\lambda$ and $\mu$ such that $\alpha = \lambda - \mu$. Substituting
this result into the first equation of (4.40) and using the formula for the
multiplicitly of the orbit $[\tp]$ of (4.39), we find that the equation is
automatically satisfied for all roots.
It remains to verify that the second condition in (4.40) holds for 
$p=1,2$ when projected onto a vector $u_\sigma$ for an arbitrary weight
$\sigma \in {\bf S}$. Using again (3.3) to evaluate this product, we find
$$
\alpha \cdot \sigma ~ m_{\alpha ^2} ^2  = \sum _{\alpha = \lambda - \mu}
C_{\lambda , \mu }^2 (\delta _{\lambda , \sigma} - \delta _{\mu , \sigma}).
\eqno (4.42)
$$
Since $\alpha $ is a root of $B_n$, and $\sigma$ is a weight of ${\bf S}$ as 
in
(4.33), we can only have the values $\alpha \cdot \sigma = 0, \pm \alpha ^2 
/2$. 
When $\alpha \cdot \sigma =0$, $\sigma \pm \alpha$ is not a root, and all 
terms in (4.42) vanish separately. When $\alpha \cdot \sigma = \pm \alpha ^2 
/2$,
$\sigma \mp \alpha $ is a weight while $\sigma \pm \alpha $ is not.
Thus, (4.42) is satisfied in all cases. In fact, a Lax pair may be found in
which all square roots of the relation (4.41) are taken with the same sign, 
and
this gives rise to the first equation in (4.35).

\medskip  

It remains to satisfy condition (3) of Theorem 2, i.e. equation (3.19) : 
$$
s C_{\lambda, \mu} d \cdot (u_\lambda - u _\mu)
= \sum _{\kappa \not= \lambda , \mu}
C_{\lambda , \kappa} C_{\kappa, \mu} \{ 
\wp ((\lambda - \kappa) \cdot x) - \wp ((\kappa - \mu)\cdot x) \}.
\eqno (4.43)
$$
Here, the weights $\lambda,~\kappa,~\mu$ all belong to {\bf S}. Because
$C_{\lambda, \mu}$ is non-zero only when $\lambda - \mu$ is a root of
$B_n$, we may replace the summation over $\kappa $ in (4.43) by a
summation over roots, $\delta = \lambda - \kappa$ in the first sum, and
$\delta = \kappa - \mu$ in the second sum. We also separate the summation
over the roots according to the value of $C_{\lambda, \mu}$ and express the 
corresponding conditions on $\delta$ in terms of inner products $\lambda 
\cdot
\delta$ and $\mu \cdot \delta$.   
$$
\eqalign{
s C_{\lambda, \mu} d \cdot (u_\lambda - u _\mu)
=
 & \sum _{{\delta ^2=1;~ 
       2 \lambda \cdot \delta = 1; 
         \atop 
       2 \mu \cdot \delta = 2 -  (\lambda - \mu) ^2}} 
            {1 \over \sqrt 2} m_1 m_2 \wp (\delta \cdot x) \
+  \sum _{{\delta ^2=2;~ 
       \lambda \cdot \delta = 1; 
         \atop 
       2 \mu \cdot \delta = 1 -  (\lambda - \mu) ^2}} 
            {1 \over \sqrt 2} m_1 m_2 \wp (\delta \cdot x) \
\cr
 & + \sum _{{\delta ^2=1;~ 
       2 \lambda \cdot \delta = 1; 
         \atop 
       2 \mu \cdot \delta = 1 -  (\lambda - \mu) ^2}} 
            \half m_1 ^2 \wp (\delta \cdot x) \
+  \sum _{{\delta ^2=2;~ 
       \lambda \cdot \delta = 1; 
         \atop 
       2\mu \cdot \delta = 2 -  (\lambda - \mu) ^2}} 
             m_2 ^2 \wp (\delta \cdot x) \
-  (\lambda \leftrightarrow \mu)
\cr}
\eqno (4.44)
$$
We analyze this equation for each possible orbits of $\lambda - \mu \in
[\tp]$ with value of $(\lambda - \mu )^2 = p$, $p=1,\cdots , n$. Since
in each of the above sums, $\delta$ is a root, it is clear from the form of 
the
weights of {\bf S} in (4.33) that $\lambda \cdot \delta$ and $\mu \cdot
\delta$ can only take on the values $0,~\pm \delta ^2 /2$. For $p\geq 4$, 
there are no roots in any of the 8 sums in (4.44) that satisfy the inner 
product
relations. As a result, for $p\geq 4$, the r.h.s. of (4.44) vanishes, and
in view of the first equation in (4.35), the l.h.s. vanishes as well. For 
$p=3$,
the third and fourth sums in (4.44) vanish since the inner product relations
cannot be satisfied for them. The first and second sums (and their $\lambda
\leftrightarrow \mu$ contribution) precisely cancel one another, the r.h.s. 
of
(4.44) vanishes, and so does the l.h.s. in view of the first equation in 
(4.35).
Finally, the remaining equations for $p=1$ and $p=2$ are found to be
proportional to one another and to
$$
s  d \cdot (u_\lambda - u_\mu) 
=   \sum _{{\delta ^2=2;~ \lambda \cdot \delta =1;
\atop \mu \cdot \delta =0}} 
      m_2\wp (\delta \cdot x)
-\sum _{{\delta ^2=2;~ \lambda \cdot \delta =0;
\atop \mu \cdot \delta =-1}} 
      m_2\wp (\delta \cdot x). 
\eqno (4.45)
$$
But, (4.45) is easily solved and we recover the second equation in (4.35).
 
\medskip

The proof of the Lax pair in (4.36) in one of the two Weyl spinor
representations of $D_n$ is analogous to the case of $B_n$, so we only give 
an
outline here. The conjugates of the spinor representations ${\bf S}_\pm $ 
are obtained by reversing the sign of the weights in (4.34) and
corresponds to ${\bf S}_\pm $ when $n$ is even, but to ${\bf S}_\mp$ when $n$ 
is
odd. Henceforth, we concentrate on the case $+$. The tensor decomposition 
${\bf
S}_+ \otimes {\bf S}_+ ^*$ is given by
$$
{\bf S}_+ \otimes {\bf S}_+ ^*
=\sum _{p=0} ^{[{n \over 2}]} 2^{n-2p-1} [{\bf T} _{{\bf 2p}}]
\eqno (4.46)
$$
Here, $[\tp]$ are again the Weyl orbits of the anti-symmetric tensor
representations of rank $p$, but this time of $D_n$. Their weight system and
length are just as given in (4.38). We now find $s^2 = 2^{n-3}$ and solve
conditions (1) and (2) of Theorem 2 by $C_{\lambda , \mu }^2 = m_2 ^2$, and 
by
taking square roots with the same sign, we recover the first equation in 
(4.36).
Condition (3) reduces to (4.44) but now with $m_1=0$, and is solved 
analogously.

\bigskip
\bigskip

\centerline{{\bf V. THE EXCEPTIONAL LIE ALGEBRAS : UNTWISTED CASES}}

\bigskip

In this section, we apply the construction of \S III, and obtain a Lax pair
with spectral parameter and one independent coupling for each of the five
(untwisted) elliptic Calogero-Moser systems for exceptional Lie algebras. 
The Lax pairs are built out of the following representations : for $E_6$,
$E_7$ and $E_8$, in the representations of dimensions 27, 56 and 248 
respectively. For the case of $E_8$, the analysis is complete only up to the 
determination of certain sign assignments which we have not constructed 
explicitly.  
For $G_2$, we construct Lax pairs
in the representations of dimension 7 (the fundamental) and 8 (the 
fundamental
plus a singlet). For $F_4$, we find a Lax pair in a 27-dimensional
representation which is the direct sum of the fundamental and a singlet. 
For conventions and general information on the group theory used here, we 
refer 
to Appendix \S A.

\bigskip

\noindent
{\bf (a) Untwisted Elliptic Calogero-Moser System for $E_6$}

\medskip

We start by embedding the 27-dimensional representation of highest weight
$(100000)$ into the fundamental representation of $GL(27, {\bf C})$. This
representation is denoted by {\bf 27} for short; its complex conjugate,the
${\bf 27 }^*$ has highest weight $(000010)$. The weights of the {\bf 27} are
given in terms of 6 orthonormal vectors $e_i$, $i=1,\cdots , 6$ by
$$
\lambda  = \left \{ 
\eqalign{
& +{2 \over \sqrt 3} e_6 \cr
& +{1 \over 2 \sqrt 3} e_6 - \half \sum _{i=1} ^5 \epsilon _i e_i
\qquad {\rm with } \qquad \prod _{i=1} ^5 \epsilon _i = 1 \cr
& - {1 \over \sqrt 3} e_6 \pm e_i. \cr} \right .
\eqno (5.1)
$$
Here, $\epsilon _i =\pm 1$. All weights belong to a single Weyl orbit of 
$E_6$, 
denoted by $[100000]$ and
have the same length $\lambda ^2 = 4/3$.

\medskip

\noindent
{\bf Theorem 6 : Lax pair for E$_6$ in the 27}

The untwisted elliptic Calogero-Moser Hamiltonian for $\G=E_6$ admits a Lax 
pair with spectral parameter and one independent coupling in the {\bf 27} of 
$E_6$, given
by (3.5), (3.6), (3.16) and 
$$
C_{\lambda , \mu } =  \left \{ \matrix{ 
m_2 & (\lambda - \mu )^2 =2 \cr
  0 & {\rm otherwise,} \cr} \right .
\qquad \qquad
\sqrt 6 d\cdot u_\lambda =  \sum _{ \lambda \cdot \delta =1} 
      m_2\wp (\delta \cdot x).
\eqno (5.2)
$$

\medskip

To prove this Theorem, we show that the conditions of Theorem 2 are 
satisfied.
The weights of the fundamental representation of $GL(27, {\bf C})$ are
denoted by 27 orthogonal vectors $u_\lambda $, where $\lambda $ runs over the
weights of the {\bf 27} of $E_6$. The decomposition (3.1) is given by $s^2 
= 6$ and
$$
\sqrt 6 u_\lambda = \lambda + v_\lambda,
\eqno (5.3)
$$
where the vectors $v _\lambda$ are orthogonal to the weight space of $E_6$. 
The
roots of $GL(27, {\bf C})$ decompose under $E_6$ as follows
$$
\sqrt 6 (u_\lambda - u _\mu) = \lambda - \mu + v_\lambda - v _\mu
\qquad \lambda \not= \mu.
\eqno (5.4)
$$
Under the Weyl group of $E_6$, these roots transform in different orbits,
which are precisely the Weyl orbits occurring in the tensor product ${\bf 27}
\otimes {\bf 27}^*$ of $E_6$. They are given by
$$
\matrix{
&{\rm Weyl ~ Orbit} & {\rm Multiplicity} & \# {\rm ~ Weights } & 
{\rm Length }^2 \cr
&&&&\cr
{\bf 27} \otimes {\bf 27}^* 
\ : \  & \ [000000] & 27 & 1 & 0 \cr
    & \ [000001] & 6  & 72 & 2 \cr
    & \ [100010] & 1  & 270 & 4 \cr
  }
\eqno (5.5)
$$
The orbits $[000000]$, corresponding to the trivial representation of $E_6$,
do not actually occur in (5.4), since we are restricting to the
off-diagonal elements for which $\lambda \not=\mu$. 

\medskip

Condition (4) of Theorem 2 applies to the weights $\lambda - \mu$ in orbit
$[100010]$, and readily implies that $C_{\lambda , \mu} = 0 $ when
$(\lambda - \mu )^2 =4$.
Conditions (1) and (2) for the roots $\alpha = \lambda - \mu$ in orbit 
$[000001]$, are given by a sum over the 6 possible orbits in which $\alpha$ 
can 
lie 
:
$$
\eqalign{
6 m_2 ^2 = &  \sum _{\lambda - \mu = \alpha} C_{\lambda , \mu }^2 \cr
0        = &  \sum _{\lambda - \mu =\alpha} C_{\lambda , \mu }^2
           (v_\lambda - v _\mu). \cr}
\eqno (5.6)
$$
Taking the inner product of the second equation with an arbitrary vector
$u _\sigma$ with $\sigma \in \R (E_6)$ yields the equation
$$
\alpha \cdot \sigma m_2 ^2 = \sum _{\lambda - \mu = \alpha} C_{\lambda , \mu
}^2 (\delta _{\lambda , \sigma} - \delta _{\mu ,\sigma}).
\eqno (5.7)
$$
Now, since $\alpha$ is a root, and $\lambda$ is in one of the fundamental
representations of $E_6$, the combination $\alpha \cdot \sigma$ can take on 
only
the values $\alpha \cdot \sigma = -1,~ 0,~ 1$. When $\alpha \cdot \sigma =0$,
$\sigma \pm \alpha$ are not weights of the {\bf 27}; thus all terms in (5.7)
vanish separately. When $\alpha \cdot \sigma = \pm 1$, $\sigma \mp \alpha$ is 
a
weight (but $\sigma \pm \alpha $ is not), so that (5.7) yields $C_{\lambda,
\mu} ^2 = m_2 ^2$ when $(\lambda - \mu )^2 =2$. In fact, we shall find that a
solution exists where all square roots have the same sign, and we thus 
recover
the first equation in (5.2).

\medskip

It remains to satisfy condition (3) of Theorem 2, i.e. (3.19). When $\alpha =
\lambda - \mu$ belongs to the orbit $[100010]$ (for which $\alpha ^2 =4$ and
$\lambda \cdot \mu = -2/3$), (3.19) becomes
$$
0=  \sum _{\kappa \cdot \lambda = \kappa \cdot \mu = 1/3}
m_2 ^2 \bigl \{ \wp ((\lambda - \kappa )\cdot x) - \wp ((\kappa - \mu) \cdot 
x).
\bigr \}
\eqno (5.8)
$$
By changing variables $\delta = \kappa - \lambda$ in the first sum and
$\delta = \kappa - \mu$ in the second sum, this equation may be rewritten in
terms of a sum over roots $\delta$ 
$$
0=  \sum _{ \delta \cdot \lambda =-\delta \cdot \mu =-1}
m_2 ^2 \wp ( \delta \cdot x) -
 \sum _{ \delta \cdot \lambda =-\delta \cdot \mu =1}
m_2 ^2 \wp ( \delta \cdot x). 
\eqno (5.9)
$$
Since the argument of the above sum is even in $\delta$, this relation is
automatically satisfied.

\medskip

When $\alpha = \lambda - \mu$ belongs to the orbit $[000001]$ (for which
$\alpha ^2 =2$ and $\lambda \cdot \mu = 1/3$), (3.19) becomes
$$
\sqrt 6 m_2 d \cdot (u _\lambda - u_\mu)  
=  \sum _{\kappa \cdot \lambda = \kappa \cdot \mu = 1/3}
m_2 ^2 \bigl \{ \wp ((\lambda - \kappa )\cdot x) - \wp ((\kappa - \mu) \cdot 
x)
\bigr \}.
\eqno (5.10)
$$
By changing variables to $\delta = \kappa - \lambda$ in the first sum and to
$\delta = \kappa - \mu$ in the second sum, this equation may be rewritten in
terms of a sum over roots $\delta$ 
$$
\sqrt 6 m_2 d \cdot (u_\lambda - u _\mu)
=  \sum _{ \delta \cdot \lambda =1;~\delta \cdot \mu =0}
m_2 ^2 \wp ( \delta \cdot x) -
 \sum _{ \delta \cdot \lambda =0;~ \delta \cdot \mu =1}
m_2 ^2 \wp ( \delta \cdot x). 
\eqno (5.11)
$$
The solution to this equation is readily obtained by extending the first sum
to include $\delta \cdot \mu =\pm1$ and the second sum to include $\delta
\cdot \lambda =\pm 1$ without affecting the left hand side. Doing so, we
recover the second equation in (5.2).

\bigskip

\noindent
{\bf (b) Untwisted Elliptic Calogero-Moser System for $E_7$}

\medskip

We start by embedding the 56-dimensional representation of $E_7$ with highest
weight $(0000010)$ (denoted {\bf 56} for short) into the fundamental
representation of $GL(56, {\bf C})$. The weights of the {\bf 56} are given in
terms of 7 orthonormal vectors $e_i$, $i=1,\cdots ,7$ by
$$
\lambda _I = \left \{ 
\eqalign{
& \pm {1 \over 2} \sum _{i=1} ^6 \epsilon _i e_i
\qquad \qquad \qquad {\rm with } \qquad \qquad  
\prod _{i=1} ^6 \epsilon _i = 1
\cr & \pm  ({1 \over \sqrt 2} e_7 + e_i), 
\quad \pm  ({1 \over \sqrt 2} e_7 - e_i) 
\qquad \qquad i=1,\cdots ,6. \cr} \right .
\eqno (5.12)
$$
All weights belong to a single Weyl orbit of $E_7$, denoted by $[0000010]$
and have the same length $\lambda ^2 = 3/2$.

\medskip

\noindent
{\bf Theorem 7 : Lax pair for E$_7$ in the 56}

The untwisted elliptic Calogero-Moser Hamiltonian for $E_7$ admits a Lax pair 
with spectral parameter and one independent coupling in the {\bf 56} of 
$E_7$, 
given by (3.5), (3.6), (3.16) and 
$$
C_{\lambda , \mu } =  \left \{ \matrix{ 
m_2 & (\lambda - \mu )^2 =2 \cr
  0 & {\rm otherwise,} \cr} \right .
\qquad \qquad
\sqrt {12} d\cdot u_\lambda =  \sum _{ \lambda \cdot \delta =1} 
      m_2\wp (\delta \cdot x).
\eqno (5.13)
$$

The proof is completely analogous to the case of $E_6$. 
The weights of the fundamental representation of $GL(56, {\bf C})$ are
denoted by 56 orthonormal vectors $u _\lambda$, where $\lambda $ runs over
the weights of the {\bf 56} of $E_7$. The decomposition (3.1) is given by
$s^2 = 12$ and 
$$
\sqrt {12} u_\lambda  = \lambda + v_\lambda,
\eqno (5.14)
$$
where the vectors $v_\lambda$ are orthogonal to the weights space of the {\bf
56} of $E_7$. The roots of $GL(56, {\bf C})$ decompose under $E_7$ as follows
$$
\sqrt {12} (u_\lambda - u _\mu) = \lambda - \mu + v_\lambda - v_\mu
\qquad \lambda \not= \mu.
\eqno (5.15)
$$
Under the Weyl group of $E_7$, these roots transform under the different
Weyl orbits that occur in the tensor product ${\bf 56} \otimes {\bf 56}$ of
$E_7$. They are given by
$$
\matrix{
&{\rm Weyl ~ Orbit} & {\rm Multiplicity} & \# {\rm ~ Weights } & 
{\rm Length }^2 \cr
&&&&\cr
{\bf 56} \otimes {\bf 56} 
\ :\  & \ [0000000] & 56 & 1 & 0 \cr
    & \ [1000000] & 12  & 126 & 2 \cr
    & \ [0000100] & 2  & 756 & 4 \cr
    & \ [0000020] & 1 & 56 & 6 \cr
  }
\eqno (5.16)
$$
Again, the orbits $[0000000]$ do not occur in (5.15).

\medskip

Applying (4) of Theorem 2, we readily have $ C_{\lambda , \mu} =0$,
whenever $(\lambda - \mu)^2 \not=2$.
Conditions (1) and (2) for roots $\alpha = \lambda - \mu$ in the remaining 
12 orbits $[1000000]$ with $\alpha ^2 =2$ and $\lambda \cdot \mu = 1/2$ give
$$
\eqalign{
12 m_2 ^2 = &  \sum _{\lambda - \mu = \alpha} C_{\lambda , \mu } ^2 \cr
0= &  \sum _{\lambda - \mu = \alpha} C_{\lambda , \mu } ^2 (v_\lambda - v
_\mu). \cr}
\eqno (5.17)
$$
Taking the inner product with an arbitrary vector $u _\sigma$ with $\sigma
\in {\bf 27}$ yields the equation
$$
\alpha \cdot \sigma m_2 ^2 = \sum _{\lambda - \mu = \alpha} C_{\lambda , \mu 
}
^2 (\delta _{\lambda, \sigma} - \delta _{\mu, \sigma}).
\eqno (5.18)
$$
Since $\alpha$ is a root, and $\sigma$ is a weight of a fundamental
representation, the product $\alpha \cdot \sigma$ can take on only the values
$\alpha \cdot \sigma = -1, ~ 0, ~+1$. When $\alpha \cdot \sigma =0$, $\sigma 
\pm
\alpha$ are not weights of the {\bf 56}; thus all terms in (5.18) vanish
separately. When $\alpha \cdot \sigma = \pm 1$, $\sigma \mp \alpha$ are 
weights
of the {\bf 56}, while $\sigma \pm \alpha$ are not. Hence, (5.18) yields
$C_{\lambda , \mu }^2 = m_2 ^2$ whenever $(\lambda - \mu )^2 =2$. In fact, we
shall find a solution where all the square roots may be taken with a positive
sign, so that we recover the first equation in (5.13).

\medskip

It remains to satsify condition (3) of Theorem 2, i.e. (3.19). The cases 
$\alpha  = \lambda - \mu$ with either $\alpha ^2 =4$ or $\alpha ^2=6$ are
satisfied by arguments analogous to (5.8) of $E_6$.
The remaining equation for roots $\alpha = \lambda - \mu$  is given by
$$
\sqrt {12} m_2 d\cdot (u_\lambda - u _\mu)
= \sum _{\kappa \cdot \lambda = \kappa \cdot \mu = 1/2}
m_2 ^2 \bigl \{ \wp ((\lambda - \kappa ) \cdot x) - 
\wp ((\kappa - \mu ) \cdot x) \bigr \}.
\eqno (5.19)
$$
It is easily solved without any further restrictions, and we find recover the
second equation in (5.13).

\bigskip

\noindent
{\bf (d) Untwisted Elliptic Calogero-Moser System for $E_8$ }

\medskip

The lowest dimensional representation of $E_8$ is the adjoint of dimension 
248,
denoted by {\bf 248} for short. Its weights are given in terms of 8
orthonormal vectors $e_i$, $i=1,\cdots , 8$, as follows. There are 8 zero
weights, and 240 (non-zero) roots, given by
$$
\lambda  = \left \{ 
\eqalign{
& \pm e_i \pm e_j, \qquad ~\quad {\rm with} \qquad \qquad i\not=j \cr
& {1 \over 2} \sum _{i=1} ^8 \epsilon _i e_i
\qquad \qquad  {\rm with } \qquad \qquad  
\prod _{i=1} ^8 \epsilon _i = 1
 \cr} \right .
\eqno (5.20)
$$
All roots belong to a single Weyl orbit of $E_8$, denoted by $[10000000]$
and have the same length $\lambda ^2 = 2$.
We embed the {\bf 248} of $E_8$ into the fundamental representation of
$GL(248,\C)$, whose weights are an orthonormal set of vectors $u_I$,
$I=1,\cdots ,248$. It is convenient to labels the first 240 of these weights 
by
the 240 roots $\lambda$ of $E_8$ : $u_\lambda$, and the
last 8 by an index $a=1,\cdots ,8$ which distinguishes the zero weights of
$E_8$.

\medskip

The presence of the 8 zero weights in the {\bf 248} gives rise to serious
complications in the construction of the Calogero-Moser Lax pair :
the centralizer in $GL(248, \C)$ of the Cartan subalgebra of $E_8$ is now 
larger
than the Cartan subalgebra of $GL(248, \C)$. As a result, the space
$GL_0$ of roots of $GL(248, \C)$ that restrict to 0 in $E_8$ is 
non-trivial, with  $\dim GL_0 =56$, and on general grounds,  we are led to
include the term $\Delta \in GL_0$ in the construction of the Lax pair for
$E_8$ in (3.5), (3.6). However, a Lax pair still involves only a single
function $\Phi$, as in (3.16), and symmetric constants $C_{I,J} =
C_{J,I}$. 

\medskip

The $E_8$ root system contains a maximal set of 8 mutually orthogonal roots,
which we shall denote by $\beta _a$, $a=1,\cdots , 8$, with $\beta _a \cdot
\beta _b= 2 \delta _{a,b}$. This set specifies a maximal $SU(2) ^8$
subalgebra of $E_8$ that will play a special role in the sequel.  

\medskip

\noindent
{\bf Theorem 8 : Lax pair for E$_8$ in the 248}

The untwisted elliptic Calogero-Moser system for $E_8$ admits a Lax pair with 
spectral parameter and one independent coupling $m_2$ given by (3.5), (3.6), 
(3.16) and 
$$
\eqalignno{ 
 C_{\lambda , \mu } 
= & \left \{ \matrix{ m_2 c(\lambda , \mu) & \lambda \cdot \mu =1 
&\qquad \qquad c(\lambda , \mu ) = \pm 1\cr
0 & {\rm otherwise} & \cr} \right .
&  (5.21a) \cr  
C_{\lambda, c}
=& \left \{ \matrix{ \sum _{a=1}^8 \half (\lambda \cdot \beta _a) ~ c(\lambda 
, 
\beta _a (\lambda \cdot \beta _a)) C_{\beta _a, c} & \lambda \not=\pm \beta 
_b
\cr \pm C_{\beta _b,c} & \lambda = \pm \beta _b \cr} \right . 
& (5.21b) \cr 
\sqrt {60} d \cdot u _\lambda = &
\sum _{\delta \cdot \lambda =1} m_2 \wp (\delta \cdot x) 
+ 2 m_2 \wp (\lambda \cdot x) & (5.21c)\cr
\Delta _{a,b} 
= & 
\sum _{{\delta \cdot \beta _a =1 \atop \delta \cdot \beta _b =1}}
{m_2\over 2}
\big ( c(\beta _a, \delta) c(\delta, \beta _b) + c(\beta _a, \beta _a -
\delta) c(\beta _a -\delta , - \beta _b) \big ) \wp (\delta \cdot x) & \cr
- & \!
\sum _{{\delta \cdot \beta _a =1 \atop \delta \cdot \beta _b =-1}} 
{m_2\over 2} \big ( c(\beta _a,
\delta) c(\delta, -\beta _b) + c(\beta _a, \beta _a -
\delta) c(\beta _a -\delta ,  \beta _b) \big ) \wp (\delta \cdot x) 
& (5.21d) \cr
\Delta _{aa}
= & \sum _{\beta _a \cdot \delta =1} m_2 \wp (\delta \cdot x) + 2 m_2 \wp
(\beta _a \cdot x), & (5.21e) 
\cr}
$$
provided there exists a solution to the following $\pm 1$ valued cocycle 
factors $c(\lambda, \mu) $ 
$$
\eqalignno{
c(\lambda , \lambda - \delta) c(\lambda - \delta , \mu)
= & ~ c(\lambda , \mu + \delta ) c (\mu + \delta , \mu) &  \cr
&{\rm when} ~ \delta \cdot \lambda =-\delta \cdot \mu =1, ~\lambda \cdot \mu 
=0
& (5.22a)
\cr
c(\lambda, \mu) c(\lambda -\delta , \mu) = & ~ c(\lambda, \lambda -\delta )  
&
\cr &{\rm when} ~ \delta \cdot \lambda =\lambda \cdot \mu =1, \delta \cdot 
\mu 
=0
& (5.22b)
\cr
c(\lambda, \mu) c(\lambda , \lambda -\mu) = &- c(\lambda - \mu, -\mu)  &
\cr &{\rm when} ~ \lambda \cdot \mu =1.
& (5.22c)
\cr}
$$
We {\it conjecture} that a solution exists to these cocycle condition of 
(5.22). (The matrix $C_{\beta _b,c}$, $b,c=1,\cdots ,8$ is proportional to an 
arbitrary 8$\times$8 orthogonal matrix, as will be discussed below.)

\medskip

To prove this Theorem, we use the fact that only a single function $\Phi $
is involved by (3.16), and that $C_{I,J}$ is symmetric. Thus, we may resort 
to
the simplifications of conditions (1), (2) and (4) of Theorem 2, since these 
are independent of $\Delta$. However, as argued previously, $\Delta \not=0$ 
now, 
so we need to keep condition (3) of Theorem 1. 

\medskip

We begin by decomposing the weights of $GL(248, \C)$ under $E_8$ according to
(3.1); we find $s^2 =60$ and 
$$
\eqalign{
\sqrt {60} u_\lambda = & ~\lambda + v _\lambda \qquad \lambda \in \R (E_8) 
\cr
\sqrt {60} u _a = & ~ v_a \qquad a=1,\cdots ,8 \cr}
\eqno (5.23)
$$
The vectors $v_\lambda$ are orthogonal to $E_8$ roots, while vectors $v_a$ 
are
orthogonal to both roots of $E_8$ and to all $v_\lambda$. The roots of 
$GL(248,
\C)$ decompose under $E_8$ as follows
$$
\eqalign{
\sqrt {60} (u_\lambda - u_\mu) = & ~ \lambda - \mu + v_\lambda - v _\mu, 
\qquad \lambda \not=0 \cr
\sqrt {60} (u_\lambda - u_a) = & ~ \lambda + v_\lambda - v_a \cr
\sqrt {60} (u_a - u_\mu) = & -\mu + v_a - v_\mu \cr
\sqrt {60} (u_a - u_b) = & ~ v_a - v_b, \qquad a\not=b, \cr}
\eqno (5.24)
$$ 
where $\lambda , ~ \mu \in \R (E_8)$ and $a,b =1,\cdots ,8$. The last line of
roots $u_a - u_b$ in (5.24) span the space $GL_0$, discussed previously. 
Under
the Weyl group of $E_8$, the roots of (5.24) transform in the orbits 
occurring 
in the tensor product ${\bf 248} \otimes {\bf 248}$. They are given by
$$
\matrix{
&{\rm Weyl ~ Orbit} & {\rm Multiplicity} & \# {\rm ~ Weights } & 
{\rm Length }^2 \cr
&&&&\cr
{\bf 248} \otimes {\bf 248} 
\ :\  & \ [00000000] & 304 & 1    & 0 \cr
    & \ [10000000] & 72  & 240  & 2 \cr
    & \ [00000010] & 14  & 2160 & 4 \cr
    & \ [01000000] &  2  & 6720 & 6 \cr
    & \ [20000000] &  1  & 240  & 8 \cr
  }
\eqno (5.25)
$$
The orbits $[00000000]$ do not occur in (5.24).

\medskip

As argued previously, we may apply condition (4) of Theorem 2 to this case. 
We
readily deduce that $C_{\lambda, \mu}=0$ whenever $(\lambda - \mu )^2 
\not=2$,
and this yields the second line in (5.21a). Conditions (1) and (2) of Theorem 
2
for roots $\alpha = \lambda - \mu$ in the remaining 72 orbits $[10000000]$ 
with
$\alpha ^2=2$ and $\lambda \cdot \mu =1$ are given by
$$
\eqalign{
60 m_2 ^2 = & \sum _{\lambda - \mu = \alpha} C_{\lambda , \mu }^2
+  \sum _{b=1} ^8 \{ C_{\alpha, b}^2 + C_{b, -\alpha }^2 \}  \cr
0 = & \sum _{\lambda - \mu = \alpha } C_{\lambda , \mu } ^2 (v_\lambda - 
v_\mu)
+ \sum _{b=1} ^8 \{ C_{\alpha , b} ^2 ( v_\alpha - v _b)
+  C_{b, - \alpha} ^2 ( v_b - v_{- \alpha}) \}  \cr}
\eqno (5.26)
$$
Using linear independence of the vectors $v_b$ from $v_\lambda$, and symmetry
of the coefficients $C_{I,J}$, we find that
$$
C_{-\alpha , b}^2 = C_{\alpha , b}^2
\eqno  (5.27)
$$
Taking the inner product with a vector $u_\sigma$ for an arbitrary root
$\sigma$ of the second equation in (5.26), using (3.3) and combining with the
first equation in (5.26), we obtain
\footnote{${}^\#$}{Henceforth, we shall assemble the 8 components of 
$C_{\lambda,a}$, for $a=1,\cdots,8$ into an 8-dimensional vector, simply 
denoted by $C_\lambda$. Inner products then stand for $C_\lambda \cdot C_\mu 
=
\sum _{a=1} ^8 C_{\lambda ,a} C_{\mu, a}$.}
$$
\alpha \cdot \sigma m_2 ^2
= \sum _{\lambda - \mu = \alpha} C_{\lambda , \mu} ^2 
(\delta _{\lambda , \sigma} - \delta _{\mu, \sigma} )
+ (\delta _{\alpha , \sigma} - \delta _{-\alpha , \sigma})
C_\alpha \cdot C_\alpha. 
\eqno (5.28)
$$
We solve this equation for each of the possible values of $\alpha \cdot
\sigma$. Since both $\alpha$ and $\sigma$ are roots, the allowed values are
$ \alpha \cdot \sigma = 0, \pm 1, \pm 2$. When $\alpha \cdot \sigma =0$,
$\alpha \pm \sigma$ are not roots, so that all sides of (5.28) vanish
separately. 
If $\alpha \cdot \sigma =\pm 1$, $\alpha \mp \sigma$ is a root, and we
find
$$
C_{\lambda , \mu } ^2 = m_2 ^2,
\eqno (5.29)
$$
which yields the first line in (5.21a), upon taking the square root, and 
introducing some as yet unspecified sign factors $c(\lambda, \mu)$.
Finally, if $\alpha \cdot \sigma = \pm 2$, then $\sigma = \pm \alpha $ and
(5.28) reduces to $C_\alpha \cdot C_{\alpha } = 2m_2 ^2$.
With these results, the first equation of (5.26) is automatically satisfied,
using the fact that a given root can be written as the difference between two
(non-zero) roots in 56 different ways !

\medskip

It remains to satisfy condition (3) of Theorem 1. Amongst other relations, 
this 
will give rise to a number of conditions on the inner products $C_\lambda 
\cdot 
C_\mu$, and it is convenient to record those here,
$$
C_\lambda \cdot C_\mu 
= \left \{ \matrix{
\pm 2 m_2 ^2 & \lambda \cdot \mu = \pm 2 \cr
\pm c(\lambda , \mu) m_2 ^2 & \lambda \cdot \mu = \pm 1\cr
0 & \lambda \cdot \mu =0. \cr} \right .  
\eqno (5.30)
$$
The first line of (5.30) with the + sign was already obtained above.

\medskip

Condition (3) of Theorem 1 reduces to two sets of equations 
$$
\eqalignno{
sC_{\lambda , \mu}  d \cdot (u _\lambda - u _\mu)
= & 
\sum _{\kappa \not= \lambda , \mu} 
C_{\lambda , \kappa} C_{\kappa, \mu}
\big \{ \wp ((\lambda - \kappa ) \cdot x) - \wp ((\kappa - \mu ) \cdot x) 
\big
\} & \cr 
&  \qquad 
+ C _\lambda \cdot C _\mu \big \{ \wp (\lambda \cdot x) - \wp (\mu \cdot x)
\big \} & (5.31a) \cr
&&\cr
s C_{\lambda , b} d \cdot u _\lambda 
- \sum _a C_{\lambda ,a} \Delta _{ab}
=& \sum _{\kappa \not=\lambda } C_{\lambda ,\kappa} C_{\kappa, b}
\big \{ \wp ((\lambda - \kappa ) \cdot x) - \wp (\kappa \cdot x) \big \}
& (5.31b) \cr}
$$

\medskip

To solve (5.31a), we recast the sums in terms of $\delta = \lambda - \kappa$ 
in
the first and $\delta = \kappa - \mu$ in the second sum. As a result, we have
$$
\eqalign{
sC_{\lambda , \mu}  d \cdot (u _\lambda - u _\mu)
= & 
\sum _{{\delta \cdot \lambda =1 \atop \delta \cdot \mu = \lambda \cdot \mu
-1}}  C_{\lambda , \lambda - \delta } C_{\lambda - \delta , \mu}
 \wp (\delta \cdot x)   
-   
\sum _{{\delta \cdot \lambda =1 - \lambda \cdot \mu \atop \delta \cdot \mu =
-1}}  C_{\lambda , \mu + \delta } C_{\mu + \delta , \mu}
 \wp (\delta \cdot x)   \cr  
 & \qquad + 
C _\lambda \cdot C _\mu \big \{ \wp (\lambda \cdot x) - \wp (\mu \cdot x)
\big \}. \cr}
\eqno (5.32)
$$
The case $\lambda \cdot \mu = 2$ is excluded, while the case $\lambda \cdot 
\mu
= -2$ is automatically satisfied, since both sides vanish identically. When 
$\lambda \cdot \mu =0$, the l.h.s. of
(5.32) vanishes, since $\lambda - \mu$ is not a root. The two sums on the 
r.h.s. of (5.32) cancel one another provided we demand (5.22b), and  
$C_\lambda 
\cdot  C _\mu =0$, which is precisely the last equation in (5.22a). When 
$\lambda \cdot \mu =-1$, the l.h.s. of (5.32) vanishes, and each sum on the 
r.h.s. reduces to
a single term, since $\delta \cdot \mu =-2$ (i.e. $\delta = - \mu$) in the 
first
and $\delta \cdot \lambda = 2$  (i.e. $\delta = \lambda$) in the second sum.
This condition reduces to
$$
C_\lambda \cdot C_\mu = m_2 ^2 ~ c(\lambda , \lambda + \mu) c(\lambda + \mu, 
\mu)
\qquad \qquad \lambda \cdot \mu =-1
\eqno (5.33)
$$ 
Finally, for $\lambda \cdot \mu =1$, (5.32) is equivalent to
$$
\eqalign{
s  d \cdot (u _\lambda - u _\mu)
= & 
\sum _{{\delta \cdot \lambda =1 \atop \delta \cdot \mu = 0 }}  
m_2 c(\lambda , \mu) c(\lambda , \lambda - \delta ) c(\lambda - \delta , \mu)
 \wp (\delta \cdot x)  \cr 
& -   
\sum _{{\delta \cdot \lambda =0  \atop \delta \cdot \mu = -1}}  
m_2 c(\lambda , \mu) c(\lambda , \mu + \delta ) c(\mu + \delta , \mu)
 \wp (\delta \cdot x)   \cr  
 & \qquad + 
{1 \over m_2} c(\lambda , \mu) C _\lambda \cdot C _\mu 
\big \{ \wp (\lambda \cdot x) - \wp (\mu \cdot x)
\big \} .\cr}
\eqno (5.34)
$$
The l.h.s. of (5.34) is a difference of the function $sd\cdot u_\alpha$,
evaluated at $\lambda = \alpha$ and $\alpha = \mu$, and the r.h.s. must also 
be
such a difference. In order to split the first two terms, it is sufficient 
(and
one can show also necessary) that the product $c(\lambda , \mu) c(\lambda ,
\lambda - \delta ) c(\lambda - \delta , \mu)$ be independent of $\lambda$ and
$\mu$. As a result, it is independent of $\delta$ as well. Its values can be
only $\pm 1$, and by choosing the sign of $m_2$, we can choose this product 
to
be $+1$ as in (5.22c). The last term must also split, so that the product
$ c(\lambda , \mu) C _\lambda \cdot C _\mu $ must also be independent of
$\lambda$ and $\mu$. Introducing a suitable new constant $m$, we may express 
the
last fact as
$$
C_\lambda \cdot C_\mu = (m^2 - m_2 ^2) c(\lambda , \mu)
\qquad \qquad
\lambda \cdot \mu =1.
\eqno (5.35)
$$
Once these conditions are fulfilled, the solution of (5.32) is readily 
obtained
$$
sd \cdot u _\lambda = \sum _{\lambda \cdot \delta =1} m_2 \wp (\delta \cdot 
x)
+ {m^2 \over m_2} \wp (\lambda \cdot x).
\eqno (5.36)
$$
It remains to solve for $m$ and to analyze (5.31b).

\medskip

To determine $m$, we make use of the fact that conditions (1) and
(2) of Theorem 2 as well as condition (3) of Theorem 1 (i.e. equations 
(5.31))
are invariant under the following constant similarity transformation of $L$ 
and
$M$, as given in (2.8), with $\eta _\lambda = \pm 1$;
$$
\eqalign{
C_{\lambda , \mu } 
&\longrightarrow \eta _\lambda \eta _\mu C_{\lambda , \mu} \cr
C_{\lambda , a} 
&\longrightarrow \eta _\lambda  \eta _{\beta _a} C_{\lambda , a} \cr
\Delta _{a,b} 
&\longrightarrow \eta _{\beta _a} \eta _{\beta _b} \Delta _{a,b}. \cr}
\eqno (5.37)
$$
It may be checked explicitly that the equations (5.21) and (5.22) of Theorem 
8
are indeed invariant under these transformations. 

\medskip

The solution of (5.27) is $C_{-\alpha, b} = \pm C_{\alpha ,b}$, and the sign 
in
general depends upon the root $\alpha$ and the zero weight label $a$. Let us
pick an orthonormal basis of roots $\beta _a$, $a=1,\cdots ,8$, discussed
before Theorem 8. We choose a basis for the vectors $C_{\beta _a}$, in which
only one component is non-vanishing. Then by the above argument, we must have
that $C_{-\beta _a} = \pm C_{\beta _a}$. Furthermore, by making a
transformation of the form (5.37), we may choose the sign for each of the 8
vectors $C_{\beta _a}$ at will. We shall choose $C_{-\beta _a} = - C_{\beta
_a}$ for all $a=1,\cdots ,8$. Let $\lambda $ be a root such that $\lambda
\cdot \beta _a=1$, for some $a$. Since the vectors $C_{\beta _a}$ and
$C_{-\beta _a}$ are now opposites of one another, we obtain a relation 
between
equations (5.33) and (5.35).
$$
\eqalign{
C_\lambda \cdot C_{- \beta _a} =  - C_\lambda \cdot C_{\beta _a} 
= & ~m_2 ^2 c(\lambda, \lambda -\beta _a) c(\lambda - \beta _a, \beta _a) \cr
= & ~(m_2 ^2 - m^2) c(\lambda , \beta _a).
\cr}
\eqno (5.38)
$$
Since $c=\pm1$, there are two possible solutions for the constant $m$ : $m^2
=0$ or $m^2 =2m_2^2$. It turns out that (5.31b) is inconsistent for $m^2=0$,
and we shall now proceed to show that the solution (5.21d,e) exists for 
$$
\eqalign{
m^2 = & \ 2 m_2 ^2 \cr
C_{-\lambda } = & - C_{\lambda} \cr
c(-\lambda, - \mu) = & \ c(\lambda , \mu). \cr}
\eqno (5.39)
$$
Consistency of (5.33) and (5.35) with (5.39) requires that condition (5.22c) 
be
satisfied, and this yields the middle equation in (5.30). Choosing
$C_{\lambda}$ to be odd in $\lambda$ restricts the symmetry of (5.37) to the
subgroup of transformations for which $\eta _\lambda = \eta _{-\lambda}$.

\medskip

Finally, we show that to satisfy (5.31b), the equations of (5.21a,b,c) and 
(5.22) suffice, and produce (5.21d,e). This in itself is remarkable since 
(5.31b) is a
set of $8 \times 240$ equations, for only $8 \times 8$ remaining unknowns
$\Delta _{a,b}$. The arguments and the calculations are lengthy and are 
deferred to Appendix \S C. 

\bigskip

\noindent
{\bf (d) Untwisted Elliptic Calogero-Moser System for $G_2$ }

\medskip

The smallest non-trivial representation of the Lie algebra $G_2$ is of
dimension 7, and denoted {\bf 7} for short. Its weights are given in
terms of 3 orthonormal vectors $e_i$, $i=1,2,3$. It is convenient to 
introduce
$e_0 = \13(e_1+e_2+e_3)$ so that
$$
\lambda _I = \left \{ \eqalign{
&\pm \alpha _i = \pm (e_i -  e_0)
\qquad (I=1,\cdots ,6) \cr
& 0 \qquad \qquad \qquad \qquad \qquad  (I=7). \cr } \right .
\eqno (5.40)
$$
The precise correspondence between the labels $\pm,~i$ and $I$ is immaterial
since the order is permuted by the Weyl group of $G_2$.

\medskip

The {\bf 7} may naturally be embedded into the fundamental of $GL(7,\C)$.
However, one way to define $G_2$ is as the subalgebra of $B_3$ that leaves 
one
of the spinor weights invariant; thus it is also natural to embed the {\bf
7}$\oplus ${\bf 1} of $G_2$ into the 8-dimensional spinor representation of
$B_3$. We shall treat both cases, and use the spinor embedding as an example 
of
a case in which the untwisted elliptic Calogero-Moser Hamiltonian and Lax 
pair 
may be obtained by
restriction to a subgroup of the Calogero-Moser Hamiltonian and Lax pair of a
larger Lie algebra.

\medskip

\noindent
{\bf Theorem 9 : Lax pairs for $G_2$ in the 7 and in the 7$\oplus$1}

The untwisted elliptic Calogero-Moser Hamiltonian for $G_2$ admits a Lax pair 
with spectral parameter and one independent coupling 

\item{(1)} in the {\bf 7} of $G_2$, given by (3.5), (3.6), (3.16) and
$$
\eqalignno{
C_{\lambda , \mu } = & \left \{ \matrix{ 
m_2 & \lambda \cdot \mu = \pm \13  \cr
  0 & {\rm otherwise,} \cr} \right .
\qquad \qquad \ \
C_{\lambda ,7} =  \sqrt 2 m_2 & (5.41a) \cr
\sqrt {2} d\cdot u_\lambda 
=  & \sum _{ {\delta ^2 =2;\atop \lambda \cdot \delta =1}} 
      m_2\wp (\delta \cdot x) + m_2 \wp (\lambda \cdot x), 
      & (5.41b)\cr
\sqrt 2 d \cdot u_7 
= & \half \sum _{\kappa ^2 = 2/3} m_2 \wp (\kappa \cdot x);
& (5.41c) \cr }
$$
\item{(2)} in the {\bf 7}$\oplus${\bf 1} of $G_2$, embedded in the spinor
representation of $B_3$, given in part (1) of Theorem 5 with
$$
m_1 =   m_2 
\qquad \qquad {\rm and } \qquad \qquad
x\cdot \lambda _s = 0 
\eqno (5.42)
$$
where $\lambda _s$ is any one spinor weight of the spinor of $B_3$.

\medskip

\noindent
{\it Proof of (1)}

We start by embedding the {\bf 7} of $G_2$ into the fundamental 
representation 
of
$GL(7,\C)$, whose weights are 7 orthonormal vectors $u_I$, $I=1,\cdots ,7$. 
The
decomposition (3.1) is given by $s^2 =2$ and 
$$
\eqalign{
\sqrt{2} u_\lambda     = & \ \lambda + v_\lambda \qquad
{\rm for} \qquad \lambda ^2 = 2/3, \cr
\sqrt{2} u_7     = & \  v_7 . \cr}
\eqno (5.43)
$$
Here, the vectors $v_\lambda$ and $v_7$ are orthogonal to one another and to 
the
weights of (5.40). The roots of $GL(7,\C)$ decompose under $G_2$  as follows
$$
\eqalign{
\sqrt 2 (u_\lambda  - u_\mu ) 
 =& \ \lambda - \mu  + v_\lambda - v_\mu,  
  \qquad  \qquad \lambda \not= \mu  
\cr
 \sqrt 2 (u_\lambda - u _7)=& \ \lambda - v_7 
\cr 
\sqrt 2 (u_7 - u_\mu) = & - \mu - v_\mu + v_7 
 \cr }
\eqno (5.44)
$$
Under the Weyl group of $G_2$, these roots transform in the Weyl orbits
occurring in the tensor product ${\bf 7} \otimes {\bf 7}$ of $G_2$, given by
$$
\matrix{
&{\rm Weyl ~ Orbit} & {\rm Multiplicity} & \# {\rm ~ Weights ~in ~Orbit} & 
{\rm
Length }^2 \cr
&&&&\cr
{\bf 7} \otimes {\bf 7} 
\ :\  & \ [00] & 7 & 1 & 0 \cr
  & \ [01] & 4 & 6 & 2/3 \cr
  & \ [10] & 2 & 6 & 2 \cr
  & \ [02] & 1 & 6 & 8/3 \cr}
\eqno (5.45)
$$
The orbits $[00]$ do  not occur in (5.43).
Using (4) of Theorem 2, and the fact that weights of the orbit $[02]$ are not
roots of $G_2$, we readily find that $C_{\lambda ,\mu } =0$ whenever $\lambda
\cdot \mu = -2/3$, i.e. $\mu =- \lambda$. The remaining orbits correspond to
roots, and we have two independent couplings : $m_2$ for long roots, and
$m_{2/3}$ for long roots.    

\medskip

Conditions (1) and (2) for the weights $\alpha = \lambda - \mu$ in the two 
Weyl orbits of type $[10]$ and length $\alpha ^2=2$ are given by
$$
\eqalign{
2m_2 ^2 = & ~ C_{\lambda , \mu}^2 + C _{-\mu, - \lambda}^2 \cr
0       = & ~ C_{\lambda , \mu}^2 (v _\lambda - v _\mu) 
            + C_{-\mu, - \lambda}^2 (v_{-\mu} - v_{-\lambda}).
\cr}
\eqno (5.46)
$$
Taking the inner product of the second equation with an arbitrary vector $u
_{\sigma}$ for $\sigma ^2 = 2/3$, this set of equations
reduces to
$$
\alpha \cdot \sigma  m_2 ^2 =  C_{\lambda , \mu}^2 (\delta _{\lambda ,
\sigma} - \delta _{\mu, \sigma}) + C _{-\mu, - \lambda}^2 (\delta _{-\mu,
\sigma} - \delta _{-\lambda , \sigma}).
\eqno (5.47)
$$
Since $\alpha$ is a root with $\alpha ^2=2$, the combination $\alpha \cdot
\sigma$ can only assume integer values; since $|\alpha \cdot \sigma |
<2$, only three possible values  $\alpha \cdot \sigma = -1, ~0, ~+1$ are
allowed. When $\alpha \cdot \sigma =0$, all terms in (5.47) vanish 
separately.
When $\alpha \cdot \sigma = \pm 1$, then $\sigma \mp \alpha$ is also a weight
and (5.47) reduces to $C_{\lambda , \mu} ^2 = m_2^2$ when $ (\lambda - \mu 
)^2
=2$. It turns out that a Lax pair may be found in which all the square roots
are taken with the same signs, which gives the statement of (5.41a) for
$\lambda \cdot \mu = -1/3$.

\medskip

Conditions (1) and (2) for the weights $\alpha = \lambda - \mu$ with $\lambda
\cdot \mu = 1/3$ or $\alpha = \lambda$, in the four Weyl orbits of type 
$[01]$
and length $\alpha ^2=2/3$ lead to the equations
$$
\eqalign{
2m_{2/3}^2 = & ~C_{\lambda, \mu} ^2 + C_{-\mu , -\lambda}^2 
             + ~C_{\alpha , 7} ^2  + C_{7, -\alpha } ^2\cr
0 = & ~C_{\lambda , \mu} ^2  (v_\lambda - v_\mu) 
    + C_{-\mu, -\lambda} ^2 (v_{-\mu} - v _{-\lambda}) \cr
     &+ C_{\alpha, 7}^2 (v_\alpha - v_7) + C_{7, -\alpha} ^2 (v_7 -
v_{-\alpha}).
\cr  }
\eqno (5.48)
$$
The projection of the last equation in (5.48) onto $v_7$ yields
$C_{7,-\alpha} ^2 = C_{\alpha ,7} ^2$.
The remaining equations are analyzed by taking the inner product with a
general vector $u_\sigma$, with $\sigma ^2 = 2/3$ 
$$
\eqalign{
 \alpha \cdot \sigma m_{2/3}^2
= & ~ C_{\lambda, \mu} ^2 (\delta _{\lambda, \sigma} - \delta _{\mu,\sigma}) 
     +C_{-\mu, -\lambda} ^2 
( \delta _{-\mu, \sigma} - \delta _{-\lambda, \sigma}) \cr
&    +C_{\alpha, 7}^2 (\delta _{\alpha, \sigma} - \delta _{-\alpha , 
\sigma}).
\cr}
\eqno (5.49)
$$
Since $\alpha$ is a root with $\alpha ^2 = 2/3$, the combination $3
\alpha \cdot \sigma$ can take on only integer values; since $|3 \alpha
\cdot \sigma | \leq 2$, only the values $3\alpha \cdot
\sigma=-2,-1,0,1,2$ are allowed. For $\alpha \cdot \sigma=0$, all terms in
(5.49) vanish separately since $\sigma \not= \pm \alpha$ and  
$\sigma \pm \alpha$ are not weights in $[01]$. Taking
respectively $3\alpha \cdot \sigma =\pm 2$ and  $3\alpha \cdot \sigma =\pm 
1$,
we find for $\lambda \cdot \mu = 1/3$ and $\alpha ^2 = 2/3$ that
$$
m_{2/3}^2 = 3 C_{\lambda , \mu} ^2 = {3 \over 2} C_{\alpha , 7}^2.
\eqno (5.50)
$$

\medskip

Finally, we must satisfy condition (3) of Theorem 2, i.e. equation (3.19).
We may do this again Weyl orbit by Weyl orbit for the weights $\alpha =
\lambda _I - \lambda _J$ in (3.19). For the orbit $[02]$ with $\alpha ^2 =
8/3$, (3.9) is satisfied automatically. The remaining equations are as
follows. For $\lambda \cdot \mu = -1/3$ (the two orbits $[10]$), we have
$$
\eqalign{
sm_2 d \cdot (u _\lambda - u _\mu)
= & 
\sum _{\kappa \cdot \lambda = \kappa \cdot \mu = \pm 1/3}
m_2 ^2 \bigl \{ \wp ((\lambda - \kappa)\cdot x) - \wp ((\kappa - \mu)\cdot x)
\bigr \} \cr
& + {2 \over 3} m_{2/3}^2 \bigl \{ \wp (\lambda \cdot x) - \wp (\mu \cdot
x)\bigr \}.
\cr }
\eqno (5.51)
$$
Here, the sum over $\kappa$ is restricted to $\kappa \cdot \lambda = \kappa
\cdot \mu = \pm 1/3$ for the following reasons. In general, $\kappa \cdot
\lambda$ (and analogously $\kappa \cdot \mu$) can take on the values $-2/3,~
-1/3, ~1/3,~ 2/3$. The values $2/3$ is ruled out since from (3.19), $\kappa
\not= \lambda$; similarly, $\kappa\cdot \lambda = -2/3$ is ruled out since
the associated coupling $C_{\lambda, -\lambda}$ that would appear on the
right and side of (5.51) vanishes in view of (5.41a). Now, since in this 
case,
we have $\lambda \cdot \mu= -1/3$, the cases $\kappa \cdot \lambda = - \kappa
\cdot \mu= \pm 1/3$ have no solutions $\kappa$ that are weights of the ${\bf
7}$ of $G_2$. Thus, their contribution was dropped from the sums in (5.51).

\medskip

For $\lambda \cdot \mu = 1/3$ (2 of the four orbits $[01]$) we have
$$
\eqalign{
s \sqrt {\13} m_{2/3} d \cdot (u _\lambda - u _\mu)
= & 
\sum _{\kappa \cdot \lambda =-  \kappa \cdot \mu = \pm 1/3}
\sqrt {\13} m_{2/3} m_2 \bigl \{ \wp ((\lambda - \kappa)\cdot x) - \wp
((\kappa - \mu)\cdot x) \bigr \} \cr
& + {2 \over 3} m_{2/3}^2 \bigl \{ \wp (\lambda \cdot x) - \wp (\mu \cdot
x) \bigr \}.
\cr }
\eqno (5.52)
$$
For reasons analogous to the ones explained after (5.50), the sums above have
been restricted to $\kappa \cdot \lambda = - \kappa \cdot \mu = \pm 1/3$. For
the remaining 2 orbits $[01]$ we have
$$
s \sqrt {\23} m_{2/3} d \cdot (u _\lambda - u _7)
=  
\sum _{\kappa \cdot \lambda = \pm 1/3}
\sqrt {\23} m_{2/3} m_2 \bigl \{ \wp ((\lambda - \kappa)\cdot x) 
- \wp (\kappa \cdot x) \bigr \}. 
\eqno (5.53)
$$

\medskip

The general solutions of (5.51) and (5.52) are respectively 
$$
\eqalign{
s d \cdot u_\lambda 
= & d_0 + {1 \over 3 m_2} m_{2/3}^2 \wp (\lambda \cdot x) 
+ \sum _{\kappa \cdot \lambda = -1/3} m_2 \wp ((\lambda - \kappa)\cdot x) \cr
s d \cdot u_\lambda 
= & d_0 + ({2 \over \sqrt 3}  m_{2/3} - m_2) \wp (\lambda
\cdot x) + \sum _{\kappa \cdot \lambda = -1/3} m_2 
\wp ((\lambda - \kappa)\cdot x),
\cr}
\eqno (5.54)
$$
where $d_0$ is independent of $\lambda$. These solutions agree provided the
coefficients of the $\wp (\lambda \cdot x)$ terms agree and  
$m_{2/3} = \sqrt 3 m_2$. Under those conditions, (5.53) is compatible with
(5.54). Combining these results with (5.50), we find all equations of
(5.41a) and (5.41b).

\medskip

\noindent
{\it Proof of (2)}

Instead of repeating the construction using Theorem 2,
we obtain the Lax pair by restricting the Lax pair of $B_3$ in the spinor 
representation (as derived in part (1) of Theorem 5) to the $G_2$ subgroup
of $B_3$. This restriction is easy to carry out, since $G_2$ is 
the subgroup of $B_3$ that leaves any one of the spinor weights of $B_3$
invariant. We shall choose the weight
$$
\lambda _s = \half (  e_1 + e_2 + e_3).
\eqno (5.55)
$$

\medskip

The Lax pair of $B_3$ in the spinor representation is given in
terms of two independent couplings $m_1$ and $m_2$ by
$$
C_{\lambda , \mu} = \left \{ 
\matrix{ m_1 & (\lambda - \mu)^2 =1 \cr
         m_2 & (\lambda - \mu)^2 =2 \cr
          0  & {\rm otherwise} \cr}
          \right .
\eqno (5.56)
$$
The restriction of $B_3$ to $G_2$ corresponds to $x$ and $p$
orthogonal to the weight  $\lambda _s = \half (e_1 + e_2 + e_3)$. It is easy 
to
analyze under which conditions this restriction is consistent with the
Hamilton-Jacobi equations of the Calogero-Moser system. Consistency
requires that when $\lambda _s \cdot x=0$, and thus $\lambda _s \cdot p=
\lambda _s \cdot \dot p =0$, the right hand side of  (2.1) be orthogonal to
$\lambda _s$ as well, so that
$$
0= \sum _{\alpha \in \R (B_3)} m_{|\alpha |} ^2 ~
\lambda _s \cdot \alpha ~ \wp ' (\alpha \cdot x).
\eqno (5.57)
$$
The sum over $\alpha$ in (5.57) is even in $\alpha$,
and may be restricted to range over positive roots only. Furthermore, the
roots $e_i - e_j$ of $B_3$ are orthogonal to $\lambda _s$ and
do not contribute to (5.57). The remaining sum reduces to
$$
\eqalign{
0= & ~ m_2 ^2 \bigl [ \wp ' (x_1 + x_2) + \wp ' (x_2 + x_3)
+ \wp ' (x_3 + x_1) \bigr ] \cr
& + m_1 ^2 \bigl [ \wp ' (x_1) + \wp '(x_2) + \wp ' (x_3) \bigr ]
\cr}
\eqno (5.58)
$$
Since $0=\lambda _s \cdot x = x_1 + x_2 + x_3$, the right hand 
side of (5.58) cancels for all $x$ when  $m_1 = m_2$, which is precisely the
condition (5.42) ot Theorem 9.

\bigskip

\noindent
{\bf (e) Untwisted Elliptic Calogero-Moser System for $F_4$}

\medskip

The smallest non-trivial representation of $F_4$ is of dimension 26, and
denoted {\bf 26} for short. This representation has 2 zero weights. As a
result, when the {\bf 26} of $F_4$ is embedded into the fundamental
representation of $GL(26,\C)$, the centralizer of the Cartan subalgebra of
$F_4$ is larger than the Cartan subalgebra of $GL(26,\C )$. The space
$GL_0$ has dimension 2, and the quantity $\Delta $ in (3.5) may not vanish.
 This situation presents serious complications, just as it did in the case of
$E_8$. Fortunately, for $F_4$, there is an alternative where no such
complications appear.

\medskip

Instead, we consider the 27-dimensional representation {\bf 26 }$\oplus${\bf 
1}
of $F_4$, which has 3 zero weights, and which may be viewed as the
restriction of the 27-dimensional representation of $E_6$ to its $F_4$
subalgebra. It is a remarkable fact (see Appendix \S A) that the 24 long 
roots 
of
$F_4$ form a $D_4$ subalgebra of $F_4$, and that the 24 short roots
of $F_4$ (which precisely coincide  with the non-zero weights of the {\bf 26} 
of
$F_4$), may be viewed as the weights of the direct sum of the three 
8-dimensional
distinct (but equivalent) representations $8^v$, $8^s$ and $8^c$ of $D_4$. 
Thus,
it makes sense to group the 24 short roots (i.e. the 24 non-zero weights of 
the
{\bf 26}) into classes -- which we shall call ``8-classes" of $F_4$ -- and 
which
we shall denote by $8^v$, $8^s$ and $8^c$. The 8-class of a non-zero weight
$\lambda$ of the {\bf 26} will be denoted by $[\lambda ]$. This underlying
$D_4$ structure of $F_4$ will turn out to play a crucial role in the
construction of the $F_4$ Lax pair. 

\medskip

The weights of the {\bf 26 }$\oplus${\bf 1} together with their 8-class
assignments are
$$
\eqalign{
8^v \qquad & 
\pm e_i \qquad \qquad  i=1,2,3,4 \cr
8^s \qquad 
& \half \sum _{i=1} ^4 \epsilon _i e _i \qquad \prod _{i=1} ^4 \epsilon _i 
=+1
\cr
8^c \qquad 
& \half \sum _{i=1} ^4 \epsilon _i e _i \qquad \prod _{i=1} ^4 \epsilon _i 
=-1
\cr
{\rm zero} \qquad & 0 \qquad \qquad \qquad  a=v,s,c.
\cr}
\eqno (5.59)
$$
We shall make use of the following equivalences
$$
\eqalign{
[\lambda ] = [\mu] 
\quad \Longleftrightarrow \quad
& \lambda \cdot \mu = 0 , ~ {\rm or} ~\pm 1 \cr
[\lambda ] \not= [\mu] 
\quad \Longleftrightarrow \quad
& \lambda \cdot \mu = \pm \half ,\cr}
\eqno (5.60)
$$
which are readily establised by inspection of (5.59).

\medskip

\noindent
{\bf Theorem 10 : Lax pair for F$_{\bf 4}$ in the 26$\oplus$1}

 The untwisted elliptic Calogero-Moser Hamiltonian for $F_4$ admits a Lax 
pair 
with spectral parameter and one independent coupling, given by (3.5), (3.6), 
(3.16) and
$$
\eqalignno{
C_{\lambda , \mu } = & \left \{ \matrix{ 
m_2 & \lambda \cdot \mu = 0,  \half  \cr
  0 & {\rm otherwise,} \cr} \right .
\qquad \qquad \ \
\matrix{ 
C_{\lambda ,a} =& ~ m_2 (1- \delta _{[\lambda ],a}) \cr
C_{a,b} = & 0\cr}  & (5.61a) \cr
\sqrt {6} d\cdot u_\lambda 
=  & ~ 2m_2 \wp (\lambda \cdot x) 
+ \sum _{ {\delta ^2 =2;\atop \lambda \cdot \delta =1}} 
      m_2\wp (\delta \cdot x)
-\half m_2 \sum _{\kappa  \in [\lambda]} 
 \wp (\kappa \cdot x) 
& (5.61b) \cr
\sqrt 6 d \cdot v_a = &
- m_2 \sum _{[\kappa]=a} \wp (\kappa \cdot x) + \sum _\kappa \half m_2 \wp
(\kappa \cdot x)  & (5.61c) \cr }
$$
The notations are explained in detail below.

\medskip

To prove this Theorem, we begin by embedding the \f27 into the
fundamental representation of $GL(27, \C)$, whose weights are 27 orthonormal
vectors $u_I$, $I=1,\cdots ,27$. The decomposition (3.1) is given by $s^2 =6$
and
$$
\eqalign{
\sqrt 6 u_\lambda = & ~\lambda + v_\lambda \qquad \lambda ^2 =1 \cr
\sqrt 6 u_a = & ~v_a \qquad \qquad a=v,s,c \cr}
\eqno (5.62)
$$
Here, the vectors $v_\lambda $ and $v_a$ are orthogonal to one another and to
the weight space of $F_4$. The roots of $GL(27, \C)$ decompose under $F_4$ as
follows
$$
\eqalign{
\sqrt 6 ( u_\lambda - u_\mu ) = & ~ \lambda - \mu + v_\lambda - v_\mu, 
\qquad \lambda \not= \mu \cr 
\sqrt 6 ( u_\lambda - u_a) = & ~ \lambda + v_\lambda - v_a \cr
\sqrt 6 ( u_a - u_\mu) = & - \mu + v_a - v_\mu \cr
\sqrt 6 (u_a - u_b ) = & ~ v_a - v_b ,
\qquad \qquad \qquad  a \not= b. \cr }
\eqno (5.63)
$$ 
Under the Weyl group of $F_4$, these roots transform in the Weyl orbits
occuring in the tensor product (\f27)$\otimes$(\f27) and are given by
$$
\matrix{
&{\rm Weyl ~ Orbit} & {\rm Multiplicity} & \# {\rm ~ Weights } & 
{\rm Length }^2 \cr
&&&&\cr
({\bf 26} \oplus {\bf 1}) \otimes ({\bf 26} \oplus {\bf 1}) 
\ :\  & \ [0000] & 33 & 1   & 0 \cr
    & \ [0001] & 14 & 24  & 1 \cr
    & \ [1000] & 6  & 24  & 2 \cr
    & \ [0010] & 2  & 96  & 3 \cr
    & \ [0002] & 1  & 24  & 4 \cr}
\eqno (5.64)
$$
The orbits $[0000]$ do not occur in (5.63). Using (4) of Theorem 2, and the
fact that only the orbits $[0001]$ and $[1000]$ are roots, we readily find 
that
$C_{\lambda  , \mu}=0$ whenever $(\lambda - \mu )^2 \not=1,2$ and recover the
result in (5.61a) for $\lambda \cdot \mu \not= 0,\half$.

\medskip

Conditions (1) and (2) for roots $\alpha = \lambda - \mu$ (with $\lambda 
\cdot
\mu =0$)  in the 6 Weyl orbits of type $[1000]$ are given by
$$
\eqalign{
6 m_2 ^2 = & \sum _{\alpha = \lambda - \mu} 
C_{\lambda , \mu }^2 \cr
0 = & \sum _{\alpha = \lambda - \mu }
 C_{\lambda , \mu } ^2 (v_\lambda - v _\mu )\cr}
\eqno (5.65)
$$
Taking the inner product of the second equation with $u_\sigma$ for an
arbitrary weights vector $\sigma$ of \f27, and using the first equation in
(5.65), we get
$$
\alpha \cdot \sigma m_2 ^2 
=\sum _{\lambda, \mu} C_{\lambda , \mu} ^2 
(\delta _{\lambda , \sigma} - \delta _{\mu , \sigma }).
\eqno (5.66)
$$
The combination $\alpha \cdot \sigma$ can take only the values $\alpha \cdot
\sigma = 0$, for which all sides of (5.66) vanish separately, and $\alpha 
\cdot
\sigma = \pm 1$, from which we find that $C_{\lambda , \mu}^2 = m_2 ^2$. 
Taking
square roots with a plus sign of this result, we obtain the expression 
(5.61a)
for $\lambda \cdot \mu = 0$.

\medskip

Conditions (1) and (2) for the 8 weights of the form $\alpha = \lambda - \mu$
with $\lambda \cdot \mu = \half$, and the 6 weights of the
form $\alpha = \lambda$, in orbits of type $[0001]$ with $\alpha ^2 =1$ yield
$$
\eqalign{
6 m_1 ^2 
= & 
\sum _{\alpha = \lambda - \mu} C_{\lambda , \mu}^2
+ \sum _{a} ( C_{\alpha , a} ^2 + C_{-\alpha ,a}^2 ) \cr 
0 
= & 
\sum _{\alpha = \lambda - \mu} 
C_{\lambda , \mu}^2 (v_\lambda - v _\mu)
+  \sum _{a}  \big ( C_{\alpha , a} ^2 (v_\alpha - v_a) 
+ C_{-\alpha ,a}^2 (v_a - v_{-\alpha}) \big ).
\cr}
\eqno (5.67)
$$ 
Using linear independence of $v_a$, we deduce that $C_{\alpha ,a} ^2 =
C_{-\alpha , a}^2$; it turns out that we can find a Lax pair with $C_{\alpha 
,
a} = C_{-\alpha ,a}$, which we shall henceforth assume to hold. Taking the
inner product of the remaining part of the second equation in (5.67) with $u
_\sigma$ for an arbitrary weight $\sigma$, and using the first equation in
(5.67) yields 
$$
\alpha \cdot \sigma m_1 ^2
= \sum _{\alpha = \lambda - \mu} 
C_{\lambda , \mu} ^2
+ C_\alpha \cdot C _\alpha (\delta _{\alpha, \sigma} - \delta _{-\alpha ,
\sigma}).
\eqno (5.68)
$$
Analyzing this equation according to the possible values of $\alpha \cdot
\sigma = 0, \pm \half  , \pm 1$, we find 
$$
C_{\lambda , \mu } =   {1 \over \sqrt 2} m_1 
\qquad \qquad 
C _\alpha \cdot C_\alpha = m_1 ^2. 
\eqno (5.69)
$$
With these values, the first equation in (5.67) is satisfied automatically.
Henceforth, we shall assume that $m_1 \not=0$, since otherwise the system
reduces to that of a $D_4$ algebra.

\medskip

To complete the proof of Theorem 10, it remains to satisfy condition (3) of
Theorem 2 for this case. Henceforth, we shall set $C_{a,b}=0$ and 
$\Delta _{a,b} =0$, since a consistent solution exists under these
assumptions. Conditions (3) may then be split into two parts. 
$$
\eqalignno{
C_{\lambda , \mu} s d \cdot (u_\lambda - u_\mu)
= & \sum _{\kappa \not= \lambda , \mu}
C_{\lambda , \kappa} C_{\kappa, \mu} 
\big ( \wp ((\lambda - \kappa ) \cdot x) - \wp ((\kappa - \mu )\cdot x) \big 
)
&\cr 
& \qquad + C_\lambda \cdot C_\mu \big ( \wp (\lambda \cdot x) - \wp (\mu 
\cdot 
x)
\big ) & (5.70a) \cr
&&\cr
C_{\lambda , a} s d \cdot (u _\lambda - u_a)
= & \sum _{\kappa \not= \lambda} C_{\lambda , \kappa} C_{\kappa , a}
\big ( \wp ((\lambda - \kappa )\cdot x) - \wp (\kappa \cdot x) \big ).
& (5.70b) \cr  
}
$$

\medskip

We analyze (5.70a) according to the value of $(\lambda - \mu )^2$, i.e. the
orbit type which $\lambda - \mu$ belongs to. For $(\lambda - \mu )^2 =4$, 
both
sides of (5.70a) manifestly cancel. For $(\lambda - \mu )^2 =3$, the l.h.s.
still vanishes, the sums on the r.h.s. reduce to a single term proportional
to $\wp (\lambda \cdot x) - \wp (\mu \cdot x)$ and the resulting condition is
$$
C_\lambda \cdot C_\mu=\half m_1 ^2,
\qquad \qquad \lambda \cdot \mu = - \half.
\eqno (5.71)
$$

\medskip

For $(\lambda - \mu )^2 =2$, i.e. $\lambda \cdot \mu =0$, $\lambda $ and 
$\mu$
must belong to the same 8-class. Thus, in the sum in (5.70a), $\kappa$ either
belongs to the common 8-class $[\lambda ] = [\mu]$ or belongs to another 
class,
so that only the cases $\kappa \cdot \lambda = \kappa \cdot \mu = 0, {\rm
or}~\half$ remain. We find
$$
\eqalign{
m_2  s d \cdot (u_\lambda - u_\mu)
= & \sum _{\kappa \cdot \lambda =\kappa \cdot \mu =0}
m_2 ^2   \big ( 
\wp ((\lambda - \kappa ) \cdot x) - \wp ((\kappa - \mu )\cdot x)
\big ) \cr
& +  \sum _{\kappa \cdot \lambda =\kappa \cdot \mu =\half }
\half m_1 ^2   \big ( 
\wp ((\lambda - \kappa ) \cdot x) - \wp ((\kappa - \mu )\cdot x)
\big ) \cr
& \qquad + C_\lambda \cdot C_\mu \big ( \wp (\lambda \cdot x) - \wp (\mu 
\cdot 
x)
\big ).   \cr}
\eqno (5.72)
$$
Upon changing variables in the second sum on the r.h.s. of (5.72), to $\delta
= \lambda - \kappa$ in the first term and to $\delta = \kappa - \mu$ in the
second term, we see that the second sum in (5.72) cancels identically. The
remaining equation may be separated in terms of $\lambda$ and $\mu$ dependent
terms provided $ C_\lambda \cdot C_\mu$ is independent of $\lambda$ and $\mu$ 
whenever  $ \lambda \cdot \mu =0$. This means that $\lambda $ and $\mu$ 
belong 
to the same 8-class. If we assume (and this will be justified by the fact 
that we can find a consistent Lax pair satisfying this assumption) that the 
inner product $C_\lambda \cdot C_\mu$ should only depend upon the 8-classes 
of 
$\lambda$ and $\mu$, then this value must coincide with that of $\lambda = 
\mu$, and we have 
$$
C_\lambda \cdot C_\mu = m_1 ^2,
\qquad \qquad [\lambda ] = [\mu].
\eqno (5.73)
$$
We then have 
$$
s d \cdot u_\lambda = d_1 ([\lambda]) 
+\sum _{\delta ^2 =2;~\delta \cdot \lambda =1} m_2 \wp (\delta \cdot x)
+{m_1 ^2 \over m_2}  \wp (\lambda \cdot x).
\eqno (5.74)
$$
It will be very important to realize in the sequel that the difference 
$d\cdot
(u_\lambda - u_\mu)$ was evaluated for $\lambda $ and $\mu$ belonging to the
same 8-class. Thus, upon separating the equation as is done in (5.74), we are
left with an arbitrary function $d([\lambda])$ which depends only upon the
8-class of $\lambda$, but not upon the representative of this class.
 
\medskip

Finally, for $(\lambda - \mu )^2 =1$, $\lambda$ and $\mu$ necessarily belong 
to
different 8-classes, and (5.70a) in this case reduces to 
$$
\eqalign{
  s d \cdot (u_\lambda - u_\mu)
= & \sum _{{\kappa \cdot \lambda =0;\atop  \kappa \cdot \mu =\half}}
m_2    \big ( 
\wp ((\lambda - \kappa ) \cdot x) - \wp ((\kappa - \mu )\cdot x)
\big ) \cr
 & +  \sum _{{\kappa \cdot \lambda =\half;\atop  \kappa \cdot \mu =0}}
m_2    \big ( 
- \wp ((\kappa - \mu ) \cdot x) + \wp ((\lambda - \kappa )\cdot x)
\big ) \cr
& +  \sum _{\kappa \cdot \lambda =\kappa \cdot \mu =\half }
{1 \over \sqrt 2}  m_1    \big ( 
\wp ((\lambda - \kappa ) \cdot x) - \wp ((\kappa - \mu )\cdot x)
\big ) \cr
& \qquad 
+ { \sqrt 2 \over m_1} C_\lambda \cdot C_\mu \big ( \wp (\lambda \cdot
x) - \wp (\mu \cdot x)
\big ).   \cr}
\eqno (5.75)
$$
The second terms in the first and second sums on the r.h.s. of (5.75) cancel
one another. The remaining terms may be separated provided $C_\lambda \cdot
C_\mu$ is independent of $\lambda$ and $\mu$ as long as they satsify $\lambda
\cdot \mu = \half$, and we obtain
$$
s d \cdot u_\lambda = d_0 
+\sum _{{\delta ^2 =2;\atop \delta \cdot \lambda =1}} m_2 \wp (\delta \cdot 
x)
-\sum _{\kappa \in [\lambda ] }
{1 \over 2 \sqrt 2} m_1 \wp (\kappa \cdot x) 
+ \big ( 
{\sqrt 2 \over m_1} C_\lambda \cdot C_\mu  + { m_1 \over \sqrt 2} \big )
\wp (\lambda \cdot x).
\eqno (5.76)
$$
Here, $d_0$ is independent of $\lambda$ (and of the 8-class of $\lambda$), 
and
the second on the r.h.s. is over all $\kappa$ in the 8-class of $\lambda$.

\medskip

We are now ready to put all the conditions obtained together and to solve
them. First, the expressions  (5.74) and (5.76) must agree, so that 
$C_\lambda
\cdot C_\mu = \half m_1 ^2$ whether $\lambda \cdot \mu =-\half$ or $+\half$.
This implies that $m_1 = \sqrt 2 m_2$ and as a consequence
$$
d_1 ([\lambda ]) = d_0 - \sum _{\kappa \in [\lambda ]}
\half m_2 \wp (\kappa \cdot x).
\eqno (5.77)
$$
Putting all inner product relations together, we have 
$$
C_\lambda \cdot C_\mu = \left \{
\matrix{ 2 m_2 ^2 & [\lambda ] = [\mu] \cr
m_2 ^2 & [\lambda ] \not= [\mu] \cr} \right .
\eqno (5.78)
$$
whence the solution of (5.61a).

\medskip

It now only remains to solve (5.70b); to do so, we use all the results
already obtained so far. First, we separate the sum over $\kappa$ according
to the inner product $\kappa \cdot \lambda$ :
$$
\eqalign{
C_{\lambda , a} s d \cdot (u _\lambda - u_a)
= &  \sum _{\kappa \cdot \lambda =0} m_2 C_{\kappa , a}
\big ( \wp ((\lambda - \kappa )\cdot x) - \wp (\kappa \cdot x) \big )
\cr
 & +  \sum _{\kappa \cdot \lambda =\half } m_2 C_{\kappa , a}
\big ( \wp ((\lambda - \kappa )\cdot x) - \wp (\kappa \cdot x) \big ).
\cr}
\eqno (5.79)
$$
Upon using the fact that $C_{\kappa, a}$ only depends upon the 8-class of
$\kappa$, and (5.77) to eliminate $sd \cdot u_\lambda$, we get
$$
\eqalign{
C_{\lambda , a} (d_0 - s d \cdot u_a)
= & - \sum _{\kappa \in [\lambda]} \half m_2 C_{\lambda , a}\wp (\kappa \cdot 
x)
+ \sum _{\kappa \in \!\!\! / [\lambda]} \half m_2 C_{\lambda \pm \kappa, a}  
\wp
(\kappa \cdot x)
\cr  
& -  \sum _{\kappa \in \!\!\! /[\lambda]} \half m_2 C_{\kappa , a} \wp
(\kappa \cdot x).
\cr}
\eqno (5.80)
$$
When $[\lambda ]=a$, the l.h.s. cancels and so does the first sum on the 
r.h.s.
of (5.80). The remaining two sums cancel one another. Thus, there only 
remains
an equation for $[\lambda ] \not= a$, which may be simplified with the help 
of
(5.61), to obtain
$$
\eqalign{
d_0 - s d \cdot u_a 
= &  - \sum _{\kappa \in [\lambda]} \half m_2 \wp (\kappa \cdot x)
+ \sum _{[\kappa]=a } \half m_2  \wp
(\kappa \cdot x)
 -  \sum _{\kappa \in \!\!\! /[\lambda], [\kappa ]\not=a} \half m_2 
\wp (\kappa \cdot x).\cr
= & - \sum _{\kappa} \half m_2 \wp (\kappa \cdot x) + \sum _{[\kappa ]=a}
m_2 \wp (\kappa \cdot x). 
\cr}
\eqno (5.81)
$$
Setting $d_0=0$ in this equation, we recover (5.61c), completing the proof
of Theorem 10.

\bigskip

\noindent
{\it $F_4$ Untwisted Elliptic Calogero-Moser Hamiltonian by Restriction from 
$E_6$}

\medskip

By comparing the root systems in Table 3 of Appendix \S A of $E_6$ and 
$F_4$, it is clear that the subalgebra $F_4$ of $E_6$ is obtained by 
projection 
orthogonal to the basis vectors $e_5$ and $e_6$. The projection of the 
weights 
of the  {\bf  27} of $E_6$ is immediately seen to reproduce the 24 non-zero 
weights of
the {\bf 26} of $F_4$, together with 3 zero weights, which arise from
the  weights $2/\sqrt 3 e_6$ and $-1/\sqrt 3 e_6 \pm e_5$ of $E_6$. We may 
check
directly that this restriction is consistent with the Hamilton-Jacobi 
equations 
for  the Calogero-Moser system (2.1). Indeed, when $e_5\cdot x = e_6 \cdot 
x=0$,
the right hand side of (2.1) is orthogonal to $e_5$ and $e_6$ :
$$
0= \sum _{\alpha \in \R (E_6)} m_2 ^2 
\wp ' (\alpha \cdot x) (e_{5,6} \cdot \alpha ).
\eqno (5.82)
$$
The contributions from the roots of $E_6$ of the form $\pm e_i \pm e_j$, 
$1\leq i < j \leq 5$ are readily seen to cancel  in (5.82), leaving just
$$
0=   \sum _{\epsilon _i = \pm 1} m_2 ^2 \epsilon _5
\wp ' \bigl ( \half \sum _{i=1} ^4 \epsilon _i e _i \cdot x \bigr )
\qquad \qquad
\epsilon _5 \prod _{i=1} ^4 \epsilon _i =1 .
\eqno (5.83)
$$
The contribution for $\epsilon _5=+1$ is a sum over the weights of
the Weyl spinor $8^s$ of $D_4$, while that for $\epsilon _5=-1$ 
is over the weights of the Weyl spinor $8^c$ of $D_4$. Since both 
representations are real, their weight lattices are even 
under sign reversal. Thus, the contributions to (5.83)
from $\epsilon = \pm1$ cancel separately in (5.83), and the 
Hamiltonian-Jacobi
equations for the Calogero-Moser system of $E_6$ in the {\bf 27} projects
consistently to the Hamiltonian for $F_4$ in the {\bf 26} $\oplus $ {\bf 1} 
of
$F_4$. The relation of the couplings $m_1$ and $m_2$ is automatically
guaranteed. 

\bigskip
\bigskip

\centerline{{\bf VI. THE TWISTED ELLIPTIC CALOGERO-MOSER SYSTEMS}}

\bigskip

The twisted elliptic Calogero-Moser systems associated with simply-laced $\G$ 
are identical to the untwisted ones, which were solved for in \S IV and \S V. 
In this section we propose to solve for  the 
Lax pair conditions of Theorem 1 for non-simply laced $\G$. There are thus 
only 
four cases left. For $\G  = B_n,~C_n$ we shall derive Lax pairs of dimensions 
$2n$ and $2n+2$ respectively. These dimensions happen to coincide with the 
dimensions in which the respective associated twisted affine Lie algebras 
$(B_n 
^{(1)})^\vee = C_n ^{(2)}$ and $(C_n ^{(1)})^\vee = D_{n+1} ^{(2)}$ can be 
realized. For $\G=F_4$, we shall obtain a Lax pair of dimension 24, which, 
surprisingly, is not equal to the dimension of any representation of $F_4$, 
but 
arises in relation with the number of short roots of $F_4$. For $\G = G_2$, 
we 
have not succeeded in proving the existence of a Lax pair. We strongly 
believe 
that  there  should exist a Lax pair of dimension 6 or 8, but the elliptic 
function analysis  appears unwieldy at this point.

\medskip

\noindent
{\bf (a) Twisted Elliptic Calogero-Moser System for B$_{\bf n}$}

\medskip

The twisted elliptic Calogero-Moser Hamiltonian for $B_n$ is defined by
$$
H= \half p \cdot p 
-  \sum _{\alpha \in \R _l (B_n)} \half m_2 ^2 \wp (\alpha \cdot x) 
-  \sum _{\alpha \in \R _s (B_n)} \half m_1 ^2 \wp  _2 (\alpha \cdot x ) ,
\eqno (6.1)
$$
where the roots are divided into long and short roots of $B_n$
$$
\eqalign{
\R _l (B_n) = &\{ \pm (e_i - e_j), ~\pm (e_i + e_j), ~ 1\leq i < j \leq n \} 
\cr
\R _s (B_n) = &\{ \pm e_i \}. \cr}
\eqno (6.2)
$$
Here, we have chosen the twisted half period of $\wp _2$ of (2.4) to be
$\omega _1$ for later convenience. It is possible (by a Landen transformation 
[24]) to express this Hamiltonian as a twisted elliptic Calogero-Moser system 
for the root system of the dual Lie algebra $B_n ^\vee = C_n$. To see this, it 
suffices to use (B.12) and to perform a canonical transformation $x \to 2 x$, 
$p \to  p/2$,  
$$
H=  {1 \over 8} p \cdot p
-  \sum _{\alpha \in \R _l (B_n ^\vee )} {1 \over 8} m_1 ^2 \wp _2 ( \alpha 
\cdot x) 
-  \sum _{\alpha \in \R _s (B_n ^\vee )} \half m_2 ^2 \{ \wp _2  ( \alpha 
\cdot 
x )
+ \wp _2 ( \alpha \cdot x + \omega _2)\},
\eqno (6.3)
$$
where the roots are expressed as roots of $C_n$
$$
\eqalign{
\R _s (B_n ^\vee ) = &\{ \pm (e_i - e_j), ~\pm (e_i + e_j), ~ 1\leq i < j 
\leq 
n 
\} 
\cr
\R _l (B_n ^\vee) = &\{ \pm 2e_i \}. \cr}
\eqno (6.4)
$$

\bigskip

\noindent
{\bf Theorem 11 : Lax pair for B$_{\bf n}$ Twisted Calogero-Moser} 

\medskip

The twisted elliptic Calogero-Moser Hamiltonian for $B_n $ admits a Lax pair 
of
dimension $N=2n$, with spectral parameter and two independent couplings $m_1$ 
and $m_2$. The Lax operators are given by (3.5), (3.6), $\Delta =0$ and 
$$
\eqalignno{
\Phi _{IJ}  (x,z) = & \left \{ 
\matrix{\Phi (x,z) & I-J \not= 0,\pm n \cr
         \Lambda (x,z) & I-J = \pm n \cr } \right . 
& (6.5a) \cr
C _{I,J}   = & \left \{ 
\matrix{m_2 &  \qquad I-J \not= 0,\pm n \cr
         m_1 & \qquad I-J = \pm n \cr }  \right .
& (6.5b) \cr
d \cdot v_i = & \sum _{J-i \not= 0, n} m_2 \wp ((e_i - \lambda _J)
\cdot x) + \half m_1 \wp _2 (e_i \cdot x) 
& (6.5c) \cr}
$$
The function $\Lambda (x,z)$ is defined in (B.22) of Appendix \S B, and the
weight vectors $\lambda _I$ will be defined below. 

\medskip

To prove this Theorem, we verify that the conditions (1), (2) and (3) of
Theorem 1, with $\Delta =0$ are obeyed. The natural starting point for the 
construction of the Lax pair for the twisted elliptic Calogero-Moser system 
for 
$B_n$ appears to be the dual algebra $\G = B_n ^\vee$. We begin by defining 
the
weights $\lambda$ of Theorem 1 as the weights of the fundamental 
representation
of $B_n ^\vee$, which is of dimension $N=2n$, and which are given by
$\lambda _i = - \lambda _{n+i} = e_i$ for $ i=1,\cdots , n$.
Following Theorem 1, we embed this representation into $GL(N,\C)$ by (3.1), 
with
$s^2=2$, and
$$
\sqrt 2 u _i = e_i +v_i ,
\qquad 
\sqrt 2 u_{n+i} = -e_i + v_i
\qquad 
i=1,\cdots ,n. 
\eqno (6.6)
$$
The roots of $GL(N, \C)$ decompose into short roots of $B_n ^\vee$ 
$$
\eqalign{
\sqrt 2 (u_i - u_j)                = &  +e_i - e_j + v_i - v_j\qquad i\not=j 
\cr
\sqrt 2 (u_{\nu +j} - u_{\nu +i} ) = &  +e_i - e_j - v_i + v_j\qquad i\not=j 
\cr
\sqrt 2 (u_i - u_{\nu +j})         = &  +e_i + e_j + v_i - v_j\qquad i\not=j 
\cr 
\sqrt 2 (u_{\nu +i} - u_j)         = &  -e_i - e_j + v_i - v_j\qquad i\not=j,
\cr}
\eqno (6.7a)
$$
and long roots of $B_n ^\vee$
$$
\eqalign{ 
\sqrt 2 (u_i - u_{\nu +i})     = & +2e_i \cr
\sqrt 2 (u_{\nu +i} - u_i)     = & -2e_i. \cr}
\eqno (6.7b)
$$
Condition (2) of Theorem 1 is manifestly satisfied by the values of $C$
listed in (6.5b), because each short root of $B_n ^\vee$ has two roots of
$GL(N,\C)$ as pre-images, and they come with opposite values of $v_i - v_j$,
which automatically cancel in (3.8). Each long root of $B_n ^\vee$ has no
$v$-dependence at all and thus does not enter into (3.8).  
Satisfying condition (1) of Theorem 1 requires that the coefficients $C$
satisfy the relations of (6.5b), that the function $\Phi$ satisfy (2.10) and 
that the functions $\Lambda$ and $\wp _2$ obey
$$
\Lambda (2x,z) \Lambda ' (-2x,z) - \Lambda '(2x,z) \Lambda (-2x,z) =  \half
\wp _2 '(x). 
\eqno (6.8)
$$ 
This relation follows from the definitions and results of (B.22-25) in 
Appendix 
\S B.

\medskip

It remains to satisfy condition (3) of Theorem 1, for $\Delta =0$. Using the
anti-symmetry of the r.h.s. of condition (3) in (3.9) under $x\to -x$ and 
$I\to
J$ on the l.h.s. of (3.9) implies that $d (-x) = d(x)$. Using now this
symmetry, we may restrict attention to the cases $I<J$ in verifying (3.9).
Two cases then arise : $J= n+I$ and $J-I \not= 0,n$, which we analyze
separately. We begin with the first, for which (3.9) reduces
to\footnote{*}{Henceforth we use the abbreviation $x_i = e_i \cdot x$.}
$$
\eqalign{
m_1 \Lambda (2x_i) d \cdot 2e_i
= & \sum _{K\not= i, n+i} m_2 ^2  \{
\Phi (\alpha _{iK} \cdot x) \Phi '(\alpha _{Kn+i}\cdot x)
-\Phi '(\alpha _{iK} \cdot x) \Phi (\alpha _{Kn+i}\cdot x) \} 
\cr
= & \ m_2^2 \Phi (2x_i) \sum _{k\not=i} ^n  \{
\wp (\alpha _{ik} \cdot x) - \wp (\alpha _{k,n+i} \cdot x) \cr
{} \qquad \qquad \qquad 
& + \wp (\alpha _{i,n+k} \cdot x) - \wp (\alpha _{n+k,n+i} \cdot x) \}.
\cr
}
\eqno (6.9)
$$
The r.h.s. of (6.9) is easily seen to vanish,
which simply requires that $d\cdot e_i=0$. Next, we consider the case $J-I
\not=0,n$, for which (3.9) becomes
$$
\eqalign{
 m_2 \Phi (\alpha _{IJ} & \cdot x) s d \cdot (u_I - u_J)
 \cr
= & \sum _{{I-K \not=0, \pm n \atop K-J \not=0,\pm n}}
m_2 ^2 \{ \Phi (\alpha _{IK} \cdot x) \Phi ' (\alpha _{KJ} \cdot x)
         -\Phi '(\alpha _{IK} \cdot x) \Phi (\alpha _{KJ}\cdot x) \}
\cr
+ & \sum _{{I-K = \pm n \atop K-J \not=0,\pm n}}
m_1 m_2 \{ \Lambda (\alpha _{IK} \cdot x) \Phi ' (\alpha _{KJ} \cdot x)
         -\Lambda '(\alpha _{IK} \cdot x) \Phi   (\alpha _{KJ} \cdot x) \}
\cr
+ & \sum _{{I-K \not=0, \pm n \atop K-J =\pm n}}
m_1 m_2 \{ \Phi (\alpha _{IK} \cdot x) \Lambda ' (\alpha _{KJ} \cdot x)
         -\Phi '(\alpha _{IK} \cdot x) \Lambda   (\alpha _{KJ} \cdot x) \}
\cr
+ & \sum _{{I-K = \pm n \atop K-J =\pm n}}
m_1 ^2 \{ \Lambda (\alpha _{IK} \cdot x) \Lambda ' (\alpha _{KJ} \cdot x)
         -\Lambda '(\alpha _{IK} \cdot x) \Lambda (\alpha _{KJ}\cdot x) \}
\cr}
\eqno (6.10)
$$
The last sum in (6.10) vanishes identically, because the conditions $I-K = 
\pm
n$ and $ K-J =\pm n$ imply that $I-J=0,\pm 2n$, which is impossible since
$I\not=j$. By noticing that if $I-K=\pm n$, we have that $\lambda _I = -
\lambda _K$ for all $I$ and $K$, we can easily make the second and third sums
collapse to single terms. Thus, we obtain
$$
\eqalign{
 m_2 \Phi & (\alpha _{IJ} \cdot x) s d \cdot (u_I - u_J) \cr
 = &  \sum _{{I-K \not=0, \pm n \atop K-J \not=0,\pm n}}
m_2 ^2 \{ \Phi (\alpha _{IK} \cdot x) \Phi ' (\alpha _{KJ} \cdot x)
         -\Phi '(\alpha _{IK} \cdot x) \Phi (\alpha _{KJ}\cdot x) \}
\cr
& +  m_1 m_2 \{ 
\Lambda (2\lambda _I \cdot x) \Phi ' (-(\lambda _I+\lambda _J) \cdot x)
 -\Lambda '(2\lambda _I \cdot x) \Phi (-(\lambda _I + \lambda _K) \cdot x) \}
\cr
& +   m_1 m_2 \{ 
\Phi ((\lambda _I + \lambda _J) \cdot x) \Lambda ' (-2\lambda _J \cdot x)
 -\Phi '((\lambda _I + \lambda _J) \cdot x) \Lambda (-2\lambda _J \cdot x) 
\}.
\cr}
\eqno (6.11)
$$
We now make use of the relations (B.20) and (B.23) for the functions $\Phi$ 
and
$\Lambda$ to simplify the r.h.s. of (6.11). Omitting also an overall factor 
of
$m_2 \Phi (\alpha _{IJ} \cdot x)$, (6.11) is reduced to
$$
sd \cdot (u_I - u_J) 
= \sum _{{I-K\not=0,\pm n \atop K-J \not=0,\pm n}}
\{ \wp (\alpha _{IK} \cdot x) - \wp (\alpha _{KJ} \cdot x) \}
+\half m_1 \{ \wp _2 (\lambda _I \cdot x) - \wp _2 (\lambda _J \cdot x ) \},
\eqno (6.12)
$$
from which (6.5c) readily follows, and this concludes the proof of Theorem 
11.

\bigskip

\noindent
{\bf (b) Twisted Elliptic Calogero-Moser System for C$_{\bf n}$}

\medskip

The twisted elliptic Calogero-Moser Hamiltonian for $C_n $ is given by
$$
H=  \half p \cdot p
-  \sum _{\alpha \in \R _l (C_n)} \half m_4 ^2 \wp (\alpha \cdot x) 
-  \sum _{\alpha \in \R _s (C_n)} \half m_2 ^2 \wp  _2 (\alpha \cdot x ) ,
\eqno (6.13)
$$
where the roots are divided into long and short roots of $C_n$
$$
\eqalign{
\R _s (C_n) = &\{ \pm (e_i - e_j), ~\pm (e_i + e_j), ~ 1\leq i < j \leq n \} 
\cr
\R _l (C_n) = &\{ \pm 2e_i \}, \cr}
\eqno (6.14)
$$
and where we have chosen the twisted half period of $\wp _2$ of (2.4) to be
$\omega _1$ for later convenience. It is possible (by a Landen transformation 
[24]) to express this Hamiltonian as a twisted elliptic Calogero-Moser system 
for the root system of the dual Lie
algebra $ C_n ^\vee = B_n$ (given in (6.2)), by using (B.12),
$$
H=  \half p \cdot p
-  \sum _{\alpha \in \R _l (C_n ^\vee)} \half m_2 ^2 \wp _2 ( \alpha \cdot x) 
-  \sum _{\alpha \in \R _s (C_n ^\vee)} {1 \over 8} m_4 ^2 \{ \wp _2  ( 
\alpha 
\cdot x ) + \wp _2 ( \alpha \cdot x + \omega _2)\}.
\eqno (6.15)
$$

\bigskip

\noindent
{\bf Theorem 12 : Lax pair for C$_{\bf n}$ Twisted Calogero-Moser}

\medskip

The twisted elliptic Calogero-Moser Hamiltonian for $C_n$ admits a Lax pair 
of
dimension $N=2n+2$, with spectral parameter and one independent couplings 
$m_2$
given by (3.5), (3.6), $\Delta =0$ and 
$$
\eqalignno{
\Phi _{IJ}  (x,z) = & 
\Phi _2 (x+\omega _{IJ},z) 
& (6.16a) \cr
\omega _{IJ} = &
\left \{ 
\matrix{0 & I\not=J =1, \cdots , 2n+1 \cr
        +\omega _2 &  \quad I=1,\cdots ,2n; J=2n+2 \cr
        -\omega _2 &  \quad J=1,\cdots ,2n; I=2n+2 \cr } \right . 
& (6.16b) \cr
C _{I,J}   = & \left \{ 
\matrix{m_2 & I,J = 1,\cdots ,2n; I-J \not= \pm n \cr
        {1 \over \sqrt 2} m_4 = \sqrt 2 m_2  & I=1,\cdots ,2n; J=2n+1,2n+2; 
I\leftrightarrow J \cr
        2m_2& I=2n+1, J=2n+2; I \leftrightarrow J\cr }  \right .
& (6.16c) \cr
s d \cdot u_I  = &
 \sum _{J-I \not= 0, \pm n} m_2 \wp _2 ((\lambda _I - \lambda _J)
\cdot x) + 8 m_2 \wp  (2\lambda _I \cdot x)  ;~ I=1,\cdots ,2n 
& (6.16d) \cr
s d \cdot u_{2n+1}   = &
\sum _{J=1} ^{2n}  \wp _2 (\lambda _J\cdot x) + 2 m_2 \wp _2 (\omega _2) 
& (6.16e)  \cr
s d \cdot u_{2n+2}   = &
\sum _{J=1}^{2n} \wp _2 (\lambda _J\cdot x +\omega _2)+2m_2\wp _2 (\omega_2).   
& (6.16f)  \cr}
$$
The function $\Phi _2$ is defined in (B.22) by $\Phi _2 (x,z) = \Lambda 
(2x,z)$, 
and satisfies the differential 
equation (B.23) and (B.24), which are crucial in establishing (6.16).
The projected weight system is defined by $\lambda _i = - \lambda _{n+i} =
e_i$, $i=1,\cdots ,n$ and $\lambda _a=0$, $a=2n+1, ~ 2n+2$. Notice that 
$\omega 
_{IJ}$ as defined in (6.16b) satisfies 
$$
\eqalign{
\omega _{JI} = & - \omega _{IJ} \cr
\omega _{IJ} + \omega _{JK} & + \omega _{KI}=  0.
\cr}
\eqno (6.17)
$$

\medskip

To prove Theorem 12, we verify conditions (1), (2) and (3) of Theorem 1, with
$\Delta =0$. The algebra $\G$ in Theorem 1 is the dual algebra $\G = C_n 
^\vee 
=B_n$. The weights $\lambda _I$, $I=1,\cdots ,2n+2$, span the fundamental
representation of $B_n = C_n ^\vee $ plus a singlet, of dimension $N=2n+2$. 
We 
embed this representation into $GL(N,\C)$ by (3.1) with $s^2=2$ and 
$$
\eqalign{
\sqrt 2 u_i = & + e_i + v_i 
\qquad i=1,\cdots ,n \cr
\sqrt 2 u_{n+i} = & - e_i + v_i \cr
\sqrt u_a = & v_a 
\qquad \qquad a=2n+1, 2n+2. \cr}
\eqno (6.18)
$$
The roots of $GL(N,\C)$ decompose onto long roots of $C_n ^\vee$
$$
\eqalign{
\sqrt 2 (u_i - u_j) = +e_i - e_j + v_i - v_j
\qquad &
\sqrt 2 (u_{n+j} - u_{n+i}) = +e_i - e_j - v_i + v_j
\cr
\sqrt 2 (u_i - u_{n+j}) = +e_i + e_j + v_i - v_j
\qquad &
\sqrt 2 (u_j - u_{n+i}) = +e_i + e_j - v_i + v_j
\cr
\sqrt 2 (u_{n+i} - u_j) = -e_i - e_j + v_i - v_j
\qquad &
\sqrt 2 (u_{n+j} - u_i) = -e_i - e_j - v_i + v_j
\cr}
\eqno (6.19)
$$
and onto short roots of $C_n ^\vee $ 
$$
\eqalign{
\sqrt 2 (u_i - u_a) = +e_i  + v_i - v_a
\qquad &
\sqrt 2 (u_a - u_{n+i}) = +e_i - v_i + v_a
\cr
\sqrt 2 (u_{n+i} - u_a) = -e_i  + v_i - v_a
\qquad &
\sqrt 2 (u_a - u_i) = -e_i  - v_i + v_a.
\cr}
\eqno (6.20)
$$
and zero roots $\sqrt 2 (u_a - u_b) = v_a - v_b$. Here, we have $i,j=1,\cdots 
,2n$ and $a,b=2n+1, 2n+2$. Conditions (1) and (2) of Theorem 1 are satisfied 
provided the coefficients $C$ are such that they match the couplings $m_4$ 
and 
$m_2$ for the long and short roots respectively. This yields
$$
\eqalign{
m_2 ^2 = & C_{i,j} ^2 = C_{n+j,n+i}^2 = C_{i,n+j}^2 = C_{j,n+i} ^2 
\cr
\half m_4 ^2 = & C_{i,a} ^2 = C_{n+i,a}^2. 
\cr}
\eqno (6.21)
$$
Actually, a Lax pair exists when all square roots of the above relations are 
taken with the same sign. This gives rise to (6.16c), except for the fact 
that 
the relation between $m_4$ and $m_2$ remains to be established below. Notice 
that at this stage, the coefficient $C_{2n+1, 2n+2}$ is unconstrained.

\medskip

It remains to verify condition (3) of Theorem 1. It turns out that this 
condition may be satisfied for $\Delta =0$, which we shall henceforth assume.
With the help of the properties of $\omega _{IJ}$, condition (3) then reduces 
to
$$
C_{I,J} s d \cdot (u_I - u_J)
= \sum _{I\not=K\not=J} C_{I,K} C_{K,J}
\{ \wp _2 (\alpha _{IK} \cdot x + \omega _{IK}) 
- \wp _2 (\alpha _{KJ} \cdot x + \omega _{KJ}) \}.
\eqno (6.22)
$$
 Making use of an argument analogous to the one used to study condition (3) 
for 
(b) in the preceding subsection, we find that $d(-x) = d(x)$ and that (3.9) 
is 
anti-symmetric under the interchange of $I $ and $J$, so that we may limit 
the 
analysis to the cases $I<J$. 

\medskip

The case $I=i$, $J=n+i$, $i=1,\cdots,n$ is automatically satisfied, along the 
lines of (6.9) for the preceding case.
The case $I-J \not= 0,\pm n$ with $i,j=1,\cdots ,2n$ yields (with $K=1,\cdots 
,2n$)
$$
\eqalign{
m_2 s d \cdot (u_I - u_J)
= & \sum _{{I-K\not= 0,\pm n \atop K-J \not= 0,\pm n}}
m_2 ^2 \{ \wp _2 (\alpha _{IK} \cdot x) - \wp (\alpha _{KJ} \cdot x) \}
\cr
& + 2 m_4 ^2 \{ \wp  (2\lambda _I \cdot x)  - \wp (2\lambda _J)\} .
\cr}
\eqno (6.23)
$$
This equation is easily solved and we find
$$
sd\cdot u_I = \sum _{I-K\not=0,\pm n} \wp _2 ( \alpha _{IK} \cdot x)
+ 2 {m_4 ^2 \over m_2} \wp (2 \lambda _I \cdot x)
\eqno (6.24)
$$
up to an $I$-independent term which we omit.

\medskip

The cases $I=1,\cdots , 2n$, and $J=2n+1, 2n+2$ are reduced to
$$
\eqalign{
s d\cdot (u_I - u_{2n+1}) = &
\sum _{I-K \not=0,\pm n} m_2 \{ 
\wp _2 (\alpha _{IK} \cdot x) -  \wp _2 (\lambda _K \cdot x) \}
\cr
& + C_{2n+1, 2n+2} 
\wp _2 (\lambda _I \cdot x + \omega _2) - \wp _2 (\omega _2) \}
\cr
s d\cdot (u_I - u_{2n+2}) = &
\sum _{I-K \not=0,\pm n} m_2 \{ 
\wp _2 (\alpha _{IK} \cdot x) -  \wp _2 (\lambda _K \cdot x + \omega _2) \}
\cr
& + C_{2n+1, 2n+2} 
\wp _2 (\lambda _I \cdot x ) - \wp _2 (\omega _2) \}.
\cr}
\eqno (6.25)
$$
Substituting the result of (6.24) into (6.25), we obtain a relation between 
$m_4$ and $m_2$. To see this, we shall just look at the first equation in 
(6.25); the second one is completely analogous. We obtain
$$
\eqalign{
{m_4 ^2 \over 2 m_2} \{ 
\wp _2 (\lambda _I \cdot x) & + \wp _2 (\lambda _I \cdot x + \omega _2) \}
- s d \cdot u_{2n+1}\cr
= & - \sum _{I-K \not=0,\pm n} \wp _2 (\lambda _K \cdot x)
+ C_{2n+2,2n+1} \{ \wp _2 (\lambda _I \cdot x + \omega _2) - \wp _2 (\omega 
_2) 
\}.
\cr}
\eqno (6.26)
$$
To make the $I$-dependence match, we need to require that the terms in $\wp 
_2 
(\lambda _I \cdot x + \omega _2)$ cancel one another, and that the remaining 
$\wp _2 (\lambda _I \cdot x)$ term combine with the sum, so that all 
$I$-dependence can indeed disappear in $d\cdot u_{2n+1}$. The conditions are
$$
m_4 = 2 m_2
\qquad \qquad
C_{2n+1,2n+2} = { m_4 ^2 \over 2 m_2} = 2m_2
\eqno (6.27)
$$
and provide the missing identifications in (6.16c). The remaining equations 
for 
$d\cdot u_{2n+1}$ and $d\cdot u_{2n+2}$ yields (6.16d) and (6.16e). To 
complete 
the proof, one case remains : $I=2n+1,~J=2n+2$. Using the value for 
$C_{2n+1,2n+2}= 2m_2$ obtained in (6.27), this condition takes the form
$$
2m_2 s d \cdot (u_{2n+1} - u_{2n+2} )
\sum _{J=1} ^{2n} \{ \wp _2 (\lambda _J \cdot x) - \wp _2 (\lambda _J \cdot x 
+ 
\omega _2) \} .
\eqno (6.28)
$$
But this condition follows directly from the solutions for $d\cdot u_{2n+1}$ 
and $d \cdot u_{2n+2}$ already obtained in (6.16e) and (6.16f).

\bigskip

\noindent
{\bf (c) Twisted Elliptic Calogero-Moser System for F$_{\bf 4}$}

\medskip

The twisted elliptic Calogero-Moser Hamiltonian for $F_4$ is given by
$$
H = \half p \cdot p
- \sum _{\alpha \in \R_l (F_4)} \half m_2 ^2 \wp (\alpha \cdot x)
- \sum _{\alpha \in \R_s (F_4)} \half m_1 ^2 \wp _2 (\alpha \cdot x),
\eqno (6.29)
$$
where the long and short roots of $F_4$ are given by
$$
\eqalign{
\R _l (F_4) = & \{ \pm (e_i - e_j), ~ \pm (e_i + e_j), 1\leq i < j \leq 4 \}
\cr
\R _s (F_4) = & \{ \pm e_i; \half \sum _{i=1} ^4 \epsilon _i e_i; ~\epsilon 
_i 
= \pm 1\}.
\cr}
\eqno (6.30)
$$
Letting $x \to 2x$, $p \to p/2$ and using the duplication formula for the 
Weierstrass function of (B.12), which induces a Landen transformation [24], the 
Hamiltonian may be re-expressed in a dual form, 
$$
H = {1 \over 8}  p\cdot p
- \sum _{\alpha \in \R_l (F_4)} {1\over 8} m_2 ^2 \{ \wp _2 (\alpha \cdot x)
+ \wp _2 (\alpha \cdot x+ \omega _2) \}
- \sum _{\alpha \in \R_s (F_4)} \half m_1 ^2 \wp _2 (2\alpha \cdot x).
\eqno (6.31)
$$
Since the Lie algebra $F_4$ is selfdual, the set $\alpha \in \R_l$ plays the 
role of short roots of $F_4$, while the set $2 \alpha $ with $\alpha \in 
\R_s$ 
plays the role of long roots of $F_4$. 

\medskip

\noindent
{\bf Theorem 13 : Lax pair for $F_4$ Twisted Elliptic
Calogero-Moser}

\medskip

The twisted elliptic Calogero-Moser Hamiltonian for $F_4 $ admits a Lax pair 
of 
dimension $N=24$, with spectral parameter and two independent couplings $m_1$ 
and $m_2$, given by (3.5), (3.6), $\Delta=0$ and 
$$
\eqalignno{
\Phi _{\lambda \mu}  (x,z) = & \left \{ 
\matrix{\Phi  (x,z) & \lambda \cdot \mu =0  \cr
        \Phi _1 (x,z) & \lambda \cdot \mu = \half \cr
        \Lambda (x,z) & \lambda \cdot \mu =-1 \cr } \right . 
& (6.32a) \cr
C _{\lambda , \mu}   = & \left \{ 
\matrix{m_2 & \lambda \cdot \mu =0 \cr
        {1 \over \sqrt 2} m_1 & \lambda \cdot \mu =\half \cr
        0   & \lambda \cdot \mu =-\half \cr
        \sqrt 2 m_1 & \lambda \cdot \mu = -1 \cr }  \right .
& (6.32b) \cr
s d \cdot v_\lambda  = & 
\sum _{\delta \in \R_l;\delta \cdot \lambda =1} 
m_2 \wp (\delta \cdot x) 
- \sum _{\kappa \in [\lambda ]}  
{1 \over 2 \sqrt 2}m_1 \wp _2 (\kappa \cdot x)
+ { m_1 \over \sqrt 2} \wp _2 (\lambda \cdot x).
& (6.32c) \cr}
$$
Here, the entries of the 24-dimensional Lax pair are labeled by the 24 {\it 
non-zero} weights $\lambda$ of the {\bf 26} of $F_4$, which are also the 24 
short roots of $F_4$, as given in (6.30). As discussed in \S V (e), the 24 
short roots of $F_4$ fall into three different 8-classes $8^v$, $8^s$ and 
$8^c$, which are defined in (5.59). The 8-class of a short root $\lambda$ 
is denoted by $[\lambda]$. In this way, the second sum on the r.h.s. of 
(6.32c) 
is over all short roots $\kappa$ in the same 8-class as $\lambda$. The 
elliptic 
functions $\Phi $, $\Phi _1$ and $\Lambda$ are defined in \S B.

\medskip

Before we start proving this Theorem, some comments are in order. As is 
manifest from (6.32a,b), we shall retain the weights $2 \lambda$, which are 
obtained when $\mu  = -\lambda$, in the construction of the Lax pair.  The 
reasons for doing so are three-fold. First, since these weights are 
proportional to the short roots, it is certainly conceivable that with a 
suitable $\Phi _{IJ}$-function, these weights will enter into the Lax 
operators 
and thus into the Hamiltonian on the same footing with the short roots. 
Second, 
we have already observed this very phenomenon in the case of the twisted 
elliptic Calogero-Moser system for $B_n$ in \S VI (a). Third, this scheme 
works 
!

\medskip

To prove Theorem 13, we verify the conditions of Theorem 1, with $\Delta =0$.
The weights $\lambda$ are the short roots of $F_4$ and following Theorem 1, 
we 
embed these into $GL(24,\C)$ by (3.1). We denote the weight vectors of the 
fundamental representation of $GL(24,\C)$ by an orthonormal basis 
$u_\lambda$, 
and we have 
$$
su_\lambda  = \lambda + v_\lambda
\eqno (6.33)
$$
where $v_\lambda$ is orthogonal to all roots of $F_4$, and $s^2 =6$.
The roots of $GL(24,\C)$ decompose under $F_4$ as
$$
\sqrt 6 (u_\lambda - u_\mu) = \lambda - \mu + v_\lambda - v_\mu,
\qquad \qquad 
\lambda \not= \mu
\eqno (6.34)
$$
which correspond to long roots in $\R (F_4)$ provided $\lambda \cdot \mu=0$, 
or 
to short roots in $\R (F_4)$ provided $\lambda \cdot \mu=\half$. As discussed 
in the previous paragraph, we also retain the weights $2 \lambda$.

\medskip

We are now ready to analyze conditions (1) and (2) of Theorem 1. We begin by 
decomposing the sums over all roots in (3.7) and (3.8) into sums over long 
roots and short root. For any long root $\alpha \in \R_l$, we have 
$$
\eqalignno{
s^2 m_2 ^2 \wp '(\alpha \cdot x) = &
\sum _{\alpha = \lambda - \mu \in \R _l} 
C_{\lambda , \mu} ^2 \wp _{\lambda \mu} ' (\alpha \cdot x) 
& (6.35a) \cr
0 = &
\sum _{\alpha = \lambda - \mu \in \R _l} 
C_{\lambda , \mu} ^2 ~ (v_\lambda - v_\mu) ~ 
\wp _{\lambda \mu} ' (\alpha \cdot x), 
& (6.35b) \cr}
$$
while for any short root $\alpha \in \R_s$ (as well as their doubles 
$2\alpha$), we have
$$
\eqalignno{
s^2m_1 ^2 \wp _2 '(\alpha \cdot x) = &
\sum _{\alpha = \lambda - \mu \in \R _s} 
C_{\lambda , \mu} ^2 \wp _{\lambda \mu} ' (\alpha \cdot x) 
+ 2  C_{\alpha , - \alpha }^2 
\wp _{\alpha, -\alpha} ' (2 \alpha \cdot x)
& (6.36a) \cr
0 = &
\sum _{\alpha = \lambda - \mu \in \R _s} 
C_{\lambda , \mu} ^2 ~ (v_\lambda - v_\mu) ~ 
\wp _{\lambda \mu} ' (\alpha \cdot x)
\cr
&\qquad \qquad
 + C_{\alpha , -\alpha} ^2 ~ (v_\alpha - v_{-\alpha}) ~
\wp _{\alpha , -\alpha} ' (2 \alpha \cdot x). 
& (6.36b) \cr}
$$
 
\medskip
 
We analyze first the case of long roots $\alpha \in \R_l$ of (6.35). Taking 
the 
inner product of (6.35b) with $u_\sigma$ where $\sigma$ is an arbitrary short 
root, using (3.3), and (6.35a) gives 
$$
m_2 ^2 \alpha \cdot \sigma ~\wp '(\alpha \cdot x)
=\sum _{\alpha = \lambda - \mu} 
C_{\lambda , \mu } ^2 (\delta _{\lambda , \sigma} - \delta _{\mu, \sigma})
\wp ' _{\lambda \mu} (\alpha \cdot x).
\eqno (6.37)
$$
Since $\alpha \in \R_l$, the product $\alpha 
\cdot \sigma$ can take only the values $ -1,0,+1$. For $\alpha \cdot \sigma 
=0$, both sides of (6.37) vanish separately, while for $\alpha \cdot \sigma = 
\pm 1$, only one term remains from the sum on the right. Setting $C_{\lambda 
,\mu} = m_2$, as in (6.32b), we obtain (omitting an irrelevant constant upon 
integrating)
$$
\wp  (x) = \wp  _{\lambda \mu} (x)
\qquad \qquad
\lambda - \mu \in \R_l (F_4).
\eqno (6.38)
$$

\medskip

For the short roots, we proceed analogously and take the inner product of 
(6.36b) with $u_\sigma$ for any short root $\sigma$. We find  
$$
\eqalign{
m_1 ^2 ~\alpha \cdot \sigma ~\wp _2 ' (\alpha \cdot x)
= \sum _{\alpha = \lambda - \mu \in \R _s}
& C_{\lambda , \mu} ^2 (\delta _{\lambda , \sigma} - \delta _{\mu , \sigma})
\wp ' _{\lambda \mu} (\alpha \cdot \sigma) \cr
& + C_{\alpha , -\alpha } ^2 
(\delta _{\alpha , \sigma} - \delta _{-\alpha , \sigma})
\wp ' _{\alpha ,-\alpha} (2\alpha \cdot x).\cr}
\eqno (6.39)
$$
Since $\alpha \in \R _s$, the product $\alpha \cdot 
\sigma$ can take only the values $-1,-\half,0,+\half,+1$. For $\alpha \cdot 
\sigma=0$, both sides of (6.39) vanish separately. For $\alpha \cdot \sigma = 
\pm \half$, we have $\sigma \not= \pm \alpha$, $\sigma \mp \alpha$ is a short 
root, while $\sigma \pm \alpha$ is not a root. Thus, only the sum on the 
r.h.s. 
of 
(6.39) contributes. By choosing $C_{\lambda , \mu}= m_1 /\sqrt 2$, as in 
(6.32b), $\wp _{\lambda \mu}$ is given in (6.40a) below.
For $\alpha \cdot \sigma = \pm 1$, we have $\sigma = \pm \alpha$. We shall 
set 
$C_{\alpha, -\alpha }= \sqrt 2 m_1$, as in (6.32b), so that
$$
\eqalignno{
\wp _2 (x) = & \wp _{\lambda \mu} (x)
\qquad \qquad
\lambda - \mu \in \R _s (F_4)
& (6.40a) \cr
\wp _2  ( x) = & \wp _{\alpha, -\alpha} (2x)
\qquad \qquad 
\alpha \in \R_s (F_4), 
& (6.40b) \cr}
$$
where we have again omitted irrelevant additive integration constants.
The functions $\Phi _{\lambda \mu}$ are related to the Weierstrass functions 
$\wp _{\lambda \mu}$ by (3.10). We conclude from (6.38) and (6.40) that there
should be three different kinds of $\Phi$-functions, $\Phi $, $\Phi _1$ and 
$\Lambda$, as in (6.32a), which are precisely those defined in Appendix \S B 
: 
(B.16), (B.26) and (B.22) respectively.

\medskip

It remains to satisfy condition (3) of Theorem 1. This condition holds 
provided 
the functions $\Phi $, $\Phi _1$ and $\Lambda$ obey the relations (B.21), 
(B.23), (B.24) (B.27) and (B.28).
To see this, we notice that condition (3) of Theorem 1 reduces to
$$
\eqalign{
C_{\lambda, \mu}  \Phi _{\lambda \mu} ((\lambda - \mu)\cdot x) 
s d\cdot (u_\lambda - u_\mu)&  \cr
 = 
\sum _{\kappa \not= \lambda, \mu}
C_{\lambda ,\kappa} C_{\kappa, \mu}
\{ &\Phi _{\lambda \kappa} ((\lambda - \kappa)\cdot x )
\Phi _{\kappa \lambda} ' ((\kappa - \mu)\cdot x) \cr
& -
\Phi _{\lambda \kappa} ' ((\lambda - \kappa)\cdot x)
\Phi _{\kappa \mu} ((\kappa - \mu)\cdot x) \} .\cr}
\eqno (6.41)
$$
Here, $\lambda, ~ \mu$ and $\kappa$ run over $\R _s(F_4)$.
We analyze this equation according to the values of $\lambda \cdot \mu 
=-1,-\half,0,+\half$; the value $\lambda \cdot \mu=+1$ is excluded since 
$\lambda \not=\mu$. 

\medskip

For $\lambda \cdot \mu =-1$, we have $\mu = -\lambda$, and $\lambda$ and 
$\mu$ 
belong to the same 8-class. Thus the sum on the r.h.s. of (6.41) for this 
case 
can receive contributions only from $\kappa \cdot \lambda =\kappa \cdot \mu = 
0,\half$. However, the contributions $\kappa\cdot \lambda = \kappa \cdot 
\mu=\half$ vanish since then at least one of the coefficients $C$ must 
vanish. 
The remaining  r.h.s. of (6.41)
vanishes by requiring equation (B.24). The l.h.s. then requires that $d\cdot 
u_{-\lambda} = d \cdot u_\lambda$.

\medskip

For $\lambda \cdot \mu= - \half$, the l.h.s. of (6.41) vanishes, since 
$C_{\lambda, \mu}=0$ then. The contributions to the sum over $\kappa$ for 
which 
$\kappa \cdot \lambda = \kappa \cdot \mu=0$ cannot contribute since this 
condition would imply that $\kappa$ belongs both to the 8-class of $\lambda$ 
and of $\mu$. But, by $\lambda \cdot \mu = -\half$, $\lambda $ and $\mu$ 
belong 
to different 8-classes. The contributions to the sum from $\kappa \cdot 
\lambda 
= \half,~\kappa \cdot \mu=0$ and $\kappa \cdot \lambda =0,~\kappa \cdot \mu= 
\half$ cancel one another. Making use of (B.27a) on the remaining sum with 
$\kappa \cdot \lambda = \kappa \cdot \mu = \half$ and of
$$
\sum _{\kappa \cdot \lambda = \kappa \cdot \mu = \half}
 \{ 
  \wp _2 ((\lambda - \kappa)\cdot x) 
- \wp _2 ((\kappa  - \mu)\cdot x) \}
=  \{ \wp _2 (\mu \cdot x) - \wp _2 (\lambda \cdot x) \},
\eqno (6.42)
$$
we find
$$
\eqalign{
\half & m_1 ^2 \{ \wp _2 (\lambda \cdot x)  - \wp _2 (\mu \cdot x) \}
\Phi _1 ((\lambda -\mu) \cdot x) \cr
= & ~ m_1 ^2\{
\Lambda (2 \lambda \cdot x) \Phi _1 ' (-(\lambda + \mu)\cdot x)
- \Lambda ' (2 \lambda \cdot x) \Phi _1 (-(\lambda + \mu) \cdot x) \cr
& \qquad 
-\Lambda  (-2 \mu \cdot x) \Phi _1 ' ((\lambda + \mu)\cdot x)
+ \Lambda ' (-2 \mu \cdot x) \Phi _1 ((\lambda + \mu) \cdot x)\}.
\cr}
\eqno (6.43)
$$
This equation will be satisfied when $\Phi _1$ and $\Lambda$ obey (B.28b).

\medskip

For $\lambda \cdot \mu=0$, $\lambda$ and $\mu$ belong to the same 8-class, so 
the sum in (6.43) reduces to contributions from $\kappa \cdot \lambda = 
\kappa 
\cdot \mu=0$, $\kappa = -\lambda$, $\kappa = -\mu$ and $\kappa \cdot \lambda 
= 
\kappa \cdot \mu = \half$. The latter 
sum is proportional to
$$
\eqalign{
\sum _{\kappa \cdot \lambda = \kappa \cdot \mu = \half} &
 \{ 
\Phi _1 ((\lambda -\kappa)\cdot x) \Phi _1 ' ((\kappa -\mu)\cdot x)
-\Phi _1 '((\lambda -\kappa)\cdot x) \Phi _1 ((\kappa -\mu)\cdot x) \}
\cr
&=
\Phi _1 ((\lambda - \mu) \cdot x)
\sum _{\kappa \cdot \lambda = \kappa \cdot \mu = \half}
\{
  \wp _2 ((\lambda - \kappa)\cdot x) 
- \wp _2 ((\kappa  - \mu)\cdot x) \},
\cr }
\eqno (6.44)
$$
and is easily seen to vanish by changing variables in the sum on the r.h.s. 
to 
$\delta = \lambda -\kappa$ for the first term, and $\delta = \kappa - \mu$ 
for 
the second term. In the remaining terms on the r.h.s. of (6.41), we use 
(B.24) 
for $\Phi $ and $\Lambda$, we simplify by the common overall factor of $m_2 
\Phi  ((\lambda - \mu ) \cdot x)$ and obtain
$$
\eqalign{
s d \cdot (u_\lambda - u_\mu)
= & \sum _{\delta \in \R_l;\delta \cdot \lambda =1}
m_2 \wp (\delta \cdot x)
-\sum _{\delta \in \R_l;\delta \cdot \mu =1}
m_2 \wp (\delta \cdot x)
\cr
& + {1 \over \sqrt 2} m_1 \{ \wp _2 (\lambda \cdot x) - \wp _2 (\mu \cdot x) 
\}.
\cr}
\eqno (6.45)
$$
This equation may be solved for $d\cdot u_\lambda$, but care must be taken of 
the fact that in (6.45), $\lambda$ and $\mu$ belong to the same 8-class. 
Thus, 
(6.45) separates and is determined up to a function that only depends upon 
the 
8-class of $\lambda$. We find
$$
s d \cdot u _\lambda 
= d_0 ([\lambda]) + 
\sum _{\delta \in \R_l;\delta \cdot \lambda =1} m_2\wp (\delta \cdot x)
+ { 1\over \sqrt 2} m_1 \wp _2 (\lambda \cdot x)
\eqno (6.46)
$$

\medskip

Finally, for $\lambda \cdot \mu = \half$, we cannot have $\kappa = -\lambda$ 
of 
$\kappa =-\mu$ since then $\kappa \cdot \mu =-\half$ and $\kappa 
\cdot \lambda = -\half $ respectively, but the coefficients $C$ corresponding 
to these values cancel. Also, the sums $\kappa \cdot \lambda = \kappa 
\cdot \mu=0$ cannot occur since $\lambda$ and $\mu$ are in different 
8-classes.
In the remaining sums, we make use of (B.27) and (B.28), we simplify by a 
common factor of ${1 \over \sqrt2 } m_1 \Phi _1 ((\lambda - \mu )\cdot x)$ 
and 
we obtain
$$
\eqalign{
s d \cdot (u_\lambda - u_\mu)
= & \sum _{\delta \in \R _l; \delta \cdot \lambda =1}
m_2 \wp (\delta \cdot x)
-\sum _{\delta \in \R _l; \delta \cdot \mu =1}
m_2 \wp (\delta \cdot x)\cr
& + \sum _{\kappa \cdot \lambda = \kappa \cdot \mu = \half}
{1 \over \sqrt 2} m_1 \{ \wp _2 ((\lambda - \kappa )\cdot x) 
- \wp _2 ((\kappa - \mu)\cdot x) \}
\cr}
\eqno (6.47)
$$
The last equation is solved by (6.46) with 
$$
d_0 ( [ \lambda ]) = - \sum _{\kappa \in [\lambda]} {1 \over 2 \sqrt 2} m_1  
\wp _2 (\lambda \cdot x)
\eqno (6.48)
$$ 
which together with (6.46) yields (6.32c). This completes the proof of 
Theorem 
13.

\bigskip

\noindent
{\bf (d) Twisted Elliptic Calogero-Moser System for G$_{\bf 2}$}

\medskip

The twisted elliptic Calogero-Moser Hamiltonian for $G_2$ is given by
$$
H = \half p \cdot p
- \sum _{\alpha \in \R_l (G_2)} \half m_2 ^2 \wp (\alpha \cdot x)
- \sum _{\alpha \in \R_s (G_2)} \half m_{2/3} ^2 \wp _3 (\alpha \cdot x),
\eqno (6.49)
$$
where the long and short roots of $G_2$ are given by
$$
\eqalign{
\R _l (G_2) = & \{ \pm (e_i - e_j),  1\leq i < j \leq 3 \}
\cr
\R _s (G_2) = & \{ \pm (e_i - e_0); \ i=1,2,3, \ e_0 = \13 (e_1 + e_2 +e _3 
)\}.
\cr}
\eqno (6.50)
$$

\medskip

We have only partially succeeded in solving for the Lax pair of the twisted 
elliptic Calogero-Moser system associated with $G_2$. The difficulty arises 
principally from : (1) the fact that the dimension $N$ of the Lax pair 
representation is not a priori known (even though some eductaed guesses may 
be 
made as to what $N$ should be), (2) the fact that several different and 
unknown 
elliptic type functions should enter, (3) the fact that the unknown elliptic 
functions satisfy many coupled non-linear differential equations. Below we 
shall 
briefly discuss the equations involved and our best conjecture for what the 
solution should look like.

\medskip

The dimension $N$ of the Lax pair has two natural candidates : $N=8$, as 
the dimension in which $(\G_2 ^{(1)})^\vee = D_4 ^\vee$ can be realized and 
$N=6$, as the number of short roots, by analogy with $F_4$. Indeed, group 
theoretically, $F_4$ and $G_2$ appear to be very similar in some respects. 
The 
set of long roots on the one hand and the set of short roots on the other 
hand 
each form the root system of mutually isomorphic subalgebras : $D_4$ for 
$F_4$ 
and $A_2$ for $G_2$. In either case, the representation on which the Lax pair 
is 
built will contain the short roots (which coincide with the non-zero weights 
of 
the {\bf 7} of $G_2$), and we shall leave the number $\nu$ of zero weights 
undetermined. Notice that $\nu$ will be allowed to vanish ! Thus, the weights 
$\lambda _I$, $I=1,\cdots, 6+\nu$ are given 
by $\lambda _i= - \lambda _{3+i} = \alpha _i$, $i=1,2,3$ and $\lambda _a=0$, 
$a=7,\cdots ,6+\nu$ and the embedding into $GL(6+\nu,\C)$ as usual by (3.1) : 
$ 
su_I = \lambda _I + v_I$ with $s^2=2$.
The roots of $GL(6+\nu, \C)$ decompose under $G_2$ as follows
$$
\eqalign{
s(u_\lambda - u_\mu) & = \lambda - \mu + v_\lambda - v_\mu \cr
s(u_\lambda - u_a  ) & = \lambda       + v_\lambda - v_a   \cr
s(u_a       - u_\mu) & =         - \mu + v_a       - v_\mu \cr
s(u_a       - u_b  ) & =  v_a - v_b. \cr}
\eqno (6.51)
$$
Long roots of $G_2$ arise in the first line when $\lambda \cdot \mu = -1/3$, 
while short roots arise in the second and third lines, and when $\lambda 
\cdot 
\mu = 1/3$. The double short roots $2 \lambda$ arise when $\mu = - \lambda$, 
i.e. $\lambda \cdot \mu = - 2/3$.

\medskip

For long roots $\alpha =\lambda - \mu$ with $\lambda \cdot \mu = -1/3$, 
conditions (1) and (2) of Theorem 1 are 
$$
\eqalign{
2m_2 ^2 \wp ' (\alpha \cdot x)
& =
\sum _{3\lambda \cdot \mu =-1} 
C_{\lambda, \mu}^2 \wp ' _{\lambda \mu} (\alpha \cdot x) \cr
0 & =
\sum _{3\lambda \cdot \mu =-1} 
C_{\lambda, \mu}^2 \wp ' _{\lambda \mu} (\alpha \cdot x) (v_\lambda - 
v_\mu).\cr}
\eqno (6.52)
$$
This set of equations is simply solved as follows : whenever $\lambda \cdot 
\mu 
= -1/3$, we have
$$
\eqalign{
C_{\lambda , \mu} & = m_2 \cr
\Phi _{\lambda \mu} (x,z) & = \Phi (x,z) \cr
\wp _{\lambda \mu} (x) & = \wp (x). \cr }
\eqno (6.53)
$$

\medskip

For short roots $\alpha = \lambda - \mu$ with $\lambda \cdot \mu = 1/3$, 
those 
arising from the second and third lines in (6.51), and those arising from the 
double short roots, we have
$$
\eqalign{
2m_{2/3} ^2 \wp _3 ' (\alpha \cdot x)
 = & 
\sum _{3\lambda \cdot \mu =1} 
C_{\lambda, \mu}^2 \wp ' _{\lambda \mu} (\alpha \cdot x)
+ 2 C_{\alpha, - \alpha} ^2 \wp _{\alpha, -\alpha} ' (2 \alpha \cdot x) \cr
& + \sum _{a=7} ^{6+\nu} \{ 
C_{\alpha, a} ^2 \wp _{\alpha, a} ' (\alpha \cdot x) +
C_{-\alpha, a} ^2 \wp _{-\alpha, a} ' (\alpha \cdot x) \}
 \cr
0 ~ = &
\sum _{3\lambda \cdot \mu =1} 
C_{\lambda, \mu}^2 \wp ' _{\lambda \mu} (\alpha \cdot x) (v_\lambda -v_\mu)
+ C_{\alpha, - \alpha}^2 \wp ' _{\alpha, -\alpha} (2 \alpha \cdot x) 
(v_\alpha 
- 
v_{-\alpha}) \cr
& + \sum _{a=7} ^{6+\nu} \{ 
C_{\alpha, a} ^2 \wp _{\alpha, a} ' (\alpha \cdot x) (v_\alpha - v_a) +
C_{-\alpha, a} ^2 \wp _{-\alpha, a} ' (\alpha \cdot x) (v_a - v_{-\alpha}) \}
.\cr}
\eqno (6.54)
$$
Using the linear independence of the vectors $v_a$ in the second line of 
(6.54), 
we readily find
$$
C_{\alpha ,a } ^2 \wp _{\alpha, a} (x) = C_{-\alpha ,a} ^2 \wp _{-\alpha ,a} 
(x).
\eqno (6.55)
$$
Projecting the second equation of (6.54) on the remaining $u_\lambda$ 
vectors, 
we obtain two sets of equations. Whenever $\lambda \cdot \mu = 1/3$, the 
equations are solved by
$$
\eqalign{
C_{\lambda , \mu} & = {1 \over \sqrt 3} m_{2/3} \cr
\Phi _{\lambda \mu} (x,z) & = \Phi _3 (x,z) \cr
\wp _{\lambda \mu} (x) & = \wp _3 (x). \cr }
\eqno (6.56)
$$
The functions $\wp _3$ and $\Phi _3$ are defined in (B.13) and (B.29).
We also obtain a set of coupled equations mixing the contributions from the 
short roots arising from the second and third lines in (6.51) and from the 
double short roots. These equations cannot be readily solved, unfortunately, 
so 
we shall leave them in their original form,
$$
{2 \over 3} m_{2/3}^2 \wp ' _3 (\alpha \cdot x)
= C_{\alpha , - \alpha} ^2 \wp ' _{\alpha , -\alpha} (2\alpha \cdot x)
+ \sum _a C_{\alpha , a} ^2 \wp ' _{\alpha , a} (\alpha \cdot x).
\eqno (6.57)
$$
To proceed further, we use $G_2$ Weyl invariance to set
$$
C_{\alpha, -\alpha} =m,
\qquad \qquad
\Phi _{\alpha, -\alpha} (x,z) = \psi (x,z),
\eqno (6.58)
$$
where $m$ and $\psi$ remain to be determined.

\medskip

It remains to work out condition (3) of Theorem 1. For general values of $\nu 
\geq 1$, this condition splits into three sets : (1) $I=\lambda, ~ J=\mu$, 
$I=\lambda,~ J=b$ (and its symmetric image $I=b$, $J=\lambda$), and 
$I=a,~J=b$, 
where $\lambda$ and $\mu$ are short roots of $G_2$ and $a,b=7,\cdots ,6+\nu$ 
label the extra zero weights. The equations in set (2) are linear in the 
coefficients $C_{\lambda, a}$, while the equations in set (3) are at least 
linear in the coefficients $C_{\lambda,a} $ and $C_{a,b}$. For the minimal 
value 
$\nu=0$, only the conditions in set (1) remain. (This value of $\nu$ is 
equivalent to setting $C_{\lambda,a}=C_{a,b}=0$.)

\medskip

Pursuing the analogy with $F_4$, we shall concentrate on the 6-dimensional 
Lax 
pair, with $\nu=0$, which involves by far the smallest number of unknown 
elliptic functions as well as the smallest number of equations, given by (for  
$\lambda \not= \mu$)
$$
\eqalign{
s C_{\lambda, \mu}  \Phi _{\lambda \mu} (\alpha \cdot x)   
d\cdot (u_\lambda - u_\mu) 
= \sum _{\kappa \not= \lambda, \mu} C_{\lambda , \kappa}  & C_{\kappa, \mu}
\{ 
\Phi _{\lambda \kappa} ((\lambda - \kappa ) \cdot x)
\Phi ' _{\kappa \mu} ((\kappa - \mu) \cdot x) \cr
& - \Phi _{\lambda \kappa}' ((\lambda - \kappa ) \cdot x)
\Phi  _{\kappa \mu} ((\kappa - \mu) \cdot x).  \}
\cr}
\eqno (5.59)
$$
This set of equations may be separated into the parts corresponding to 
$\lambda 
\cdot \mu = \pm 1/3$. It is convenient to partially solve these equations by 
setting
$$
s d\cdot u_\lambda (x) = b (\lambda \cdot x) + \sum _{{\delta ^2 =2; \atop 
\delta \cdot \lambda =1}} m_2 \wp (\delta \cdot x).
\eqno (5.60)
$$
Then, making use of the differential equations satisfied by $\Phi$ and $\Phi 
_3$, given in (B.30), we are left with just two equations to be obeyed by 
$\psi$ 
and by $b$. From the $\lambda \cdot \mu=1/3$ part, we have
$$
\eqalign{
\Phi _3 (x-y) (b(x) - b(y))
= ~ {m_2 \over 3} \{  
[\wp _3 (y) - \wp _3 (x)] \Phi _3 (x-y) + &
[\wp _3 ^- (y) - \wp _3 ^+ (x)] \Phi _3 ^+ (x-y)
\cr  + &
[\wp _3 ^+ (y) - \wp _3 ^- (x)] \Phi _3 ^- (x-y) 
\} \cr
 +  { \sqrt 3 m m_2 \over m_{2/3}} 
\{\psi (2x) \Phi ' (-x-y) - & \psi ' (2x) \Phi (-x-y) \cr
 -\psi (-2y) \Phi ' (x+y) + & \psi ' (-2y) \Phi (x+y) \},
\cr}
\eqno (5.61)
$$
where $\Phi _3 ^\pm$ and $\wp _3 ^\pm$ are defined in Appendix \S B.
{}From the $\lambda \cdot \mu = -1/3$ part, we have 
$$
\eqalign{
m_2 \Phi  (x-y) (b(x) - b(y))
= & ~ {m_{2/3} ^2 \over 3}  [\wp _3 (y) - \wp _3 (x)] \Phi _3 (x-y)  \cr
 & +  { m m_{2/3} \over \sqrt 3} 
\{\psi (2x) \Phi _3 ' (-x-y) -  \psi ' (2x) \Phi _3 (-x-y) \cr
& {} \qquad \qquad 
 -\psi (-2y) \Phi _3 ' (x+y) +  \psi ' (-2y) \Phi _3 (x+y) \}.
\cr}
\eqno (5.62)
$$
It is possible to show that condition (5.62) is a consequence of (5.61), 
provided we assume that the functions $b(x)$ and $\psi (x)$ are periodic with 
third period $2 \omega _1 /3$. We suspect that without such an assumption, 
(5.61) and (5.62) will be contradictory.  To establish this fact, it suffices 
to 
shift the arguments $x$ and $y$ by third periods, and to use the definitions 
of 
$\Phi _3$-functions and $\wp _3$-functions given in Appendix \S B.

\medskip

Thus, there remains a single equation (5.61) for $\psi (x)$ and $b(x)$. We 
believe, but we have not succeeded in proving, that a solution with the usual 
analyticity and monodromy properties exists.

\vfill\eject

\noindent
{\bf A. APPENDIX : LIE ALGEBRA THEORY}

\bigskip

In Table 1, we give the Dynkin diagrams for the finite dimensional simple Lie
algebras; for the untwisted affine Lie algebras (left
column) and for the twisted affine Lie algebras (right column). The simple
roots are labeled following Dynkin notation, and are given in an orthonormal
basis in Table 2, where we also list the dimension, the Coxeter and dual
Coxeter numbers (to be defined below). We list the set of all roots in Table 
3, and of the highest roots in Table 4. Below we provide additional notations 
and definitions [21,25].

\medskip

Let $\G$ be one of the finite dimensional simple Lie algebras of rank $n$,
let $\alpha _i$, and $\alpha _i ^\vee \equiv 2\alpha _i /\alpha _i ^2$,
$i=1,\cdots, n$ be its simple roots and coroots respectively. 
The coroot $\alpha ^\vee$ of any root is defined by $\alpha ^\vee = 2 \alpha 
/\alpha ^2$. Any (co-)root admits a unique decomposition into a sum of simple 
(co-)roots, with integer coefficients $l_i$ and $l_i ^\vee$.
$$
\alpha = \sum _{i=1} ^n l_i \alpha _i
\qquad \qquad
\alpha ^\vee = \sum _{i=1} ^n l_i ^\vee \alpha _i ^\vee.
\eqno (A.1)
$$
The coefficients $l_i$ and $l_i ^\vee$ are either all positive or all 
negative 
according to whether $\alpha$ (or $\alpha ^\vee$) is positive or negative 
respectively. They are related by
$$
l_i ^\vee = { \alpha _i ^2 \over \alpha ^2} l_i,
\qquad i=1,\cdots,n.
\eqno (A.2)
$$
The {\it highest root} $\alpha _0$ and co-root $\alpha _0 ^\vee$ play special 
roles. The extension of the simple root system of an algebra $\G$ by $\alpha 
_0$ generates the {\it untwisted affine Lie algebra} $\G ^{(1)}$, while the 
extension of the simple coroot system of $\G$ by $\alpha _0 ^\vee$ generates 
the {\it dual affine Lie algebra} $(\G ^{(1)})^\vee$. When $\G$ is non-simply 
laced, $(\G ^{(1)})^\vee$ coincides with one of the {\it twisted affine Lie 
algebras}. The Dynkin diagrams of these various Lie algebras are given in 
Table 
1. The decompositions of $\alpha _0$ and $\alpha _0 ^\vee$ onto roots or 
coroots 
$$
\alpha _0 
= \sum _{i=1} ^n a_i \alpha _i 
\qquad \qquad 
\alpha _0 ^\vee
= \sum _{i=1} ^n a _i ^\vee  \alpha _i ^\vee.
\eqno (A.3)
$$
define the {\it marks} $a_i$ and the {\it comarks} $a_i ^\vee$, which are 
given 
in Table 4.
The {\it Coxeter number} $h_\G$ and the {\it dual Coxeter number} $h_\G 
^\vee$
are defined by
$$
h_\G = 1 + \sum _{i=1} ^n a_i
\qquad \qquad
h _\G ^\vee = 1 + \sum _{i=1} ^n a_i ^\vee,
\eqno (A.4)
$$
and their values are given in Table 2.
For simply laced Lie algebras, for which all roots have the same length 
(normalized to $\alpha _i ^2=2$), we have $a_i ^\vee = a_i$ and $h_\G = h_\G 
^\vee$. The dual Coxeter number equals the {\it quadratic Casimir} 
operator in the adjoint representation, $h_\G ^\vee = C_2 (\G)$.

\medskip

The highest weight vectors $\lambda _j$, $j=1, \cdots ,n$ of the {\it 
fundamental
representations}  (also called fundamental weights) of $\G$ are defined by
$$
\alpha _i ^\vee \cdot \lambda _j = \delta _{ij}.
\eqno (A.5)
$$
The highest weight vector $\lambda$ of any finite dimensional representation
$\Lambda $ of $\G$ is then uniquely specified by positive or zero integers 
$q_i$, $i=1,\cdots ,n$, with
$$
\Lambda \equiv  (q_1, \cdots, q_n)
\qquad \qquad 
\lambda = \sum _{i=1} ^n q _i \lambda _i
\eqno (A.6)
$$
The Weyl orbit of the highest weight vector of $(q_1, \cdots,
q_n)$ is denoted by $[q_1, \cdots , q_n]$.

\medskip

The {\it level} ~$l(\lambda)$ and the {\it co-level} ~$l^\vee (\alpha)$ are 
defined by
$$
\eqalign{
l(\lambda ) = & ~ \lambda \cdot \rho ^\vee, 
\qquad  \qquad  
l(\alpha _i ) =  ~1, \qquad i=1,\cdots ,n; \cr
l^\vee (\lambda ) = & ~ \lambda \cdot \rho,  
\qquad  \qquad  
l ^\vee (\alpha _i ^\vee ) =  ~1, \qquad i=1,\cdots ,n. \cr
}
\eqno (A.7)
$$
Here, the {\it level vector} $\rho ^\vee$ is related to the {\it Weyl vector} 
$\rho$ 
by exchanging weights $\lambda _i$ and coweights $\lambda _i ^\vee = 2\lambda 
_i / \alpha _i ^2$. Both are uniquely determined by the above normalization, 
and may be expressed in terms of the fundamental weights and coweights by
$$
\eqalign{
\rho = & \sum _{i=1} ^ n \lambda _i = \half \sum _{\alpha \in \R _+ (\G)} 
\alpha
\cr
\rho ^\vee = & \sum _{i=1} ^ n \lambda _i ^\vee  = \half \sum _{\alpha ^\vee 
\in \R _+ (\G) ^\vee } \alpha ^\vee.
\cr}
\eqno (A.8)
$$
Here, we have provided the relation between the Weyl vector and 
the half sum of all positive roots, and its dual relation.
It is clear from (A.3), (A.4) and (A.8) that the Coxeter and dual Coxeter
numbers are related to the level of the highest root and the co-level of the 
highest coroot
$$
\eqalign{
h_\G =      & ~1+ \alpha _0 \cdot \rho ^\vee = 1+ l(\alpha _0)\cr
h_\G ^\vee =& ~1+ \alpha _0 ^\vee  \cdot \rho= 1+ l^\vee (\alpha _0 ^\vee) .  
\cr}
\eqno (A.9)
$$
As $\alpha$ (resp. $\alpha ^\vee$) ranges through $\R (\G)$ (resp. $\R (\G) 
^\vee$), the maxima of $l(\alpha)$ and $l^\vee (\alpha ^\vee)$ are $h_\G -1$ 
and 
$h^\vee _\G -1$ respectively.

\medskip

The {\it exponents} $\gamma _i $, $i=1,\cdots ,n$ are such that the 
numbers $\gamma _i+1$ are the degrees of the independent Casimir operators of 
the algebra $\G$. Their values are also listed in Table 4.

\vfill\eject

\centerline{\epsfxsize 8.0 truein \epsfbox[-20 80 600 660]{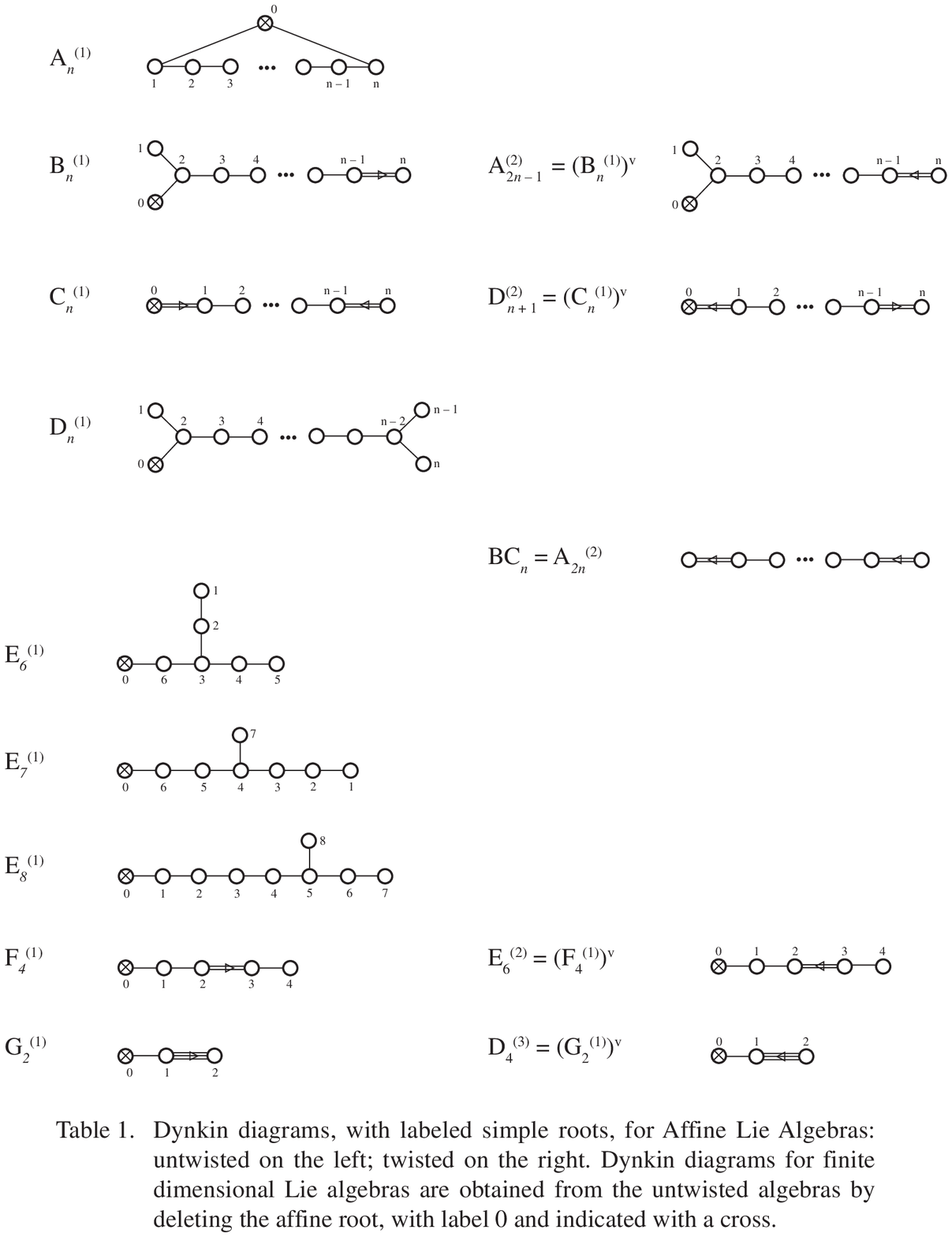}}

\vfill\eject

\noindent
{\bf  Table 2 : Basic data on finite dimensional simple Lie algebras}
 
\medskip

\settabs 8 \columns
\+ $\G$  & {\rm dim} $\G$ & $h_\G$ & $h_\G ^\vee$ &  simple  roots \cr
\+
-----------------------------------------------------------------------------
-
-------------------------------------- \cr
\+ $A_n$ & $n^2+2n$       & $n+1$  &  $n+1$      
   & $\alpha _i = e_i - e_{i+1}, i=1,\cdots, n.$ \cr
\+ $B_n$ & $2n^2+n$       & $2n$  &  $2n-1$      
   & $\alpha _i = e_i - e_{i+1}, i=1,\cdots, n-1;~\alpha _n=e_n.$ \cr
\+ $C_n$ & $2n^2+n$       & $2n$  &  $n+1$      
   & $\alpha _i = e_i - e_{i+1}, i=1,\cdots, n-1;~\alpha _n = 2 e_n.$ \cr
\+ $D_n$ & $2n^2-2n$       & $2n-2$  &  $2n-2$      
   & $\alpha _i = e_i - e_{i+1}, i=1,\cdots, n-1;~\alpha _n = e_{n-1} + e_n.$ 
\cr
\+ $E_6$ & $78$            & $12  $  &  $12  $ 
   & $\alpha _1 = \half (e_1 -e_2 - e_3 - e_4 -e_5 +\sqrt 3 e_6);$ \cr
\+ &&&     
   & $\alpha _i = e_i - e_{i-1}, i=2,\cdots, 5;~\alpha _6 = e_1 + e_2.$ \cr
\+ $E_7$ & $133$            & $18  $  &  $18  $ 
   & $\alpha _1 = \half (e_1 -e_2 - e_3 - e_4 -e_5 -e_6 +\sqrt 2 e_7);$ \cr
\+ &&&     
   & $\alpha _i = e_i - e_{i-1}, i=2,\cdots, 6;~\alpha _7 = e_1 + e_2.$ \cr
\+ $E_8$ & $248$            & $30  $  &  $30  $ 
   & $\alpha _1 = \half (e_1 -e_2 - e_3 - e_4 -e_5 - e_6 -e_7 +e_8);$ \cr
\+ &&&     
   & $\alpha _i = e_i - e_{i-1}, i=2,\cdots, 7;~\alpha _8 = e_1 + e_2.$ \cr
\+ $G_2$ & $14$            & $6  $  &  $4  $ 
   & $\alpha _1 = e_1 - e_2$; $\alpha _2 = \13 (-2 e_1 + e_2 +e_3).$ \cr
\+ $F_4$ & $52$            & $12  $  &  $9  $ 
   & $\alpha _1 = e_2 - e_3;~\alpha _2 = e_3 - e_4; ~ \alpha _3 = e_4;$ \cr
\+ &&& & $\alpha _4 = \half (e_1 -e_2 - e_3 - e_4).$ \cr

\bigskip

\noindent
The set of vectors $e_i$ forms an orthonormal basis. 

\bigskip
\bigskip

\noindent
{\bf  Table 3 : Root system of finite dimensional simple Lie algebras}
 
\medskip

\settabs 8 \columns
\+ $\G$  & all  roots \cr
\+
-----------------------------------------------------------------------------
-
-------------------------------------- \cr
\+ $A_n$      
   & $\pm (e_i - e_j), 1\leq i< j\leq n+1.$ \cr
\+ $B_n$   
   & $\pm (e_i - e_j); ~ \pm (e_i + e_j), ~1\leq i<j\leq n; 
    {} ~\pm e_i, ~1\leq i \leq n.$ \cr
\+ $C_n$       
   & $\pm (e_i - e_j); ~ \pm (e_i + e_j), ~1\leq i<j\leq n; 
    {} ~\pm 2e_i, ~ 1\leq i \leq n.$ \cr
\+ $D_n$  
   & $\pm (e_i - e_j), ~ \pm (e_i + e_j), 1\leq i<j\leq n.$ \cr
\+ $E_6$ 
   & $\pm (e_i - e_j), ~ \pm (e_i + e_j), ~1\leq i<j \leq 5;
    \pm \half (\sqrt 3 e_6 + \sum _{i=1} ^5 \epsilon _i e_i),~ \prod _i
    \epsilon _i =+1 .$\cr
\+ $E_7$ 
   & $\pm (e_i - e_j), ~ \pm (e_i + e_j), ~1\leq i<j \leq 6;
    \pm \half (\sqrt 2 e_7 + \sum _{i=1} ^6 \epsilon _i e_i),~ \prod _i
    \epsilon _i =-1 .$\cr
\+ $E_8$
   & $\pm (e_i - e_j), ~ \pm (e_i + e_j), ~1\leq i<j \leq 8;
    \half  \sum _{i=1} ^8 \epsilon _i e_i,~ \prod _i
    \epsilon _i =+1 .$\cr 
\+ $G_2$  
   & $\pm (e_i - e_j), ~ 1 \leq i<j\leq 3; ~\pm (e_i -\13 (e_1+e_2+e_3)),
    {}~ i=1,2,3.$ \cr
\+ $F_4$
   & $\pm (e_i - e_j), ~ \pm (e_i + e_j), ~1\leq i<j \leq 4;~
    \pm e_i, ~1\leq i\leq 4;
    \pm \half  \sum _{i=1} ^4 \epsilon _i e_i .$\cr

\vfill\eject

\noindent
{\bf Table 4 : Marks, Co-marks and Exponents}

\medskip

\settabs 5 \columns
\+ $\G$  & marks $(a_i)$ & comarks $(a_i ^\vee)$ & exponents $\gamma _i$ \cr
\+
-----------------------------------------------------------------------------
-
-------------------------------------- \cr
\+ $A_n$ & (1,1,1,$\cdots$,1,1)  & (1,1,1,$\cdots$,1,1)&  1,2,3,$\cdots$ ,n   
\cr
\+ $B_n$ & (1,2,2,$\cdots$,2,2)  & (1,2,2,$\cdots$,2,1)&  
1,3,5,$\cdots$,2n-1\cr
\+ $C_n$ & (2,2,2,$\cdots$,2,1)  & (1,1,1,$\cdots$,1,1)&  1,3,5,$\cdots$,2n-1 
\cr
\+ $D_n$ & (1,2,$\cdots$,2,1,1)  & (1,2,$\cdots$,2,1,1)&
1,3,5,$\cdots$,2n-3,n-1\cr
\+ $E_6$ & (1,2,3,2,1,2)       & (1,2,3,2,1,2)       &  1,4,5,7,8,11 \cr
\+ $E_7$ & (2,3,4,3,2,1,2)     & (2,3,4,3,2,1,2)     &  1,5,7,9,11,13,17 \cr 
\+ $E_8$ & (2,3,4,5,6,4,2,3)   & (2,3,4,5,6,4,2,3)   &  
1,7,11,13,17,19,23,29\cr
\+ $G_2$ & (2,3)               & (2,1)               &  1,5  \cr
\+ $F_4$ & (2,3,4,2)           & (2,3,2,1)           &  1,5,7,11\cr

\bigskip
\bigskip

\noindent
{\bf B. APPENDIX : ELLIPTIC FUNCTIONS}

\medskip

In this appendix we review some basic definitions and properties of elliptic
functions on an elliptic curve of periods $2\omega _1$ and $2\omega_2$ and
modulus $\tau = \omega _2 / \omega _1$, $\Im \tau >0$. The half periods are
$\omega _1$, $\omega _2$ and $\omega _3 = \omega _1 + \omega _2$. For a useful
source, see [24].

\bigskip

\noindent
{\bf (a) Basic Definitions and properties}

\medskip

The Weierstrass function is defined by
$$
\wp(z; 2\omega _1, 2 \omega _2) \equiv 
{1 \over z^2} + \sum _{{(m_1,m_2) \atop \not=(0,0)}}
\left \{ {1 \over (z + 2 \omega _1 m_1 + 2 \omega _2 m_2) ^2}
- {1 \over (2 \omega _1 m_1 + 2 \omega _2 m_2) ^2} \right \}
\eqno (B.1)
$$
and may alternatively be written as
$$
\wp (z; 2\omega _1 , 2 \omega _2)
= - {\eta _1 \over \omega _1} + \bigl ( { \pi \over 2 \omega _1} \bigr ) ^2
\sum _{n=-\infty} ^ \infty {1 \over \sinh ^2 {i \pi \over 2 \omega _1}
(z-2n\omega _2) },
\eqno (B.2)
$$
where 
$$
{\eta _1\over \omega _1} 
= - {1 \over 12} + \half \sum _{n=1} ^\infty
{1 \over \sinh ^2 i\pi n \tau}.
\eqno (B.3)
$$
The function $\wp $ is related to the Weierstrass functions $\zeta$ and
$\sigma$ by
$$
\wp (z; 2 \omega _1, 2 \omega _2) = - {d \over dz} \zeta (z; 2 \omega _1, 2
\omega _2) = - {d^2 \over dz^2} \log \sigma (z; 2 \omega _1, 2 \omega _2).
\eqno (B.4)
$$
These functions satisfy
$$
\eqalign{
\wp (-z) = \wp (z),
\qquad &
\wp (z+2\omega _a) = \wp (z)
\quad\qquad \qquad a=1,2,3
\cr
\zeta (-z) = - \zeta (z),
\qquad &
\zeta  (z+2\omega _a) = \zeta (z) + 2 \eta _a
\cr
\sigma (-z) = \sigma (z),
\qquad &
\sigma (z+2 \omega _a) =-\sigma (z) e^{2 \eta _a (z + 2 \omega _a)},
\cr}
\eqno (B.5)
$$
where $\eta _a = \zeta (\omega _a)$ and 
$$
\eqalign{
\sigma(z) &= z + {\cal O} (z^5) \cr
\zeta (z) &= {1 \over z} + {\cal O} (z^3) \cr
\wp (z) & = {1 \over z^2} + {\cal O}(z^2). \cr}
\eqno (B.6)
$$
The function $\sigma$ may be expressed in terms of the Jacobi
$\vartheta$-function
$$
\sigma (z; 2 \omega _1, 2 \omega _2)
=2 \omega _1 \exp \bigl ( {\eta _1 z^2 \over 2 \omega _1} \bigr )
{\vartheta _1 ({z \over 2 \omega _1} |\tau ) \over \vartheta _1 ' (0|\tau)},
\eqno (B.7)
$$
which in turn is defined in terms of
$$
q = e^{2 \pi i\tau} 
\qquad \qquad
v= {z \over 2 \omega _1}
\eqno (B.8)
$$
by
$$
\vartheta _1 (u|\tau)
=  2 q ^{1 \over 4} \sin \pi u \prod _{n=1} ^\infty
\bigl ( 1 - q ^n e^{2 \pi i u} \bigr )
\bigl ( 1 - q ^n e^{-2 \pi i u} \bigr )
\bigl ( 1 - q ^n  \bigr ).
\eqno (B.9)
$$
The function $\wp$ satisfies the differential equation
$$
\wp '(z) ^2 = 4 (\wp (z) - \wp (\omega _1) )(\wp (z) - \wp (\omega _2))
(\wp (z) - \wp (\omega _3)).
\eqno (B.10)
$$
These and additional properties of elliptic functions may be found in 
[24].

\bigskip

\noindent
{\bf (b) Half and Third Period Functions}

\medskip

Elliptic functions at half and third period (which are the only ones that we
shall need here, since the order of twisting is at most 3) are expressible in
terms of the original periods using Landen's transformations [24]. It is
convenient to make a definite choice for the period that is to be twisted; we
shall choose this period to be $\omega _1$. It is straightforward to adapt
these formulas when an arbitrary period $2 \omega _a$, $a=1,\cdots ,3$ is
twisted.

\medskip

\noindent
{\it Formulas for Twists of Order 2 : Elliptic Functions with Half Periods}

\medskip

Henceforth, we shall reserve the notation $\wp (z)$, $\zeta (z)$ and $\sigma
(z)$ for the corresponding Weierstrass functions with periods $2 \omega _1$ 
and
$2 \omega _2$, as defined in \S B (a). The elliptic functions at half period
$\omega _1$ are given by
$$
\eqalign{
\wp _2 (z) = \wp (z; \omega _1 , 2 \omega _2) 
=& \wp (z) + \wp (z+\omega _1) - \wp (\omega _1) \cr
=& ~{1 \over \wp (\omega _1)} \big [ \wp (z) \wp (z+\omega _1) -(\wp (\omega 
_1) -\wp (\omega _2) ) (\wp (\omega _1 ) - \wp (\omega _3)) \big ] \cr
\zeta _2 (z) = \zeta (z;\omega _1, 2 \omega _2) 
=& ~ \zeta (z) + \zeta (z+\omega _1) + z \wp (\omega _1) - \eta _1 \cr
\sigma _2 (z) = \sigma (z; \omega _1 , 2 \omega _2) 
=& ~ {\sigma (z) \sigma (z+ \omega _1) \over \sigma (\omega _1) }
e^{\half z^2 \wp (\omega _1) - z \eta _1} \cr}
\eqno (B.11)
$$
This gives rise to the duplication formula
$$
4\wp (2z) = \wp (z) + \wp (z+\omega _1) +\wp (z+\omega _2) + \wp (z+\omega _1 
+
\omega _2).
\eqno (B.12)
$$

\medskip

\noindent
{\it Formulas for Twists of order 3 : Elliptic Functions at Third Periods}

\medskip

Similarly, we have the following formulas for the third period elliptic
functions
$$
\eqalign{
\wp _3 (z) = \wp (z; 2\omega _1/3 , 2 \omega _2)= &
\ \wp (z) + \wp (z+2\omega _1/3) + \wp (z+4\omega _1/3) \cr 
 & - \wp (2\omega _1/3) -\wp (4 \omega _1 /3)
 \cr
\zeta _3 (z) = \zeta (z;2\omega _1/3 , 2 \omega _2) =&
\ \zeta (z) + \zeta (z+2\omega _1/3) + \zeta (z +4 \omega _1 /3) \cr 
 & + z \wp (2\omega _1/3) + z \wp (4 \omega _1 /3)  - \eta _1
\cr
\sigma _3 (z) = \sigma (z; 2\omega _1/3 , 2 \omega _2) =&
\ {\sigma (z) \sigma (z + 2\omega _1/3 ) \sigma (z + 4 \omega _1 /3) 
\over \sigma (2\omega _1/3) \sigma (4 \omega _1 /3)} 
e^{\half z^2 \wp (\omega _1) - z \eta _1}
\cr}
\eqno (B.13)
$$
This gives rise to the triplication formula
$$
9\wp (3z) 
= 
\sum _{j,k=0} ^2 
\wp (z+ j {2 \omega _1 \over 3} + k{2 \omega _2 \over 3}). 
\eqno (B.14)
$$
All of the above formulas may be established by identifying singularities in
$z$ and establishing that the remainder must be independent of $z$ by
Liouville's theorem.

\medskip

The functions at half and third periods, defined above are related to one
another in a way analogous to (B.4)
$$
\wp _\nu (z; 2 \omega _1, 2 \omega _2) 
= - {d \over dz} \zeta _\nu (z; 2 \omega _1, 2 \omega _2)
= - {d^2 \over dz^2} \log \sigma _\nu (z; 2 \omega _1, 2 \omega _2),
\eqno (B.15)
$$
where $\nu =1,2,3$.

\bigskip

\noindent
{\bf (c) The Function $\Phi$}

\medskip

We define the function $\Phi$ by
$$
\Phi (x,z) = 
\Phi (x,z; 2 \omega _1 , 2 \omega _2) =
{ \sigma (z-x) \over \sigma (z) \sigma (x) } e ^{x \zeta (z)},
\eqno (B.16)
$$
where $\sigma (z)$ and $\zeta (z)$ are the Weierstrass functions of (B.4) and 
(B.7).
As a function of $z$, $\Phi (x,z)$ is periodic with periods $2 \omega _1$ and 
$2 
\omega _2$, is holomorphic except for an essential singularity at $z=0$, and 
has 
a single zero at $z=x$. As a function of $x$, $\Phi (x,z)$ has multiplicative 
monodromy, given by
$$
\Phi (x+ 2 \omega _a,z ) = \Phi (x,z) e^{2 \omega _a \zeta (z) - 2 \eta _a 
z},
\eqno (B.17)
$$
is holomorphic in $x$ except for a simple pole at $x=0$, and has a single 
zero 
at $x=z$.
Some useful asymptotics are given as follows
$$
\eqalign{
\Phi (x,z) = & {1 \over x} - \half x \wp (z) + {\cal O}(x ^2 ) \cr
\Phi (x,z) = & \bigl \{ -{1 \over z} + \zeta (x) + {\cal O}(z) \bigr \} e^{x 
\zeta (z)}
\qquad z\to 0.
\cr}
\eqno (B.18)
$$
Products of the function $\Phi (x_\alpha, z)$, with $\sum _{\alpha =1} ^n
x_\alpha =0$, are periodic in $z$, with periods $2 \omega _1$ and $2 \omega 
_2$, 
and meromorphic in $z$ since the essential singularities cancel. As a result, 
they satisfy
$$
\prod _{\alpha =1} ^n \Phi (x_\alpha , z) = P_n[\wp (z);x_\alpha ] + \wp '(z)
Q_n [\wp (z); x_ \alpha], 
\eqno (B.19)
$$
where $P_n$ and $Q_n$ are polynomials of degrees $[{n\over 2}]$ and $[{n-3 
\over 2} ]$ in $\wp (z)$ respectively, with $x_\alpha$-dependent 
coefficients. 
The simplest case is 
$$
\Phi (x,z) \Phi (-x,z) =  \wp (z) - \wp (x) 
\eqno (B.20)
$$
In general, the polynomials $P$ and $Q$ may be determined by the fact that 
the
r.h.s. of (B.19) has a simple zero at each point $z=x_\alpha$, and that the 
pole highest order in $z$ has coefficient $(-1)^n$.

\medskip

The function $\Phi (x,z)$ satisfies a fundamental differential equation,
$$
\Phi (x,z) \Phi ' (y,z) - \Phi (y,z) \Phi '(x,z) 
=  (\wp (x) - \wp (y) ) \Phi (x+y,z), 
\eqno (B.21)
$$
where $\Phi '(x,z)$ denotes the derivative with respect to $x$ of $\Phi 
(x,z)$.

\bigskip

\noindent
{\bf (d) The Functions $\Lambda$, $\Phi _1$, $\Phi _2$ and $\Phi _3$}

\medskip

The functions $\Lambda$ and $\Phi _2$ are defined by
$$
\Lambda (2x,z)  = \Phi _2 (x,z) =
{\Phi (x,z) \Phi (x+\omega _1,z) \over \Phi (\omega _1,z)}.
\eqno (B.22)
$$
The essential singularity in $z$ and the monodromy in $x$ of $\Lambda (2x,z)$ 
coincide with those of $\Phi (x,z)$.
We shall need the following two basic differential equations,
$$
\eqalignno{
\Lambda  (2x,z) \Lambda ' (2y,z) - \Lambda '(2x,z) \Lambda (2y,z) 
& =  \half (\wp _2 (x) - \wp _2 (y) ) \Lambda (2x+2y,z), 
& (B.23a) \cr
\Phi _2 (x,z) \Phi _2 ' (y,z) - \Phi _2(y,z) \Phi _2'(x,z) 
& =  (\wp _2 (x) - \wp _2 (y) ) \Phi _2(x+y,z), 
& (B.23b) \cr}
$$
as well as differential equation that involves both $\Phi $ and $\Lambda$,
$$
\eqalign{
\Lambda (2x,z) \Phi  ' (-x-y,z) - \Lambda ' (2x,z) \Phi (-x-y,z) & \cr
-\Lambda (-2y,z) \Phi ' (x+y,z) + \Lambda ' (-2y,z) \Phi (x+y,z) &
=\half (\wp _2 (x) - \wp _2 (y) ) \Phi (x-y,z). \cr}
\eqno (B.24)
$$
By letting $y\to x$ in (B.23), and taking into account the known zeros of 
$\Lambda $, we derive another useful formula
$$
\Lambda (2x,z) \Lambda (-2x,z) = \wp _2 (z) - \wp _2 (x).
\eqno (B.25)
$$
Actually, $\Phi _2(x,z)$ may be viewed as the function $\Phi (x,z)$
associated with a torus of periods $\omega _1$ and $2 \omega _2$.

\medskip

The function $\Phi _1 (x,z)$ is defined by
$$
\eqalign{
\Phi _1(x,z) & = \Phi  (x,z) +  f(z) \Phi  (x+\omega _1,z) \cr
f(z) & = -e^{\pi i \zeta (z) +  \eta _1 z}.\cr}
\eqno (B.26)
$$
It obeys the monodromy relation $\Phi _1 (x+\omega _1, z)
= f(z) ^{-1} \Phi _1 (x,z)$, as well as the following differential equations
$$
\eqalignno{
\Phi _1 (x,z) \Phi _1 ' (y,z) - \Phi _1 ' (x,z) \Phi _1(y,z)
= & (\wp _2 (x) - \wp _2 (y) ) \Phi _1 (x+y,z) 
& (B.27a) \cr
\Phi _1 (x,z) \Phi  ' (y,z) - \Phi  (y,z) \Phi _1 '(x,z) 
= & \ \{ \wp (x+\omega _1) - \wp (y) \} \Phi _1 (x+y,z) \cr 
& + \{ \wp (x) - \wp (x+\omega _1) \} \Phi  (x+y,z), 
& (B.27b) \cr}
$$
and
$$
\eqalignno{
\Phi (2x,z) \Phi _1 ' (-x-y,z) - \Phi  ' (2x,z) \Phi _1 (-x-y,z) & \cr
-\Phi  (-2y,z) \Phi _1 ' (x+y,z) + \Phi  ' (-2y,z) \Phi _1 (x+y,z) & \cr
= (\wp  (2x) - \wp (2y) ) & \Phi _1 (x-y,z),
& (B.28a) \cr
\Lambda (2x,z) \Phi _1 ' (-x-y,z) - \Lambda ' (2x,z) \Phi _1 (-x-y,z) & \cr
-\Lambda (-2y,z) \Phi  _1 ' (x+y,z) + \Lambda ' (-2y,z) \Phi  _1(x+y,z) & \cr
=\half (\wp _2 (x) - \wp _2 (y) ) & \Phi _1 (x-y,z). 
& (B.28b) \cr}
$$

\medskip

Finally, we define functions of twist order 3. We introduce $\gamma = 2\omega 
_1/3$ so that
$$
\eqalign{
\Phi _3 (x,z) = & \Phi (x,z) + \Phi (x+\gamma,z) + \Phi (x+2\gamma,z), \cr
\Phi _3 ^\pm (x,z) = & \Phi (x,z) + e^{\mp \gamma} \Phi (x+\gamma,z) + e^{\pm 
\gamma} \Phi (x+2\gamma,z), \cr
\wp _3 (x) = & \wp (x) + \wp (x+\gamma) + \wp (x+2 \gamma) \cr
\wp _3 ^\pm  (x) = & \wp (x) + e^{\mp \gamma} \wp (x+\gamma) + e^{\pm \gamma} 
\wp (x+2 \gamma). \cr}
\eqno (B.29)
$$
These functions obey many natural differential equations, of which we shall 
only 
quote the one directly relevant here,
$$
\eqalign{
\Phi (x,z) \Phi ' _3 (y,z) - \Phi '(x,z) \Phi _3 (y,z)
=& \wp (x) \Phi _3 (x+y,z) - \13 \wp _3 (y) \Phi _3 (x+y,z) \cr
& - \13 \wp _3 ^-(y)  \Phi _3 ^+ (x+y,z) - \13 \wp _3 ^+ (y) \Phi _3 ^- 
(x+y,z).
\cr}
\eqno (B.30)
$$

\bigskip
\bigskip

\noindent
{\bf C. APPENDIX : COMPLETING THE PROOF FOR E$_{\bf 8}$}

\bigskip

In this appendix, we obtain (5.21b) and derive (5.21d) and (5.21e) from 
(5.31b) 
and the results
already obtained in the main section : (5.21a), (5.21c) and the assumed
solution to equations (5.22). Then, we show that the vastly overdetermined
system (5.31b) is satisfied by the solution of (5.21d,e).

\medskip

First, we obtain the general equations for $\Delta _{a,b}$ from (5.31b), by
substituting the solution (5.21c) into (5.31b). It is convenient to introduce
the quantities 
$$
\Delta _{\lambda , \mu} 
=  {1 \over 2 m_2 ^2} \sum _{a,b}  C_{\lambda , a} \Delta _{a,b} C_{b, \mu} 
\eqno (C.1)
$$
Equation (5.31b) may be recast in terms of $\Delta _{\lambda, \mu}$ as 
follows
$$
\eqalign{
\Delta _{\lambda , \mu} 
= & ~{1 \over 2 m_2} \biggl  ( \sum _{\lambda \cdot \delta =1}  \wp (\delta
\cdot x) + 2  \wp (\lambda \cdot x) \biggr ) C_\lambda \cdot C_\mu \cr
& \qquad  - {1 \over 2 m_2} \sum _{\kappa \cdot \lambda =1}  c(\lambda , 
\kappa)
C_\kappa \cdot C_\mu
\bigl ( \wp ((\lambda - \kappa )\cdot x) - \wp (\kappa \cdot x) \bigr ) \cr}
\eqno (C.2)
$$
which is more symmetrical in $\lambda $ and $\mu$, as can be seen from
using the symmetry properties of $C_{\lambda}$
$$
\Delta _{\lambda , \mu } 
= \Delta _{\mu , \lambda }
= \Delta _{-\lambda , - \mu} 
= - \Delta _{\lambda, - \mu} 
= - \Delta _{- \lambda , \mu}
\eqno (C.3)
$$

\medskip

Next, we make use of the basis of orthonormal roots $\beta _a$,
$a=1,\cdot , 8$, and label the zero weights of the {\bf 248} in terms of
this basis. As a result, the vectors $C_{\beta _a}$ form an orthogonal basis
of the space of $C_\lambda$. To show this, use (5.30) between different
$\beta_a$ and between any $\beta _a$ and any root $\lambda $ which does not
equal any of the $\pm \beta _b$. We then have the following inner product
relations 
$$
\eqalign{
C_{\beta _a} \cdot C_{\beta _b} = & ~ 2 m_2 ^2 \delta _{a,b} \cr
C_\lambda \cdot C_{\beta _a} = & ~ m_2 ^2 c(\lambda , \beta _a (\lambda \cdot
\beta _a)) ~ \lambda \cdot \beta _a \qquad \qquad (\lambda \not= \pm \beta 
_b) 
\cr}
\eqno (C.4)
$$
This gives us the projections of the vectors $C_\lambda$ onto the basis of
$\beta _a$, and determines $C_\lambda$ up to an undetermined vector $V$ which 
is 
orthogonal to all $C_{\beta _a}$. 
$$
C_\lambda = V_\lambda + \sum _{a=1} ^8 \half \lambda \cdot \beta _a ~ 
c(\lambda 
,
\beta _a (\lambda \cdot \beta _a)) ~ C_{\beta _a} .
\eqno (C.5)
$$
Using the relation $C_\lambda \cdot C_\lambda = 2 m_2 ^2$ in (5.30) and then
evaluating the same quantity directly from (C.5), we find that $V=0$, so that
the vectors $C_{\beta _a}$ indeed span a basis of the space of all 
$C_\lambda$.
With $V=0$, (C.5) precisely reproduces (5.21b) of Theorem 8.
Using the orthogonality of $C_{\beta _a}$ in (C.4) and equation (C.1), we
find 
$$
\Delta _{\beta _a, \beta _b} = \Delta _{a,b},
\eqno (C.6)
$$
evaluating (C.2) for $\lambda = \beta _a$ and $\mu = \beta _b$, with the help 
of
(5.30), we recover the expressions in (5.21d) and (5.21e). 
 
\medskip

It remains to show that the results of (5.21d,e) consistently solve (5.31b) 
for
all roots $\lambda$. Since the solution (5.21d,e) was derived for $\lambda = 
\pm \beta _a$ already, it suffices to study the cases $\lambda \not= \beta 
_a$,
$a=1, \cdots , 8$. The issue of consistency arises here because on the one 
hand,
$\Delta _{\lambda , \beta _b} $ may be evaluated directly from (C.1)  (we 
shall
denote this quantity by $ \Delta _{\lambda , \beta _b}$), while on the other
hand, the same quantity may be evaluated using the solution (5.21d,e) and the
expression (C.4) above (we shall denote this quantity by $\bar \Delta 
_{\lambda
,\beta _b}$, and the two need to be equal for consistency. Following this 
notation, we evaluate 
$$
\bar \Delta _{\lambda , \beta _b} 
= m_2 ^2 \lambda \cdot \beta _b ~ c(\lambda , \beta _b (\lambda \cdot \beta
_b)) ~ \Delta _{b,b}
+ m_2 ^2 \sum _{a \not= b} \lambda \cdot \beta _a ~ c(\lambda , \beta _a
(\lambda \cdot \beta _a)) ~ \Delta _{a,b}
\eqno (C.7)
$$ 
The quantity $\Delta _{\lambda , \beta _b}$, which is directly evaluated from
(C.2), equals
$$
\eqalign{
\Delta _{\lambda , \beta _b}
= & \half m_2 \lambda \cdot \beta _b ~ 
c(\lambda , \beta _b (\lambda \cdot \beta _b)) 
\biggl  ( \sum _{\lambda \cdot \delta =1}  \wp (\delta \cdot x) 
+ 2  \wp (\lambda \cdot x) \biggr ) \cr 
& - {1 \over 2 m_2} \sum _{\kappa \cdot \lambda =1} 
c(\lambda , \kappa) C _\lambda \cdot C _{\beta _b} \bigl ( \wp ((\lambda -
\kappa )\cdot x) - \wp (\kappa \cdot x) \bigr ) \cr }
\eqno (C.8)
$$
With the help of (5.22), we shall now show that (C.8) agrees with (C.7). 
Since
we restricted to $\lambda \not= \beta _a$, and using the symmetry properties
of (C.3), we are left to consider only the cases $\lambda \cdot \beta _a = 0$
and $\lambda \cdot \beta _a = 1$.

\medskip

\noindent
{\it The case $\lambda \cdot \beta _a =0$}

Eq. (C.8) may be evaluated using (5.30) and reduces to
$$
\eqalign{
\Delta _{\lambda , \beta _b}
= &  
 - \half m_2 \sum _{{\kappa \cdot \lambda =1 \atop \kappa \cdot \beta _b =1}} 
c(\lambda , \kappa)  ~ c(\kappa, \beta _b ) 
 \ \bigl ( \wp ((\lambda - \kappa )\cdot x) - \wp (\kappa \cdot x) \bigr ) 
\cr
 & +  
  \half m_2 \sum _{{\kappa \cdot \lambda =1 \atop \kappa \cdot \beta _b =-1}} 
c(\lambda , \kappa)  ~ c(\kappa, -\beta _b ) 
 \ \bigl ( \wp ((\lambda - \kappa )\cdot x) - \wp (\kappa \cdot x) \bigr ). 
\cr}
\eqno (C.9)
$$
Making a change of variables $\lambda - \kappa = \delta$ in the first term of
each sum, and letting $\kappa = \delta$ in the second, and regrouping terms, 
we
obtain
$$
\eqalign{
\Delta _{\lambda , \beta _b}
= &  
 + \half m_2 \sum _{{\delta \cdot \lambda =1 \atop \delta \cdot \beta _b =1}} 
\bigl ( c(\lambda , \delta)  ~ c(\delta, \beta _b ) 
+ c(\lambda , \lambda - \delta) ~ c(\lambda - \delta , - \beta _b) \bigr )
  \wp (\delta \cdot x)
\cr
 & -  
  \half m_2 \sum _{{\delta \cdot \lambda =-1 \atop \delta \cdot \beta _b =1}} 
\bigl ( c(\lambda , -\delta)  ~ c(\delta , \beta _b )
+ c(\lambda , \lambda + \delta) ~ c(\lambda + \delta , \beta _b) \bigr ) 
  \wp (\delta \cdot x) . 
\cr}
\eqno (C.10)
$$
The product relations between $\lambda$, $\beta _a$ and $\delta$ are 
precisely
such that we are allowed to use (5.22a) on the second term in each sum.  The
relations used are
$$
\eqalign{
c(\lambda , \lambda - \delta) ~ c(\lambda - \delta, - \beta _b)
= & c(\lambda , -\beta _b + \delta) ~ c(\beta _b + \delta , \beta _b) 
\qquad {\rm first ~ sum} \cr
c(\lambda , \lambda + \delta) ~ c(\lambda + \delta,  \beta _b)
= & c(\lambda , \beta _b - \delta) ~ c(\beta _b - \delta , \beta _b) 
\qquad {\rm second ~ sum}, \cr}
\eqno (C.11)
$$
and yield
$$
\eqalign{
\Delta _{\lambda , \beta _b}
= &  
 + \half m_2 \sum _{{\delta \cdot \lambda =1 \atop \delta \cdot \beta _b =1}} 
\bigl ( c(\lambda , \delta)  ~ c(\delta, \beta _b ) 
+ c(\lambda , \delta - \beta _b) ~ c(\beta _b - \delta ,  \beta _b) \bigr )
  \wp (\delta \cdot x)
\cr
 & -  
  \half m_2 \sum _{{\delta \cdot \lambda =-1 \atop \delta \cdot \beta _b =1}} 
\bigl ( c(\lambda , -\delta)  ~ c(\delta , \beta _b )
+ c(\lambda , \beta _b - \delta) ~ c(\beta _b - \delta , \beta _b) \bigr ) 
  \wp (\delta \cdot x) . 
\cr}
\eqno (C.12)
$$
Using (5.30) again, we rewrite the cocycle factors that involve $\lambda$ as
inner products with $C_\lambda$.
$$
\Delta _{\lambda , \beta _b}
=   
 + {1 \over 2 m_2} \sum _{ \delta \cdot \beta _b =1} 
\bigl (C _\lambda \cdot C_\delta  ~ c(\delta, \beta _b ) 
+ C_\lambda \cdot C_{\delta - \beta _b} ~ c(\beta _b - \delta , \beta _b) 
\bigr )
\wp (\delta \cdot x). 
 \eqno (C.13)
$$
In establishing equivalence of (C.12) and (C.13), we use the fact that terms 
with $\delta \cdot \lambda = \pm 2$ cannot contribute since they would imply
$\delta = \pm \lambda$, but this cannot be realized with $\lambda \cdot \beta
_b=0$ and $\delta \cdot \beta _b=1$. Next, we make use of (5.21b), and obtain
$$
\Delta _{\lambda , \beta _b} 
=\half \sum _{a=1} ^8 c(\lambda, \beta _a (\lambda \cdot \beta _a)) 
\eqno (C.14)
$$
with
$$
\Delta _{a,b}'
=   
 + {1 \over 2 m_2} \sum _{ \delta \cdot \beta _b =1} 
\bigl (C _{\beta _a} \cdot C_\delta  ~ c(\delta, \beta _b ) 
+ C_{\beta _a} \cdot C_{\delta - \beta _b} ~ c(\beta _b - \delta , \beta _b) 
\bigr )
\wp (\delta \cdot x). 
 \eqno (C.15)
$$
It remains to show that $\Delta _{a,b}'$ coincides with $\Delta _{a,b}$ of
(5.21). This is established by decomposing the sum over $\delta$ according to
the values of $\delta \cdot \beta _a$. The values $\delta \cdot \beta _a = 
\pm
2$ are not allowed, since already $\delta \cdot \beta _b=1$.
$$
\eqalign{
\Delta _{a,b} '
= &  
 + \half m_2 \sum _{{\delta \cdot \beta _a =1 \atop \delta \cdot \beta _b 
=1}} 
\bigl ( c(\beta _a , \delta)  ~ c(\delta, \beta _b ) 
+ c(\beta _a , \delta - \beta _b) ~ c(\beta _b - \delta ,  \beta _b) \bigr )
  \wp (\delta \cdot x)
\cr
 & -  
  \half m_2 \sum _{{\delta \cdot \beta _a =-1 \atop \delta \cdot \beta _b 
=1}} 
\bigl ( c(\beta _a , -\delta)  ~ c(\delta , \beta _b )
+ c(\beta _a , \beta _b - \delta) ~ c(\beta _b - \delta , \beta _b) \bigr ) 
  \wp (\delta \cdot x) . 
\cr}
\eqno (C.16)
$$
Using again the relations (5.22a), but this time for $\beta _a$, $\beta _b$ 
and
$\delta $, we see that this expression precisly reproduces (5.21d,e).

\medskip

\noindent
{\it The Case $\lambda \cdot \beta _b =1$}

The spirit of the proof of this case is completely analogous to that of the
case $\lambda \cdot \beta _b=0$. We start with (C.2) for $\mu = \beta _b$ ane
show that it is solved by (5.21d,e) for all roots $\lambda \not= \beta _a$,
$a=1,\cdots ,8$. To do so, let $\delta = \lambda - \kappa$ in the first term
of the second sum on the r.h.s. of (C.7), and $\delta = \kappa$ in the second
term, then we decompose these sums according to the values of
$\delta \cdot \beta _b$ and evaluate the inner products using (5.30).
$$
\eqalign{
\Delta _{\lambda , \beta _b} 
= & ~\half m_2 c(\lambda , \beta _b)
 \biggl  ( 
\sum _{\lambda \cdot \delta =1}   \wp (\delta \cdot x) 
+ 2  \wp (\lambda \cdot x)
-2 \wp ((\lambda - \beta _b) \cdot x) + 2 \wp (\beta _b \cdot x) \biggr )
\cr & 
- \half m_2  \sum _{{\delta \cdot \lambda =1 \atop \delta \cdot \beta _b=0}} 
c(\lambda , \lambda - \delta ) ~ c(\lambda - \delta , \beta _b) 
\wp (\delta \cdot x)
\cr & 
+ \half m_2  \sum _{{\delta \cdot \lambda =1 \atop \delta \cdot \beta _b=1}} 
c(\lambda ,  \delta ) ~ c( \delta , \beta _b) 
\wp (\delta \cdot x)
- \half m_2  \sum _{{\delta \cdot \lambda =1 \atop \delta \cdot \beta _b=-1}} 
c(\lambda ,  \delta ) ~ c( \delta , -\beta _b) 
\wp (\delta \cdot x)
\cr & 
+ \half m_2  \sum _{{\delta \cdot \lambda =1 \atop \delta \cdot \beta _b=2}} 
c(\lambda , \lambda - \delta ) ~ c(\lambda - \delta , -\beta _b) 
\wp (\delta \cdot x).
\cr}
\eqno (C.17)
$$
We now also decompose the first sum on the r.h.s. of (C.17) according to the
values of $\delta \cdot \beta _b$, 
$$
\sum _{\delta \cdot \lambda =1} \wp (\delta \cdot x)
=
\wp (\beta _b \cdot x) +
\sum _{{\delta \cdot \lambda =1, \atop \delta \cdot \beta _b =0}} \wp (\delta
\cdot x)
+
\sum _{{\delta \cdot \lambda =1, \atop \delta \cdot \beta _b =\pm 1}} \wp 
(\delta
\cdot x).
\eqno (C.18)
$$
Using relation (5.22b) on the terms woth $\delta \cdot \beta _b =0$, we find
that these terms cancel between the sums in the first and second lines in
(C.17). Rearranging the remaining terms, we find
$$
\eqalign{
\Delta _{\lambda , \beta _b} 
= & ~\half m_2 c(\lambda , \beta _b)
 \biggl  ( 
\sum _{{\lambda \cdot \delta =1, \atop \delta \cdot \beta _b = \pm 1}}   \wp
(\delta \cdot x)  
+ 2  \wp (\lambda \cdot x)
-2 \wp ((\lambda - \beta _b) \cdot x) + 2 \wp (\beta _b \cdot x) \biggr )
\cr & 
+ \half m_2  \sum _{{\delta \cdot \lambda =1 \atop \delta \cdot \beta _b=1}} 
c(\lambda ,  \delta ) ~ c( \delta , \beta _b) 
\wp (\delta \cdot x)
- \half m_2  \sum _{{\delta \cdot \lambda =1 \atop \delta \cdot \beta _b=-1}} 
c(\lambda ,  \delta ) ~ c( \delta , -\beta _b) 
\wp (\delta \cdot x).
\cr}
\eqno (C.19)
$$
The last two sums in (C.19) are rewritten with the help of (5.30) as
$$
-m_2 c(\lambda , \beta _b) \wp (\lambda \cdot x)
+ {1 \over 2 m_2} \sum _{\delta \cdot \beta _b=1}
C_\lambda \cdot C_\delta c(\delta , \beta _b) \wp (\delta \cdot x),
$$ 
so that, using (5.30) also on the first term, we have 
$$
\eqalign{
\Delta _{\lambda , \beta _b} 
= & ~{ 1 \over m_2} C_\lambda \cdot C_{\beta _b}
 \biggl  ( 
\sum _{{\lambda \cdot \delta =1, \atop \delta \cdot \beta _b = \pm 1}}   \wp
(\delta \cdot x)  
-2 \wp ((\lambda - \beta _b) \cdot x) + 2 \wp (\beta _b \cdot x) \biggr )
\cr & 
+ {1 \over 2 m_2}   \sum _{\delta \cdot \lambda =1 }
C_\lambda \cdot C_\delta ~ c(\delta , \beta _b) ~
\wp (\delta \cdot x).
\cr}
\eqno (C.20)
$$
Expressing $C_\lambda$ again with the help of (C.4), and decomposing the sums
over $\delta$ according to the values of $\delta \cdot \beta _a$, we find 
that
(C.20) reproduces (C.6), with $\Delta _{a,b}$ given in (5.21d,e). This
completes the proof of Theorem 8.

\bigskip
\bigskip

\centerline
{\bf ACKNOWLEDGEMENTS}

\bigskip

We have benefited from useful conversations with Elena Caceres, Ron Donagi 
and Igor Krichever. The first author wishes to thank Edward Witten for a 
generous invitation to the Princeton Institute for Advanced Study, where this 
research was initiated, as well as the Aspen Center for Physics. 
He would also like to acknowledge David Gross and the members of the Institute 
for Theoretical Physics in Santa Barbara for the hospitality extended to him 
while most of this work was being carried out.
Both authors would like to thank David Morrison, I.M. Singer and Edward Witten 
for inviting them to participate in the 1998 workshop on ``Geometry and 
Duality", at the Institute for Theoretical Physics.

\bigskip
\bigskip

\centerline
{\bf REFERENCES}

\bigskip

\item{[1]} Seiberg, N. and E. Witten, ``Electro-magnetic duality,
monopole condensation, and confinement in N=2 supersymmetric Yang-Mills
theory", Nucl. Phys. B 426 (1994) 19, hep-th/9407087.\hfill\break
Seiberg, N. and E. Witten, ``Monopoles, duality,
and chiral symmetry breaking in N=2 supersymmetric QCD",
Nucl. Phys. {\bf B431} (1994) 494, hep-th/9410167.

\item{[2]} Lerche, W., ``Introduction to Seiberg-Witten theory
and its stringy origin", Proceedings
of the {\it Spring School and Workshop in String Theory},
ICTP, Trieste (1996), hep-th/9611190; Nucl. Phys. Proc. Suppl. {\bf 55B} 
(1997) 83, and references therein.

\item{[3]} 
Gorski, A., I.M. Krichever, A. Marshakov, A. Mironov, A.
Morozov, ``Integrability and Seiberg-Witten Exact Solution", Phys. Lett. 
{\bf B355} (1995) 466, hep-th/9505035; 
\hfil\break
Matone, M., ``Instantons and Recursion Relations in N=2 SUSY Gauge 
Theories", 
Phys. Lett. {\bf  B357} (1996) 342, hep-th/9506102;
\hfil\break
Nakatsu, T. and K. Takasaki, ``Whitham-Toda Hierarchy and N=2 Supersymmetric 
Yang-Mills Theory", Mod. Phys. Lett. {\bf A 11}
(1996) 157-168, hep-th/9509162; ``Isomonodromic Deformations and 
Supersymmetric 
Gauge Theories", Int. J. Mod. Phys. {\bf A11} (1996) 5505, hep-th/9603069.

\item{[4]} Donagi, R. and E. Witten, ``Supersymmetric Yang-Mills
theory and integrable systems", Nucl. Phys. {\bf B460} (1996) 299-334,
hep-th/9510101.

\item{[5]} Martinec, E. and Warner, N., ``Integrable systems and
supersymmetric gauge theories", Nucl. Phys. {\bf B459} (1996) 97-112,
hep-th/9509161.

\item{[6]} Martinec, E., ``Integrable structures in supersymmetric
gauge and string theory", hep-th/9510204.

\item{[7]}
Sonnenschein, J., S. Theisen, and S. Yankielowicz, ``On the Relation between 
the Holomorphic Prepotential and the Quantum Moduli in SUSY Gauge Theories", 
Phys. Lett. {\bf B367} (1996) 145-150, hep-th/9510129.
\hfil\break
Eguchi, T. and S.K. Yang, ``Prepotentials
of N=2 supersymmetric gauge theories and soliton
equations", hep-th/9510183;
\hfil\break
Itoyama, H. and A. Morozov, ``Prepotential and the Seiberg-Witten theory", 
Nucl. Phys. {\bf B491} (1997) 529, hep-th/9512161; ``Integrability and 
Seiberg-Witten theory", hep-th/9601168; ``Integrability and Seiberg-Witten 
Theory : Curves and Periods", Nucl. Phys. {\bf B477} (1996) 855, 
hep-th/9511126;
\hfil\break
Ahn, C. and S. Nam, ``Integrable Structure in Supersymmetric Gauge Theories 
with Massive Hypermultiplets", Phys. Lett. {\bf B387} (1996) 304, 
hep-th/9603028;
\hfil\break
Krichever, I.M. and D.H. Phong, ``On the integrable geometry of 
soliton equations and N=2
supersymmetric gauge theories", J. Differential Geometry {\bf 45} (1997)
349-389, hep-th/9604199;
\hfil\break
Bonelli, G., M. Matone, ``Nonperturbative Relations in N=2 SUSY Yang-Milss 
WDVV 
Equation", Phys. Rev. Lett. {\bf 77} (1996) 4712, hep-th/9605090;
\hfil\break
Marshakov, A., A. Mironov, and A. Morozov,
``WDVV-like equations in N=2 SUSY Yang-Mills theory", Phys. Lett. {\bf B389} 
(1996) 43, hep-th/9607109;
\hfil\break
Marshakov, A. ``Non-perturbative quantum theories and integrable equations",
Int. J. Mod. Phys. {\bf A12} (1997) 1607, hep-th/9610242; 
\hfil\break
Nam, S. ``Integrable Models, Susy Gauge Theories and String Theory", Int. J. 
Mod. Phys. {\bf A12} (1997) 1243, hep-th/9612134;
\hfil\break
Marshakov, A., A. Mironov, A. Morozov, ``More Evidence for the WDVV
Equations in N=2 SUSY Yang-Mills Theories", hep-th/9701123;
\hfil\break
Marshakov, A., ``On Integrable Systems and Supersymmetric Gauge Theories",
Theor. Math. Phys. {\bf 112} (1997) 791, hep-th/9702083;
\hfil\break
Krichever, I.M. and D.H. Phong, ``Symplectic forms in the
theory of solitons", hep-th/9708170,
to appear in Surveys in Differential Geometry, Vol. {\bf III}.

\item{[8]} D'Hoker, E. and D.H. Phong, ``Calogero-Moser Systems in SU(N)
Seiberg-Witten Theory", Nucl. Phys. {\bf B513} (1998) 405, hep-th/9709053.

\item{[9]} Donagi, R., ``Seiberg-Witten Integrable Systems", alg-geom/9705010.
\hfil\break
Freed, D.S., ``Special K\"ahler Manifolds", hep-th/9712042;
\hfil\break
Carroll, R., ``Prepotentials and Riemann Surfaces'', hep-th/9802130.

\item{[10]} M. Adler and P. van Moerbeke, ``Completely integrable systems,
Euclidian Lie algebras, and curves", Advances in Math. {\bf 38} (1980) 
267-317;
\hfil\break
M. Adler and P. van Moerbeke, ``Linearization of Hamiltonian systems,
Jacobi varieties, and representation theory", Advances in Math. {\bf 38} 
(1980) 
318-379;
\hfil\break
M. Adler and P. van Moerbeke, ``The Toda lattice,
Dynkin diagrams, singularities and Abelian varieties",
Inventiones Math. {\bf 103} (1991) 223-278.

\item{[11]} Calogero, F.,
``Solution of the one-dimensional N-body problem with quadratic
and/or inversely quadratic pair potentials", J. Math. Physics {\bf 12} (1971)
419-436.
\hfill\break
Moser, J., ``Integrable systems of non-linear evolution equations",
in {\it Dynamical Systems, Theory and Applications},
J. Moser, ed., Lecture Notes in Physics 38 (1975) Springer-Verlag.

\item{[12]} Krichever, I.M., ``Elliptic solutions of the
Kadomtsev-Petviashvili equation and integrable systems
of particles", Funct. Anal. Appl. {\bf 14} (1980) 282-290.

\item{[13]} Hitchin, N., ``Stable bundles and integrable systems",
Duke Math. J. {\bf 54} (1987) 91.

\item{[14]} M.A. Olshanetsky and A.M. Perelomov, ``Classical integrable 
finite-dimensional systems related to Lie algebras'', Phys. Rep. {\bf 71C} 
(1981) 313-400.
\hfil\break
Perelomov, A.M., {\it ``Integrable Systems of Classical Mechanics
and Lie Algebras"}, Vol. I, Birkh\"auser  (1990), Boston; 
and references therein.
\hfil\break
Leznov, A.N. and M.V. Saveliev, {\it Group Theoretic Methods for Integration 
of Non-Linear Dynamical Systems}, Birkhauser 1992.

\item{[15]} D'Hoker, E. and D.H. Phong, ``Calogero-Moser and Toda Systems
for Twisted and Untwisted Affine Lie Algebras", April 1998 preprint, 
hep-th/9804125.

\item{[16]} D'Hoker, E. and D.H. Phong, ``Spectral Curves for 
Super-Yang-Mills with Adjoint Hypermultiplet for General Lie Algebras", April 
1998 preprint, hep-th/9804126.

\item{[17]} V.I. Inozemtsev, ``The finite Toda lattices'', Comm. Math. Phys. 
{\bf 121} (1989) 629-638.

\item{[18]} V.I. Inozemtsev, ``Lax representation with spectral parameter on 
a torus for integrable particle systems'', Lett. Math. Phys. {\bf 17} (1989) 
11-17.

\item{[19]} Klemm, A., W. Lerche, and S. Theisen, ``Non-perturbative
actions of N=2 supersymmetric gauge theories",
Int. J. Mod. Phys. {\bf  A11} (1996) 1929-1974, hep-th/9505150.
 
\item{[20]} D'Hoker, E., I.M. Krichever, and D.H. Phong,
``The effective prepotential for N=2 supersymmetric
SU($N_c$) gauge theories", Nucl. Phys. {\bf B 489} (1997) 179-210,
hep-th/9609041;
\hfil\break
D'Hoker, E., I.M. Krichever, and D.H. Phong,
``The effective prepotential
for N=2 supersymmetric SO($N_c$) and Sp($N_c$) gauge theories",
Nucl. Phys. {\bf B 489} (1997) 211-222, hep-th/9609145;
\hfil\break
D'Hoker, E., I.M. Krichever, and D.H. Phong,
``The renormalization group equation for N=2 supersymmetric
gauge theories", Nucl. Phys. {\bf B 494} (1997), 89-104, hep-th/9610156.
\hfil\break
D'Hoker, E. and D.H. Phong,
``Strong coupling expansions in $SU(N)$ Seiberg-Witten theory",
hep-th/9701151, Phys. Lett. {\bf B 397} (1997) 94-103.

\item{[21]} Kac, V., {\it ``Infinite-dimensional Lie algebras"},
Birkh\"auser (1983) Boston.

\item{[22]} Goddard, P. and D. Olive, ``Kac-Moody and Virasoro algebras
in relation to quantum physics", International J. of Modern Physics A,
Vol. I (1986) 303-414.

\item{[23]} Krichever, I.M., ``Elliptic solutions of the
Kadomtsev-Petviashvili equation and integrable systems
of particles", Funct. Anal. Appl. {\bf 14} (1980) 282-290.

\item{[24]} Erdelyi, A., ed., {\it ``Higher Transcendental Functions''},
Bateman Manuscript Project, Vol. II, R.E. Krieger (1981) Florida.

\item{[25]} W.G. McKay, J. Patera and D.W. Rand, ``Tables of Representations
of Simple Lie Algebras'', Vol. I : Exceptional Simple Lie algebras,
Centre de Math\'ematiques, Universit\'e de Montr\'eal, 1990.

\item{[26]} Kachru, S., C. Vafa, ``Exact Results for N=2 Compactifications 
of 
Heterotic Strings", Nucl. Phys. {\bf B450} (1995) 69, hep-th/9505105;
\hfil\break
Bershadsky, M., K. Intrilligator, S. Kachru, D.R. Morrison, V. Sadov and C. 
Vafa, ``Geometric Singularities and Enhanced Gauge Symmetries", Nucl. Phys. 
{\bf B481} (1996) 215, hep-th/9605200;
\hfil\break
Katz, S., A. Klemm, C. Vafa, ``Geometric Engineering of Quantum Field 
Theories", 
Nucl. Phys. {\bf B497} (1997) 173, hep-th/9609239;
\hfil\break
Katz, S., P. Mayr, and C. Vafa, ``Mirror symmetry
and exact solutions of 4D N=2 gauge theories", Adv. Theor. Math. Phys. {\bf 
1} 
(1998) 53, hep-th/9706110.

\item{[27]} Witten, E., ``Solutions of four-dimensional
field theories via M-theory", Nucl. Phys. {\bf B500} (1997) 3, 
hep-th/9703166.

\item{[28]} Hanany, A., and E. Witten, ``Type IIB Superstrings, BPS 
Monopoles, 
and Three-Dimensional Gauge Dynamics", Nucl. Phys. {\bf B492} (1997) 152.

\item{[29]} Brandhuber, A., J. Sonnenschein, S. Theisen and S. Yankielowicz, 
``M-Theory and Seiberg-Witten Curves : Orthogonal and Symplectic Groups",
Nucl. Phys. {\bf B504} (1997) 175, hep-th/9705232.
\hfil\break
Landsteiner, K., E. Lopez, ``New Curves From Branes", hep-th/9708118;
\hfil\break
Landsteiner, K., E. Lopez, DA. Lowe, ``N=2 Supersymmetric Gauge Theories,
Branes and Orientifolds", Nucl. Phys. {\bf B507} (1997) 197, hep-th/9705199;
\hfil\break
Uranga, A.M., ``Towards Mass Deformed N=4 SO(N) and Sp(K) Gauge Theories 
from 
Brane Configurations", hep-th/9803054;
\hfil\break
Yokono, T., ``Orientifold four plane in brane configurations and N=4 
USp(2N) and SO(2N) theory'', hep-th/9803123.

\item{[30]}
Gorskii, A., ``Branes and Integrability in the N=2 SUSY YM Theory", Int. J. 
Mod. Phys. {\bf A12} (1997) 1243, hep-th/9612238;
\hfil\break
Gorskii, A., S. Gukov and A. Mironov, ``Susy Field Theories, Integrable 
Systems 
and their Stringy/Brane Origin", hep-th/9710239;
\hfil\break
Cherkis, S.A., A. Kapustin, ``Singular Monopoles and Supersymmetric 
Gauge Theories in Three Dimensions", hep-th/9711145.

\item{[31]} 
Braden, H.W. ``R-Matrices, Generalized Inverses and Calogero-Moser-Sutherland 
Models", to appear in the {\it Proceedings of the Workshop on
Calogero-Moser-Sutherland Models}, in the CRM Series in Mathematical Physics,
Springer-Verlag; available from http://www.maths.ed.ac.uk/preprints/97-017.

\end